\documentclass{jfm}
\usepackage{graphicx}
\usepackage{natbib}
\usepackage{amsmath}
\usepackage{amssymb}
\usepackage{latexsym}
\usepackage{wasysym}
\usepackage{multirow}
\usepackage{color}
\usepackage{subfigure}
\usepackage{pdflscape}
\usepackage{multirow}

\usepackage{mathtools}
\usepackage[utf8]{inputenc}
\usepackage{float}
\usepackage[version=4]{mhchem}
\usepackage{longtable,tabularx}
\usepackage{gensymb}
\DeclareGraphicsExtensions{.eps}

\usepackage [english]{babel}
\usepackage [autostyle, english = american]{csquotes}
\MakeOuterQuote{"}

\newcommand{\bea}{\begin{equation}\begin{aligned}}
\newcommand{\eea}{\end{aligned}\end{equation}}

\shorttitle{DNS of turbulent channel flow over random rough surfaces}
\shortauthor{R. Ma, K. Alam{\'e} and K. Mahesh}

\title{Direct numerical simulation of turbulent channel flow over random rough surfaces }

\author{Rong Ma\aff{1},
  Karim Alam{\'e}\aff{1}
 \and Krishnan Mahesh\aff{1}
 \corresp{\email{kmahesh@umn.edu}}}

\affiliation{\aff{1}Department of Aerospace Engineering and Mechanics, University of Minnesota,
Minneapolis, MN 55455, USA}

\begin{document}

\maketitle

\begin{abstract}
Direct numerical simulation (DNS) of flow in a turbulent channel with a random-rough bottom wall is performed at friction Reynolds number $Re_{\tau}=400$ and $600$. The rough surface corresponds to the experiments of \cite{flack2019skin}. The computed skin friction coefficients and the roughness functions show good agreement with experimental results. The double-averaging methodology is used to investigate mean velocity, Reynolds stresses, dispersive flux, and mean momentum balance. The roll-up of the shear layer on the roughness crests is identified as a primary contributor to the wall-normal momentum transfer. The mean-square pressure fluctuations increase in the roughness layer and collapse onto their smooth-wall levels away from the wall. The dominant source terms in the pressure Poisson equation are examined. The rapid term shows that high pressure fluctuations observed in front of and above the roughness elements are mainly due to the attached shear layer formed upstream of the protrusions. The contribution of the  slow term is relatively smaller. The slow term is primarily increased in the troughs and in front of the roughness elements, corresponding to the occurrence of quasi-streamwise vortices and secondary vortical structures. The mean wall shear on the rough surface is highly correlated with the roughness height, and depends on the local roughness topography. The p.d.f of wall-shear stress fluctuations are consistent with higher velocities at roughness crests and reverse flow  in the valley regions. Extreme events are more probable due to the roughness. Events with comparable magnitudes of the streamwise and spanwise wall-shear stress occur more frequently, corresponding to a more isotropic vorticity field in the roughness layer. 

\end{abstract}

\begin{keywords}
Turbulent Channel Flow, Turbulent Boundary Layer, Random Rough Surfaces, Rough Walls, DNS  
\end{keywords}

\section{Introduction}
Wall roughness has important effects on turbulent flows, especially at high Reynolds numbers. Surfaces with riblets enable drag reduction \citep{bechert1997experiments}, sediment and vegetation canopies affect the near-bed region \citep{mignot2009double}, and "urban roughness" influences the urban climate \citep{cheng2002near}. Many processes such as erosion, pitting \citep{bons2001many} and bio-fouling \citep{kirschner2012bio}  produce complex surface topographies, which decrease the efficiency of engineering systems.

\cite{raupach1991rough} and \cite{jimenez2004turbulent} have summarized the effects of roughness  on turbulent boundary layers. They describe the offset of the mean velocity profile, the enhancement of turbulent intensities, and the modification of flow structures. The roughness function $\Delta U^+= \Delta U u_{\tau}/\nu$, where $\Delta U$ is the difference in mean velocity between smooth and rough walls in the logarithmic layer, $u_{\tau}$ is the average friction velocity and $\nu$ is the kinematic viscosity of the fluid. Based on experiments by \cite{nikuradse1933laws} for sand-grain roughness, three flow regimes are defined for rough-wall flows by expressing $\Delta U^+$ as a function of $k^+$, the roughness scale in viscous units. When $k^+$ is small, $\Delta U^+$ is nearly zero, i.e., the flow is hydraulically smooth. In this regime, the viscosity damps out the perturbations caused by the roughness. The flow becomes transitionally rough as $k^+$ increases, where the skin friction has contributions from both viscous drag and form drag. As $k^+$ further increases, the roughness function reaches a linear asymptote, and the flow is considered fully rough. 

The roughness scale is an important parameter and different definitions exist, such as the average roughness height $k_a^+$, the peak-to-valley roughness height $k_t^+$, and the equivalent sand-grain roughness height $k_s^+$. According to \cite{nikuradse1933laws}, $k_s^+$ is determined by fitting a roughness height to match the measured pressure drop in experiments. \cite{jimenez2004turbulent} and \cite{flack2010review} suggest that $k_s^+$ can provide good collapse of $\Delta U^+$ in the fully rough regime for various types of roughness. Correlations to predict the frictional drag for rough surfaces are summarized by \cite{flack2010review}, who also propose a new correlation to predict $k_s$ in the fully rough regime. However,  data of $\Delta U^+$ in the transitionally rough regime shows considerable scatter for different roughness types. \cite{flack2010review} note that the transitionally rough regime is the least understood, and the parameter ranges that determine the transitionally rough regime remain unknown for most roughness types. \cite{barros2017measurements} measured the skin friction for systematically-controlled random rough surfaces and concluded that the understanding of the frictional drag in the transitionally rough regime is poor.

Surface roughness can be categorized as being regular or irregular/random. Many experimental and computational studies \citep{schlichting1936experimentelle, bechert1997experiments, orlandi2006dns, lee2011direct} have examined the effects of  regular roughness involving ribbed, cubed, or spherical elements. Recently, more attention has been paid to irregular rough surfaces \citep{mejia2010low, cardillo2013dns, yuan2014, busse2015direct, anderson2015numerical}. The effects of irregular roughness on velocity profiles, turbulent intensities, turbulent kinetic energy, and two-point correlations, have been examined. \cite{flack2005experimental} and \cite{wu2007outer} found that for  small roughness heights (relative to the boundary layer thickness or channel half-height), "outer-layer similarity" \citep{townsend1980structure} holds and turbulent statistics in the outer layer are not directly affected by the roughness. The near-wall region where the roughness effects on the mean flow are significant is termed "roughness sublayer". \cite{busse2017reynolds} investigated $Re$ dependence of the near-wall flow in the vicinity of, and within the rough surfaces. They characterized the near-wall flow by estimating the thickness of the roughness sublayer and examining the probability distribution of the reverse flow. \cite{jelly2018reynolds} studied the dependence of the near-wall flow on higher-order surface parameters, such as skewness, by evaluating the influence on the roughness function. 

Rough surfaces produce statistically inhomogeneous flow fields in the roughness sublayer on the length-scale of the roughness. The multiple length scales present in random rough surfaces can cause the inhomogeneity to be quite complex. The "double-averaging" (DA) decomposition, first introduced by \cite{raupach1982averaging} to examine the "wake production" term within vegetation canopies, is used to describe this spatial inhomogeneity in the time-averaged flow field. Section \ref{sec:problem description} describes the DA decomposition as  used in the present work. While standard Reynolds-averaging yields the Reynolds stress, the DA decomposition yields a "dispersive stress" which represents the contribution of spatially correlated time-averaged flow to momentum transport.  Dispersive stresses  for  regular roughness were studied by \cite{cheng2002near}, \cite{coceal2007spatial} and  \cite{bailey2013turbulence}.
The ratio of maximum dispersive stress to Reynolds stress was found by
\cite{forooghi2018direct} to be highly dependent on the skewness and effective slope, for irregular rough surfaces. \cite{yuan2018topographical} note the effects of large surface scales on the dispersive stress from their DNS. \cite{jelly2019reynolds} examine the Reynolds number dependence of the dispersive stresses for irregular near-Gaussian rough surfaces.


Past work has largely focused on the mean flow and the scaling of velocity fluctuations over rough surfaces. Less is known about how roughness affects  pressure and wall-shear stress fluctuations, both of which are closely related to form and frictional drag, sound radiation, and structural vibration. Since pressure  satisfies a  global Poisson equation,  arguments based on the local length and velocity scales that  work well for velocity, do not work very well for pressure fluctuations. \cite{chang1999relationship} investigated the contributions of velocity-field sources to the fluctuating wall pressure in smooth channel flow by computing partial pressures. \cite{panton2017correlation} studied pressure fluctuations using DNS data sets at $Re_{\tau}$ ranging from $180-5200$ and observed a contribution from the low-wavenumber range to the wall pressure fluctuations for $Re_{\tau}>1000$. \cite{anantharamu2020analysis} analyzed the sources of wall-pressure in a smooth wall turbulent channel flow at $Re_{\tau}=180$, and $400$.  Using spectral POD, they identified the features of the wall-pressure sources responsible for the linear and pre-multiplied frequency peaks in the power spectrum. Their analysis revealed the importance of buffer layer sources to the high frequency/wavenumber region of the pre-multiplied spectrum, at their Reynolds numbers. For rough-wall flows, \cite{bhaganagar2007effect} performed DNS for periodic roughness elements and found that the pressure statistics are altered significantly in the inner region of the channel. \cite{meyers2015wall} studied the wall-pressure spectrum over rough walls experimentally and suggested that different scalings were needed to collapse different frequencies.

Studies in smooth wall flows show that wall-shear stress fluctuations are correlated with turbulence structures in the boundary layer. Smooth-wall experiments  by \cite{Khoo2001} note the close similarity between the probability distribution function (p.d.f) of wall-shear stress and streamwise velocity fluctuations.  The DNS of \cite{abe2004very} found that positive and negative- dominant $\tau_{yx}$ correspond to  high- and low-speed regions in the very large-scale motions in the outer layer. \cite{orlu2011fluctuating} note that  footprints of the near-wall streaks are visible in $\tau_{yx}$ as regions of higher and lower shear. \cite{diaz2017wall} examined the angle $\phi_{\tau}$  between the instantaneous wall-shear stress and the streamwise direction, and found that events with $\phi_{\tau}$ much higher than $90 \degree$ were extremely unlikely to occur, indicating that  negative  $\tau_{yx}$ is generally accompanied by high values of $\tau_{yz}$. Similar studies on rough-wall flows are limited. Roughness is expected to affect the near-wall motions and therefore, the wall shear stress.  

The complex geometry and multiple length scales present in realistic rough surfaces induce both local and global effects on the dispersive fluxes, wall pressure, and shear stress fluctuations. The direct measurement of these variables is challenging, therefore, simulations provide a useful complement to experiment.
This paper performs DNS of turbulent channel flow at two $Re_{\tau}$ over realistic random rough surfaces under the same conditions as the experiment. The simulations reproduce the skin-friction coefficient measured in the experiments. We use double-averaging to explore the effects of roughness and $Re_{\tau}$ on the mean velocity, Reynolds stresses, dispersive flux, correlations between form-induced velocity and pressure, and the mean momentum balance. We use the DNS data and the pressure Poisson equation to study how roughness affects the pressure fluctuations. We characterize the local variation and statistics of wall-shear stress fluctuations.

The numerical method and validations of the DNS solver are introduced in \S \ref{sec:simulation details}. The surface processing, problem description and grid convergence are shown in \S \ref{sec:problem setup}. The results and discussions are presented in \S \ref{sec:results}. Finally, the paper is summarized in \S \ref{sec:conclusions}. 

\section{Simulation details}\label{sec:simulation details}
\subsection{Numerical method}\label{sec:numerical method}
The governing equations are solved using the finite volume algorithm developed by \citet{mahesh2004} for the incompressible Navier-Stokes equations. The governing equations for the momentum and continuity equations are given by the Navier-Stokes equations:
\begin{equation}
\frac{\partial u_i}{\partial t} + \frac{\partial}{\partial x_j}(u_iu_j)=-\frac{\partial p}{\partial x_i} + \nu\frac{\partial^2 u_i}{\partial x_i x_j}+K_i,~~\frac{\partial u_i}{\partial x_i} = 0,
\label{eqn:nsme}
\end{equation}
where $u_i$ and $x_i$ are the $i$-th component of the velocity and position vectors respectively, $p$ denotes pressure divided by density, $\nu$ is the kinematic viscosity of the fluid and $K_i$ is a constant pressure gradient (divided by density). Note that the density is absorbed in the pressure and $K_i$.
The algorithm is robust and emphasizes discrete kinetic energy conservation in the inviscid limit which enables it to simulate high-Re flows without adding numerical dissipation. A predictor-corrector methodology is used where the velocities are first predicted using the momentum equation, and then corrected using the pressure gradient obtained from the Poisson equation yielded by the continuity equation. The Poisson equation is solved using a multigrid pre-conditioned conjugate gradient method (CGM) using the Trilinos libraries (Sandia National Labs). The implicit time advancement uses the second-order Crank-Nicolson discretization:
\begin{equation}
    \frac{\hat{u}_i-u_i^n}{\Delta t}=\frac{1}{2}[(NL+VISC)^{n+1}+(NL+VISC)^{n}] \, ,
\end{equation}
where the face normal velocities $V^{n+1}_N$ are linearized in time (time-lagged) such that $V^{n}_N$ is used instead; the linearization in time yields an error of $O(\Delta t^2)$, which is the same order as that of the overall scheme. All the terms expressed as $\hat{u}_i$ are taken to the left hand side and a system of equations is solved using SOR until convergence.


The geometry of the rough surface is generated from highly resolved Cartesian line scans obtained from the experiment.  The Cartesian line scans are first used to compute Fourier spectra, surface statistics, and p.d.fs of the surface height. Based on the spectra, the computational mesh in the $x-z$ plane is chosen and the surface is approximated on the computational mesh in the $x-z$ plane. The height distribution determines the computational mesh in the $y$ direction. The surface is represented using a mask function that is defined to be one in the fluid and zero in the solid. We ensure that the p.d.fs, statistics, and spectra of the masked surface agree acceptably with the experimental scan. The masked surface is used to perform the simulations reported in the paper. Details are presented in \S \ref{sec:surface data processing}. No-slip Dirichlet boundary conditions are enforced on the face velocities of the rough surface in the computations. Figure \ref{fig:surface_profile} shows the face velocities on the boundaries for a portion of the rough surface. Note that the face velocities are zero on the boundaries. This  methodology has been validated and used in past work to study idealized superhydrophobic surfaces \citep{li2017feature} and realistically rough superhydrophobic surfaces \citep{alame2019wall}.


\subsection{Validation}\label{sec:validation}
\label{appA}
  \begin{table}
  \begin{center}
\def~{\hphantom{0}}
    \begin{tabular}{lccccccccc}
    Turbulent channel flow & Case & $Re_{\tau}$ & $N_x\times N_y\times N_z$ & $L_x\times L_y\times L_z$ & $\Delta x^{+}$ & $\Delta z^{+}$ & $\Delta y^{+}_{min}$ & $\Delta y^{+}_{max}$\\[3pt]\hline
    Smooth wall & SW & 400 & $768\times 320\times 384$ & $2\pi\times 2\times \pi$ & $3.27$ & $3.27$ & $0.85$ & $5.48$\\
    Rod-roughened wall & RRW & 400 & $768\times 320\times 320$ & $6.528\times 2\times \pi$ & $3.40$ & $3.92$ & $0.85$ & $5.48$\\\hline
    \end{tabular}
    \caption{\label{tab:validation} Simulation parameters for the validation problems.}
    \end{center}
\end{table}

\begin{figure}
\includegraphics[height=60mm]{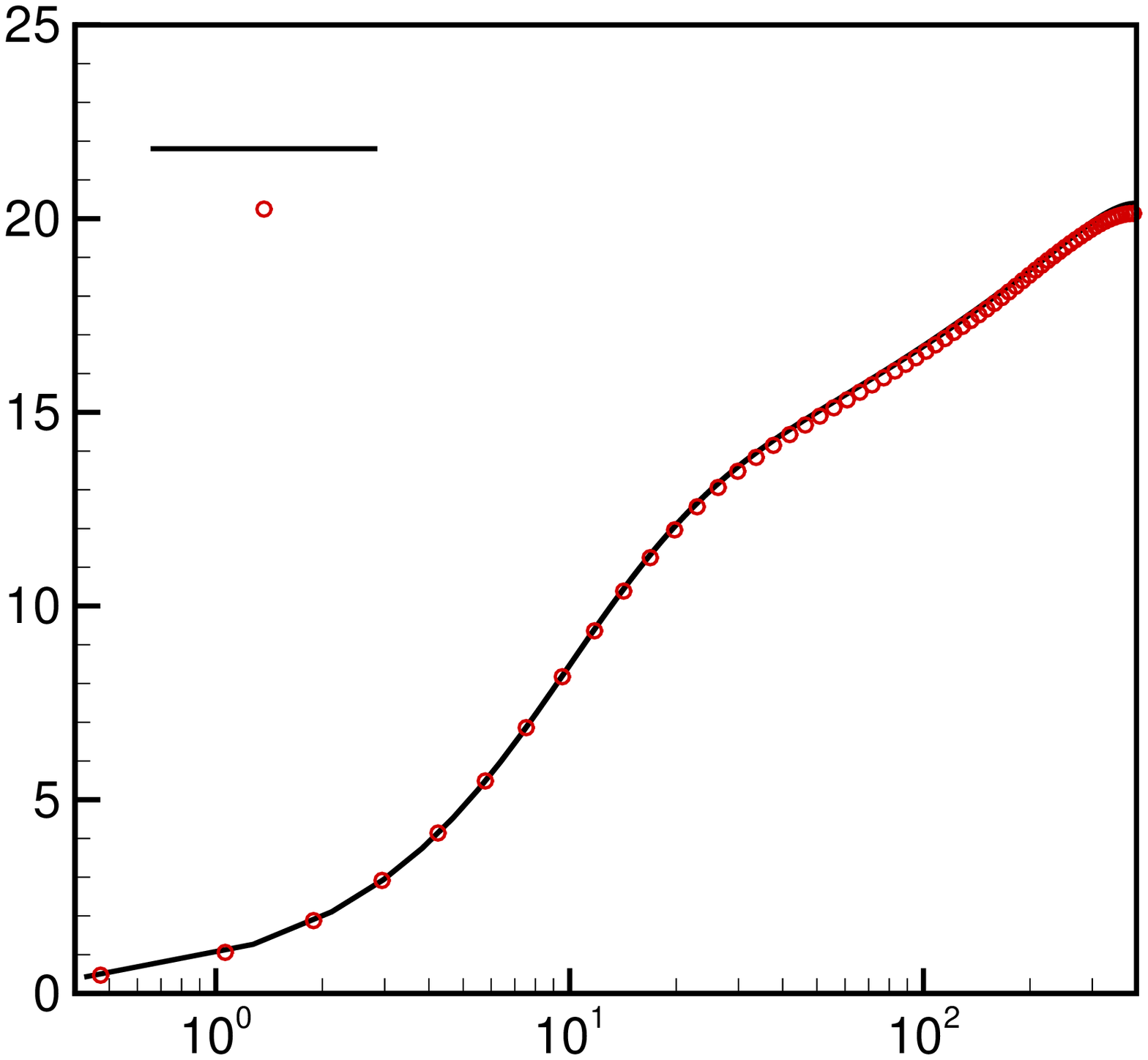}
\put(-100,160){$(a)$}
\put(-190,70){\rotatebox{90}{$U/u_{\tau}$}}
\put(-105,0){$y u_{\tau}/\nu$}
\put(-120,132){SW}
\put(-120,122){Moser et al.}
\includegraphics[height=60mm]{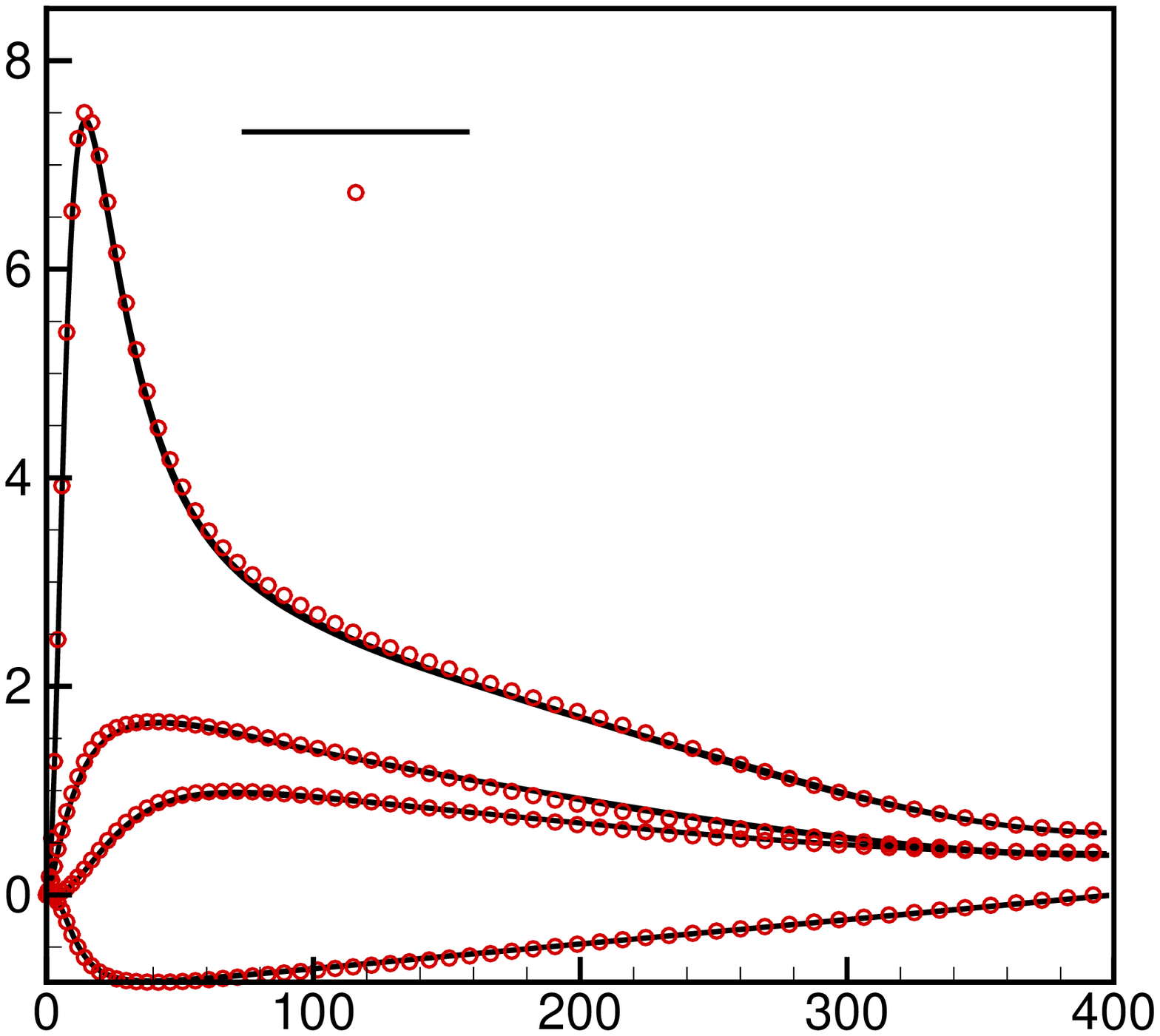}
\put(-100,160){$(b)$}
\put(-150,90){$\langle u'^{2}\rangle^+$}
\put(-160,60){$\langle w'^{2}\rangle^+$}
\put(-160,30){$\langle v'^{2}\rangle^+$}
\put(-130,25){$\langle u'v'\rangle^+$}
\put(-105,0){$y u_{\tau}/\nu$}
\put(-105,132){SW}
\put(-105,122){Moser et al.}
\caption{DNS of smooth channel flow at Re$_{\tau}=400$ compared to the DNS of \citet{moser1999}: $(a)$ mean velocity and $(b)$ Reynolds stresses in inner coordinates.}
\label{fig:validation_smoothwall}
\end{figure}  

The DNS code is validated using smooth turbulent channel flow, and a rod-roughened turbulent channel flow. The simulation details are provided in table \ref{tab:validation}. A constant pressure gradient (divided by density) $K_1$ is applied to drive the flow in the streamwise direction. The average friction velocity, $u_{\tau}=(\delta K_1)^{1/2}$ and the friction Reynolds number is $Re_{\tau}=u_{\tau} \delta/\nu$, where $\delta$ is the channel half-height. Periodic boundary conditions are used in the streamwise and spanwise directions and no-slip conditions are imposed at the solid surfaces. Non-uniform grids are used in the wall-normal direction while uniform grids are used in both streamwise and spanwise directions.

Smooth channel flow at $Re_{\tau}=400$ (Case SW) is used as the baseline, and compared to \citet{moser1999} at $Re_{\tau}=395$. The streamwise mean velocity $U$ is calculated by taking the spatial average in the streamwise and spanwise directions of the time-averaged streamwise velocity. Results are plotted  in wall units $y^+=yu_{\tau}/\nu$, where $U$ is normalized by $u_{\tau}$, and the Reynolds stresses are normalized by $u_{\tau}^{2}$. Good agreement of the mean velocity and Reynolds stresses profiles is shown in figure \ref{fig:validation_smoothwall}. 

 \begin{figure}
\includegraphics[height=50mm,trim={0cm 2.5cm 0.2cm 2cm},clip]{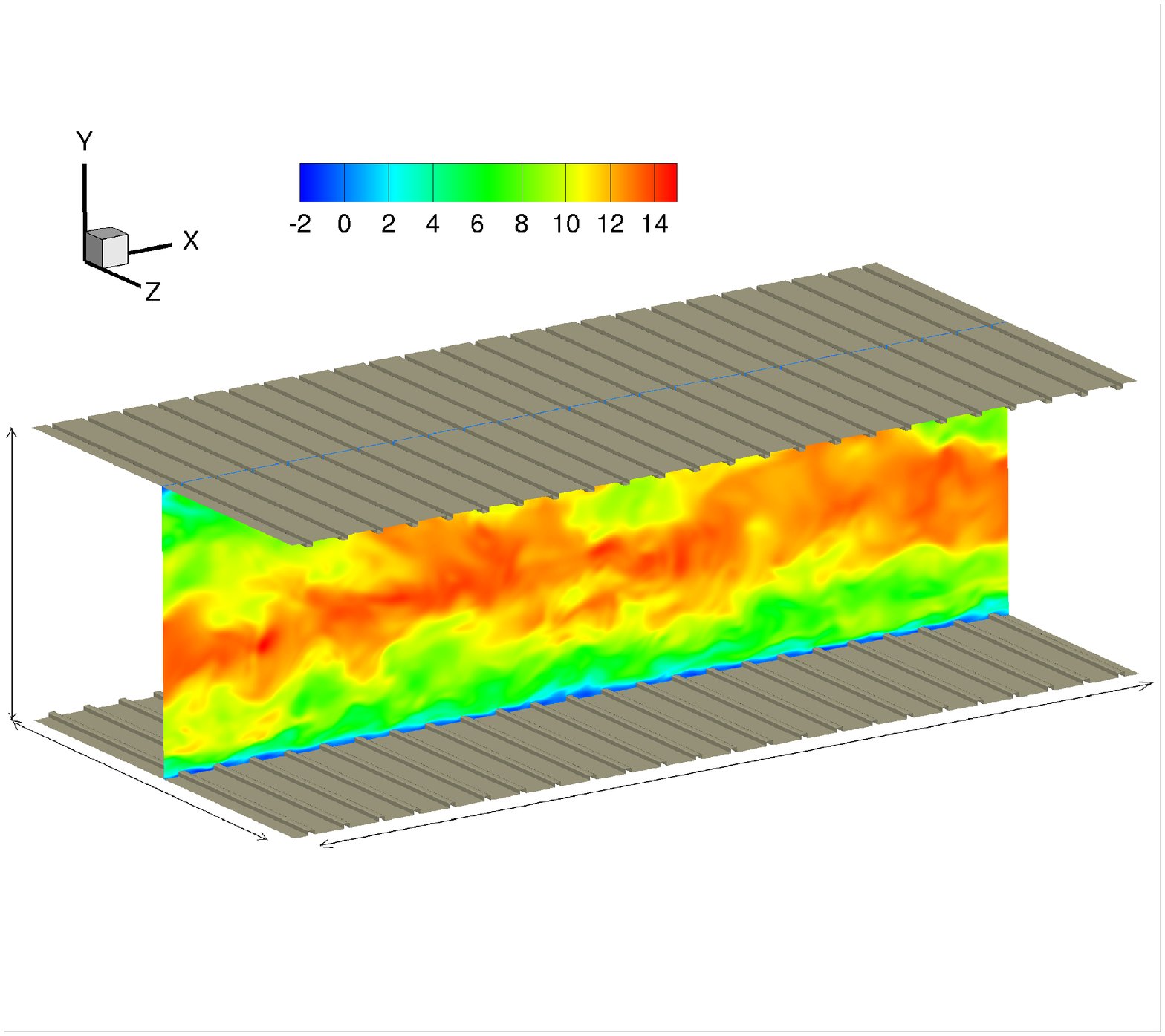}
\put(-220,130){$(a)$}
\put(-165,125){u}
\put(-80,15){$L_x$}
\put(-220,55){$L_y$}
\put(-190,15){$L_z$}
\hspace{2mm}
\includegraphics[height=30mm]{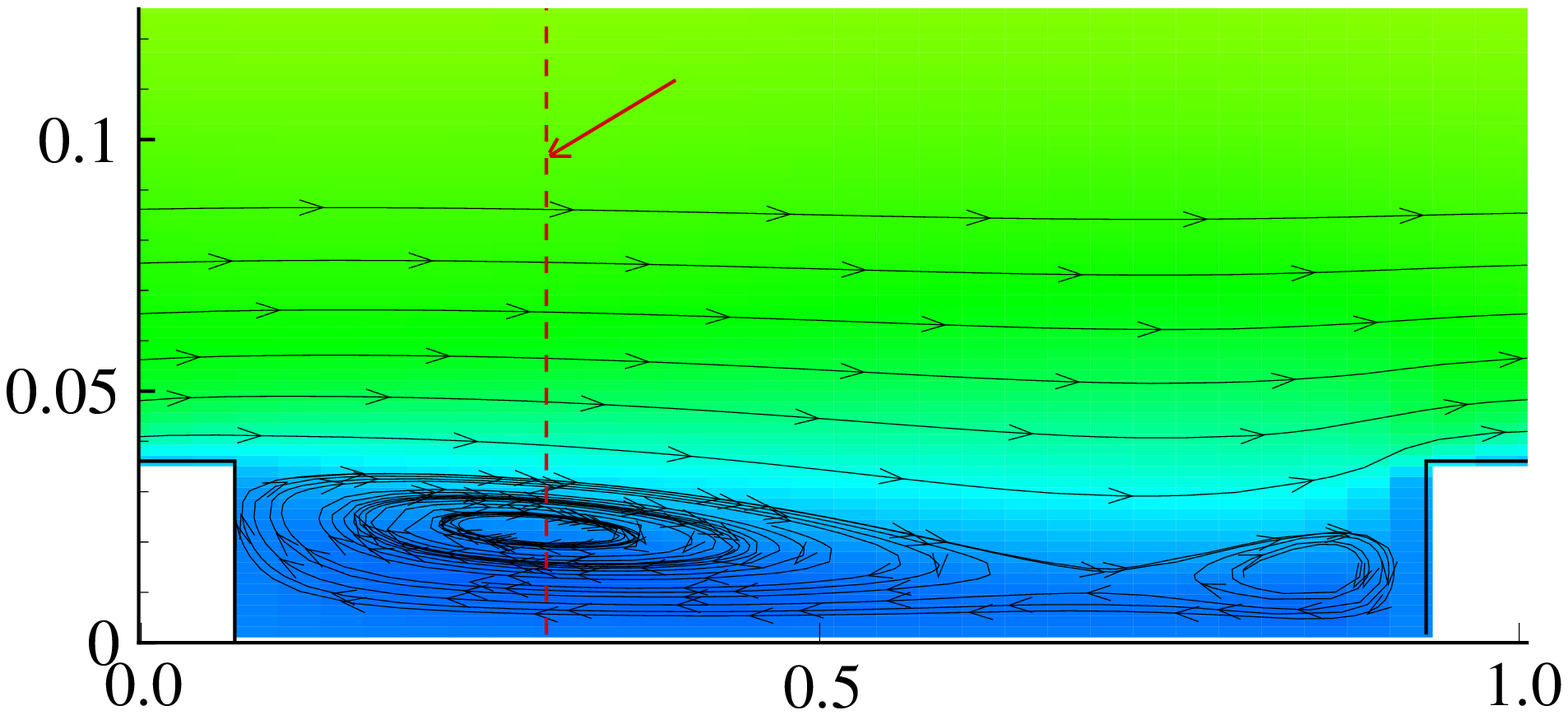}
\put(-100,90){$(b)$}
\put(-192,35){\rotatebox{90}{$y/\delta$}}
\put(-100,-5){$x/\lambda$}
\put(-105,65){$x/\lambda=0.312$}
\caption{DNS of rod-roughened channel flow at Re$_{\tau}=400$: $(a)$ geometry and instantaneous streamwise velocity  contours in the $x$-$y$ plane and $(b)$ mean streamlines averaged with respect to time and spanwise direction in the vicinity of the rods.}
\label{fig:rodroughened_configuration}
\end{figure}

 \begin{figure}
\includegraphics[height=60mm]{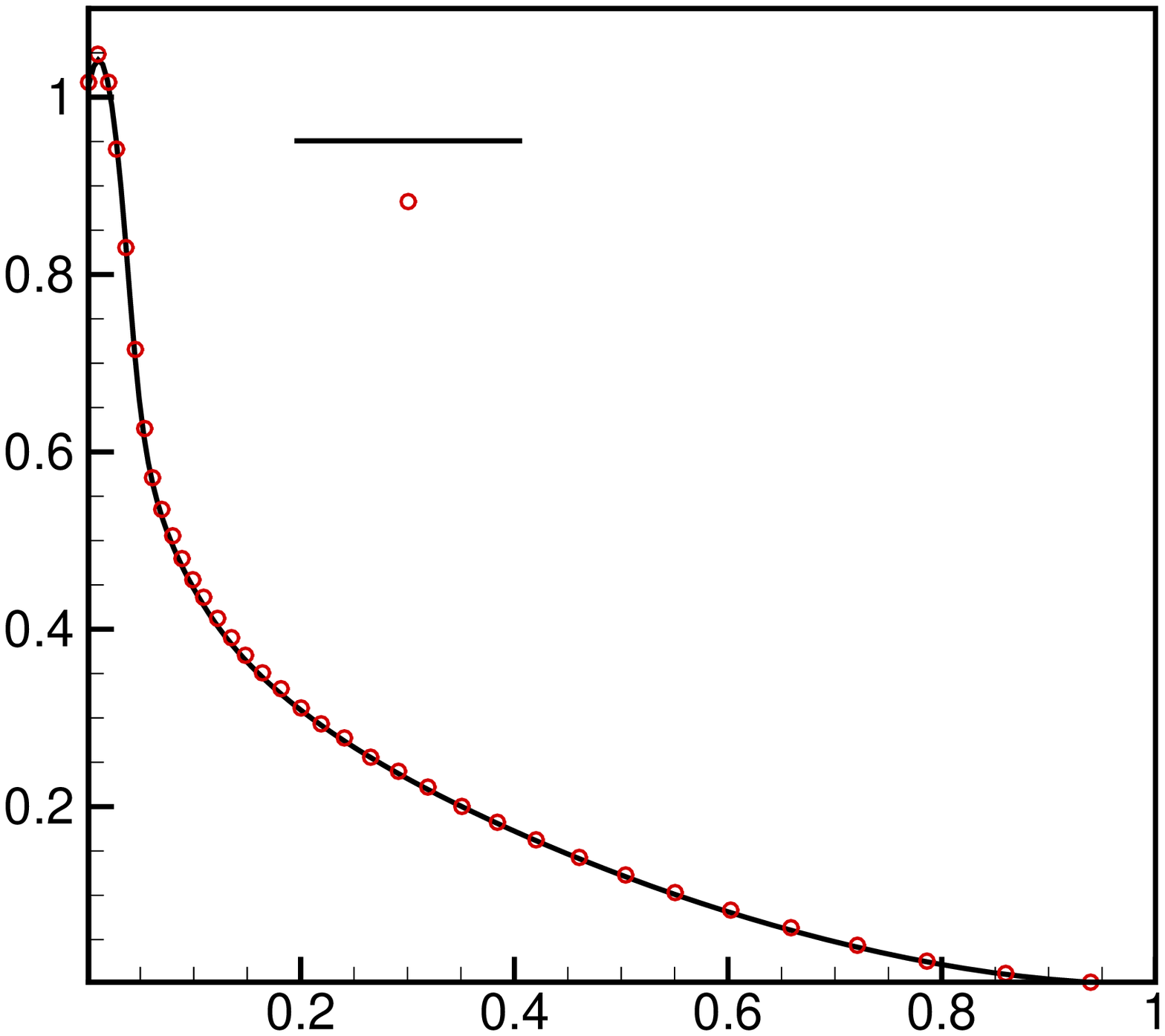}
\put(-100,160){$(a)$}
\put(-195,55){\rotatebox{90}{$(U_{0}-U)/U_{0}$}}
\put(-100,0){$y/\delta$}
\put(-100,132){RRW}
\put(-100,122){Ashrafian et al.}
\includegraphics[height=60mm]{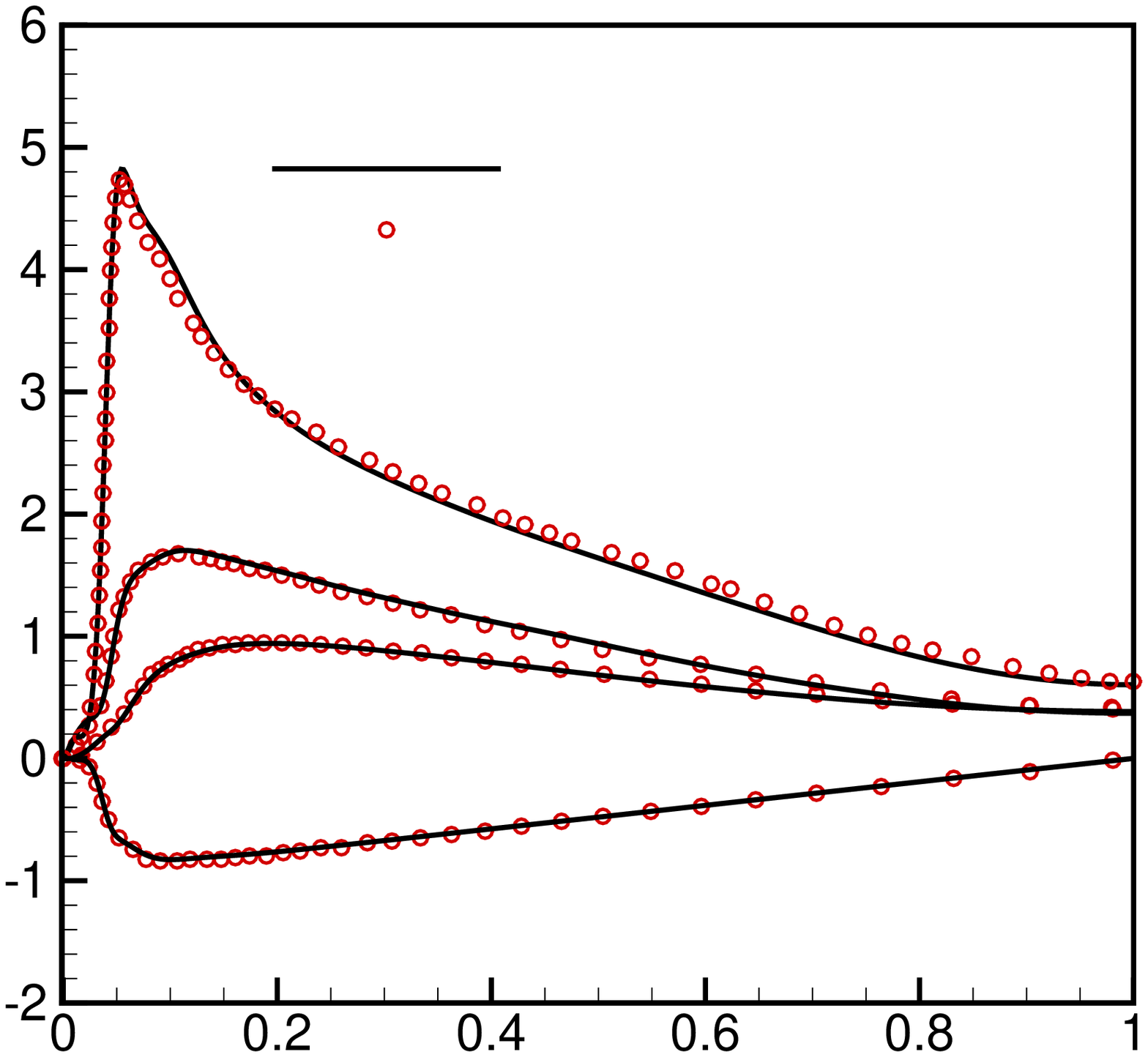}
\put(-100,160){$(b)$}
\put(-145,105){$\langle u'^{2}\rangle^+$}
\put(-160,83){$\langle w'^{2}\rangle^+$}
\put(-155,55){$\langle v'^{2}\rangle^+$}
\put(-130,30){$\langle u'v'\rangle^+$}
\put(-100,0){$y/\delta$}
\put(-100,132){RRW}
\put(-100,122){Ashrafian et al.}
\caption{DNS of rod-roughened channel flow at $Re_{\tau}=400$ compared to the DNS of \citet{ashrafian2004}: $(a)$ defect profiles scaled with centerline velocity $U_{0}$ at $x/\lambda=0.312$ and $(b)$ Reynolds stresses  at $x/\lambda=0.312$.}
\label{fig:validation_rodroughened}
\end{figure}

We then simulate the turbulent channel flow over rod-roughened walls at $Re_{\tau}=400$ (Case RRW). Both top and bottom walls are roughened by 24 square rods with a roughness height $k$ which is $1.7\%$ of the channel height (figure \ref{fig:rodroughened_configuration}$(a)$). The pitch-to-height ratio $\lambda /k$ is 8, where $\lambda$ denotes the pitch, defined as the summation of the rod width and the streamwise distance between two adjacent rods. The roughness height in viscous units is $k^{+}=13.6$. The flow regime is classified as  transitionally rough, according to \cite{ligrani1986structure}. The coordinate $x$ is aligned with the primary flow direction, $y$ is normal to the walls, and $z$ is parallel to the roughness crests. 

The cross-section $x/\lambda=0.312$ is located at the focal point of the primary re-circulation downstream of the roughness element, as shown in figure \ref{fig:rodroughened_configuration}$(b)$. The mean velocity and Reynolds stresses at $x/\lambda=0.312$ are compared to \citet{ashrafian2004} in figure \ref{fig:validation_rodroughened}. Figures \ref{fig:validation_rodroughened}$(a)$ and $(b)$ show the streamwise mean velocity and Reynolds stresses scaled with the centerline velocity $U_{0}=U|_{y=\delta}$ and $u_{\tau}^{2}$ respectively. 
The results show good agreement with \citet{ashrafian2004}. 


\section{Problem setup}\label{sec:problem setup}
\subsection{Implementation of the rough surface}\label{sec:surface data processing}

 \begin{figure}
\includegraphics[height=55mm,trim={1.5cm 1.5cm 1.5cm 1.5cm},clip]{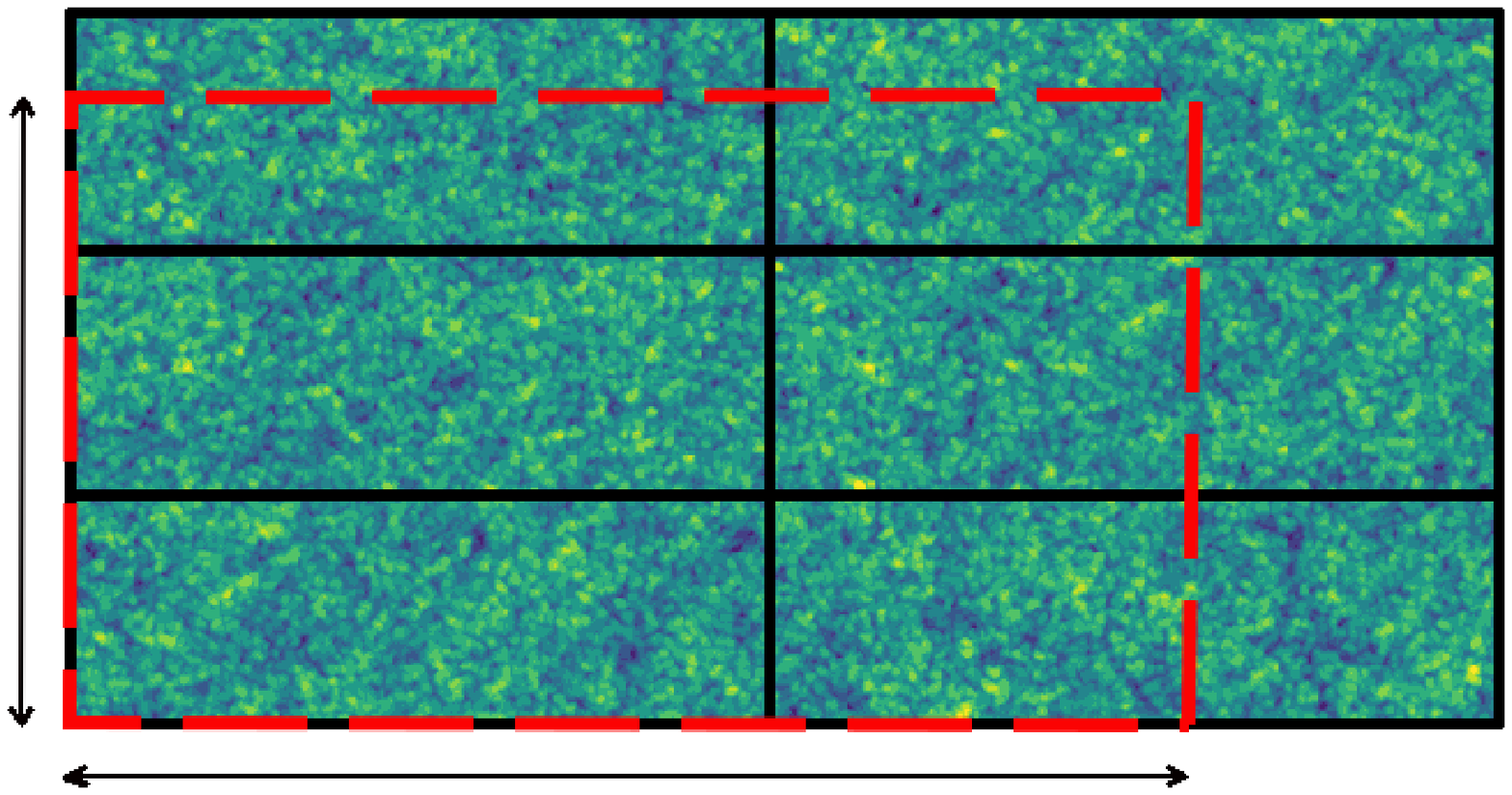}
\put(-97,130){$(a)$}
\put(-185,75){$L_z$}
\put(-115,25){$L_x$}
\includegraphics[height=50mm]{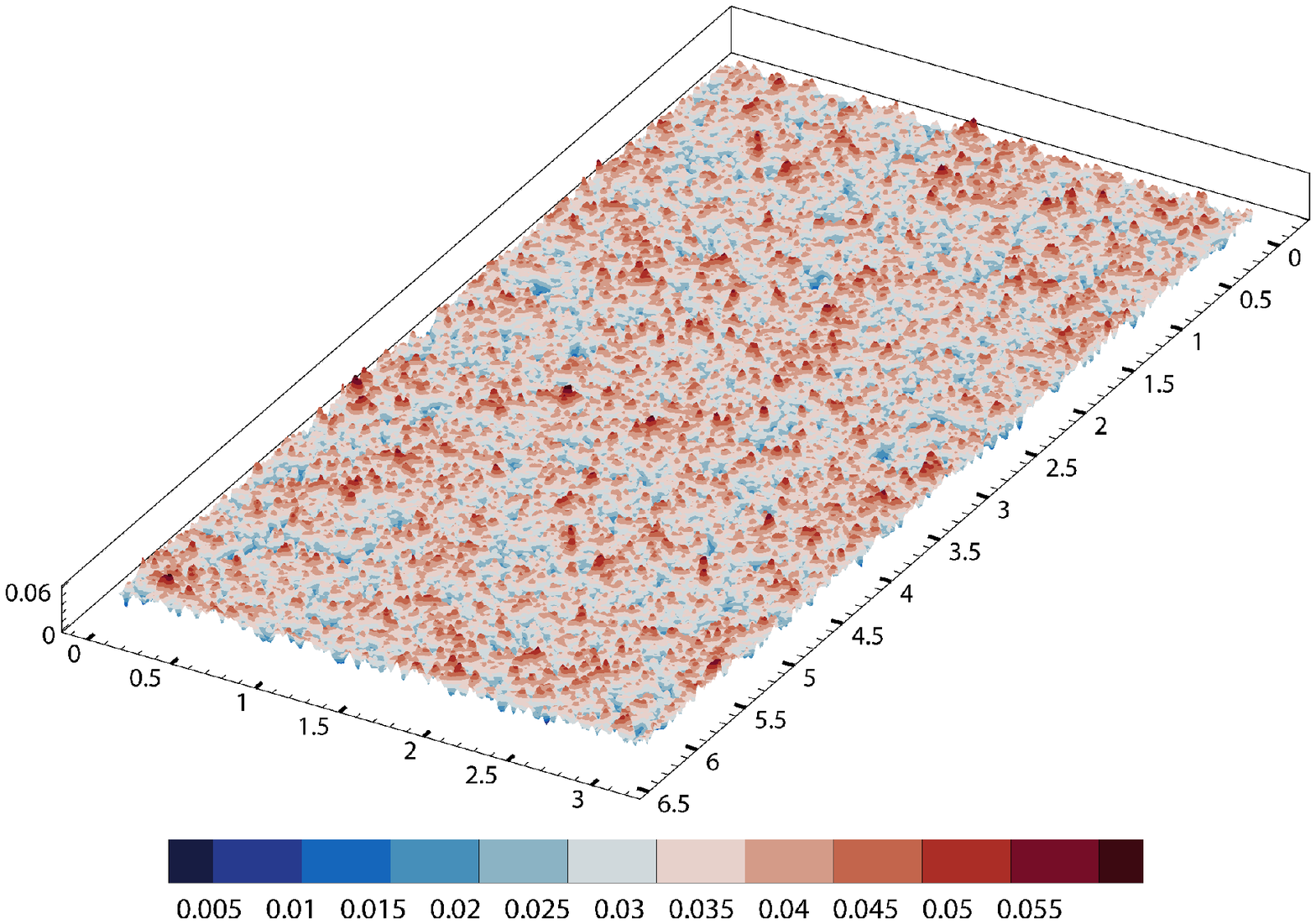}
\put(-190,130){$(b)$}
\put(-55,55){\scriptsize{$x/\delta$}}
\put(-170,25){\scriptsize{$z/\delta$}}
\put(-217,50){\scriptsize{$h/\delta$}}
\put(-200,13){\scriptsize{$h/\delta$}}
\caption{$(a)$ The tiled surfaces with their physical boundaries (solid black) and turbulent channel domain (dashed red); $(b)$ Illustration of the rough surface height.}
\label{fig:tile_surface}
\end{figure}  

The random rough surfaces investigated in this work are processed from   tiles that are scanned and provided by Flack and Schultz (personal communication). The experimental tiles were generated from prescribed power law Fourier spectra and random phases.   \citet{barros2017measurements} and \cite{flack2019skin} discuss the details of this approach which permits systematic control over surface statistics such as the root-mean-square roughness height $k_{rms}$ and skewness. 


Each experimental tile is a rectangular rough patch with a $k_{rms}$ approximately equal to $88 \mu m$. 
The dimensions of the tiles are $50 mm$ by $15 mm$, which are not large enough to cover the bottom wall. Therefore, several rough tiles have to be combined and randomly rotated to minimize directional bias, to span the extent of the channel wall. This yields, in non-dimensional units the required domain size of $2\pi \delta \times \pi \delta$, where $\delta$ is the channel half-height (figure \ref{fig:tile_surface}$(a)$). The interfaces between the rough tiles are set equal to the average value of the adjacent roughness heights, and periodicity is enforced in streamwise and spanwise directions. The test section height is 25 $mm$ in the turbulent channel facility, as mentioned by \citet{flack2019skin}.  The length, width, and height of the rough surface  are all therefore scaled by the channel half-height of 12.5 $mm$. The rough surface is visualized in figure \ref{fig:tile_surface}$(b)$. 

\begin{figure}
\includegraphics[height=40mm]{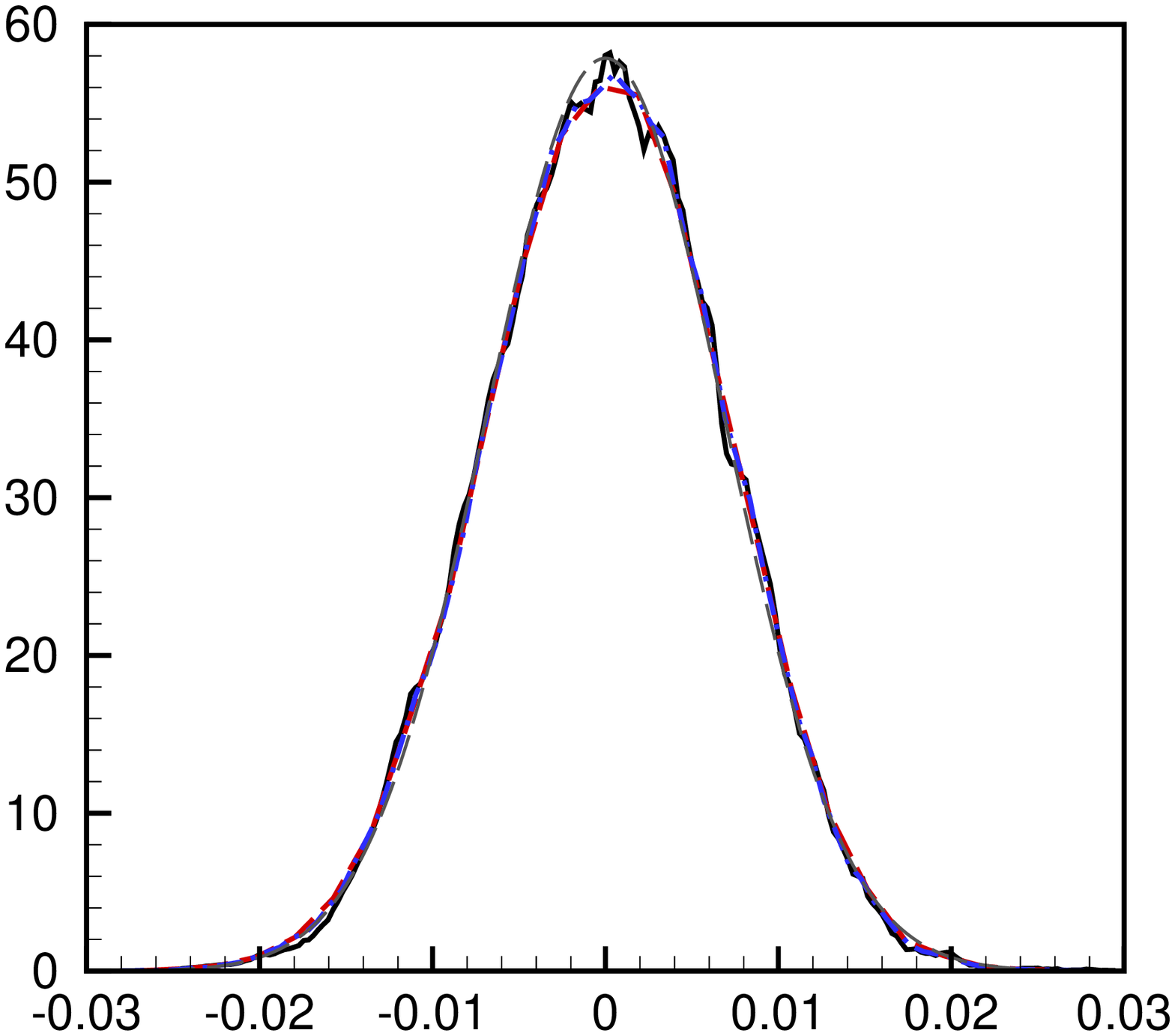}
\put(-70,110){$(a)$}
\put(-132,48){\rotatebox{90}{$p.d.f$}}
\put(-75,-3){$h/\delta$}
\includegraphics[height=40mm]{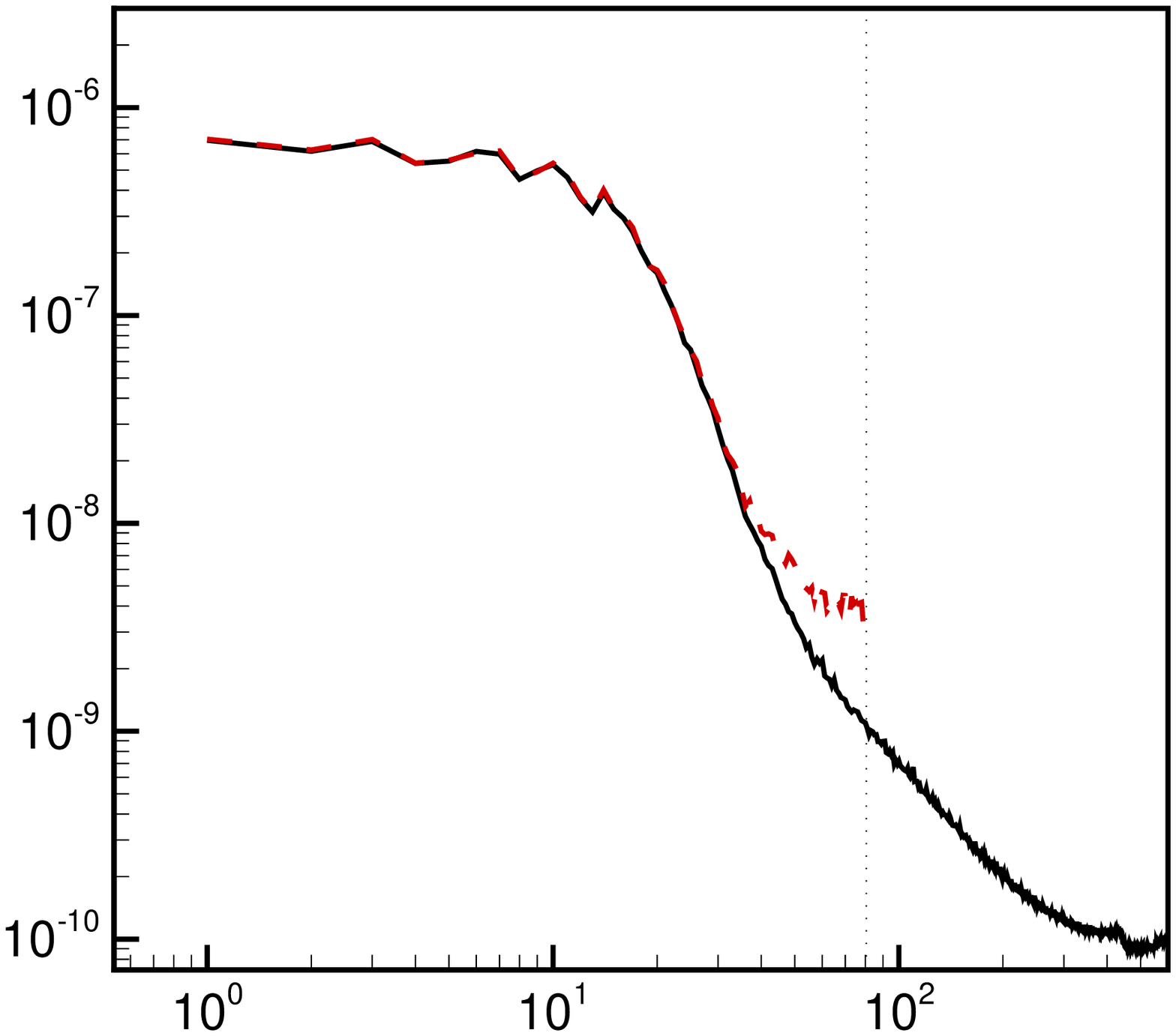}
\put(-70,110){$(b)$}
\put(-132,48){\rotatebox{90}{$E(k)$}}
\put(-70,-3){$k$}
\includegraphics[height=40mm,trim={0.2cm 0.2cm 0.2cm 0.2cm},clip]{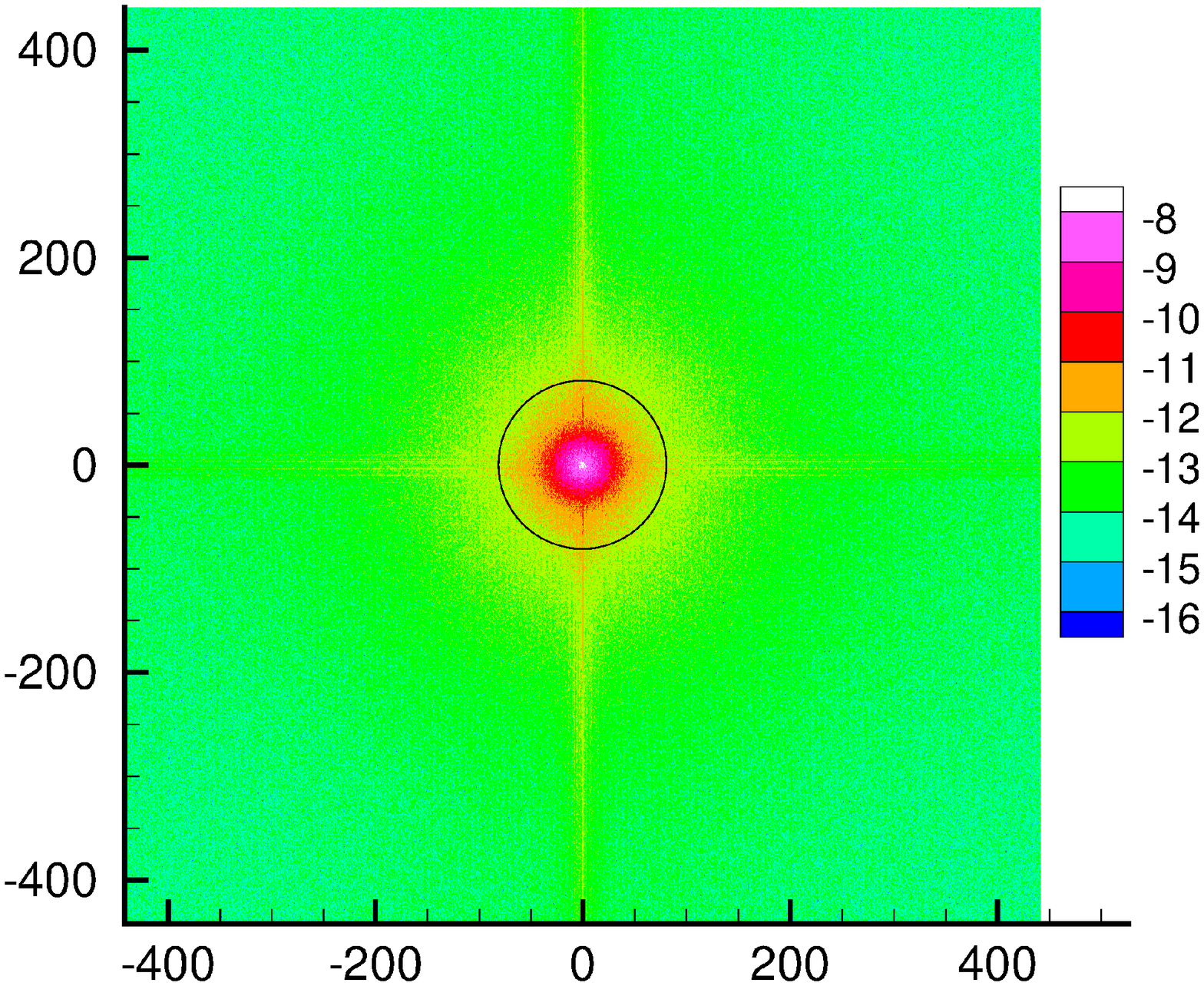}
\put(-75,110){$(c)$}
\put(-132,48){\rotatebox{90}{$k_z$}}
\put(-70,-3){$k_x$}
\put(-25,90){\scriptsize{$\log_{10}(psd)$}}
\caption{$(a)$ The probability density function (p.d.f) of the roughness height for the processed rough surfaces of Case R400 (dashed red) and R600 (dash dot blue), compared to the original surface tiles (solid black) and a Gaussian (long dash grey) with the same $k_{rms}$. $(b)$ Energy spectrum of the processed surface compared to the original tile. $(c)$ Two--dimensional spectrum of the original tile. The circle indicates the cut-off radial wavenumber after processing the surface. }
\label{fig:surface}
\end{figure}  

At the beginning of a simulation, the experimental roughness heights are interpolated to the  computational mesh in the $x-z$ plane. These interpolated roughness heights are then resolved in the $y$ direction to obtain the final rough surface used in the simulation. All cells that share a face with a fluid cell are tagged as  boundary cells. Boundary cells can either be an edge cell (if the boundary cell borders exactly one fluid cell) or a corner cell (if the boundary cell shares a corner with two or more fluid cells). The momentum equations are solved inside the fluid domain while the pressure is solved in both fluid and solid domains. No-slip boundary conditions are applied at the faces of boundary cells. As a result, face-normal velocities are set to zero at the boundaries independent of the cell center value. This ensures that the pressure values inside the solid domain do not affect the pressure values in the fluid domain. 

  \begin{table}
 \begin{center}
\begin{tabular}{lcccccccc}
Parameter  & Description & Tile 1 & Tile 2 & Tile 3 & Tile 4 & Average & Rough surface \\
\hline
$k_{a}$      & Average Roughness Height & 0.0691 & 0.0698 & 0.0700 & 0.0718 & 0.0702 & 0.0692 \\
$k_{rms}$      & RMS Roughness Height & 0.0862 & 0.0875 & 0.0878 & 0.0892 & 0.0877 & 0.0865  \\
$k_{t}$      & Maximum Peak to Valley Height& 0.703 & 0.692 & 0.696 & 0.729 & 0.705 & 0.745\\
$Sk$     & Skewness  &  -0.021 & -0.083 & -0.073 & -0.080 & -0.064 & -0.053\\
$Ku$     & Kurtosis (Flatness) & 2.932 & 2.985 & 2.974 & 2.812 & 2.926 & 2.933 \\
$ES_x$     & Effective Slope of Roughness in $x$& 0.360 & 0.390 & 0.372 & 0.370 & 0.373 & 0.265 \\
$ES_z$     & Effective Slope of Roughness in $z$&  0.359 & 0.392 & 0.370 & 0.371 & 0.373 & 0.265 \\
\hline
\end{tabular}
\caption{Surface statistics of the computational rough surface in Case R400, compared to the experimental tiles. The roughness height is in $mm$.}
\label{tab:stats_surface}
\end{center}
\end{table}

 \begin{figure}
\includegraphics[height=60mm]{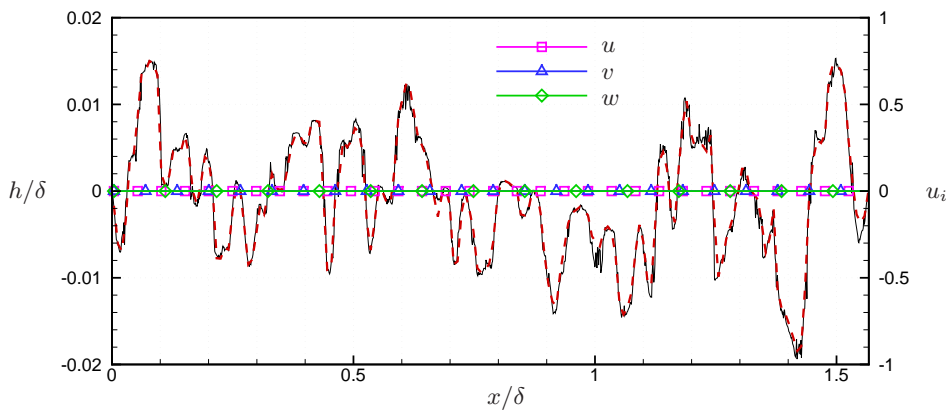}
\put(-190,5){$x/\delta$}
\put(-370,82){$h/\delta$}
\put(-25,82){$u_i$}
\put(-147,137){$u$}
\put(-147,127){$v$}
\put(-147,117){$w$}
\caption{
Comparison of a portion of the computational surface to the experimental scan. The solid black line shows the experimental surface, and the dashed red line shows the computational rough surface for Case R400. The symbols show the face velocities on the boundaries.} 
\label{fig:surface_profile}
\end{figure}  

The characteristic parameters of the rough surface in Case R400  are compared to those of the original tiles in table \ref{tab:stats_surface}, and good agreement is observed. The p.d.f of the processed rough surface in the simulations is compared to the original tiles in figure \ref{fig:surface}$(a)$. Good agreement is observed in the p.d.f, which is well-approximated by a  Gaussian distribution. The extent of discrepancies at small scales due to the surface processing is illustrated by comparing  energy spectra of the original rough tile to that of the processed surface for Case R400 in figure \ref{fig:surface}$(b)$. The radial wavenumber  $k=\sqrt{k_x^2+k_z^2}$, where $k_x$ and $k_z$ are the streamwise and spanwise wavenumbers, normalized by the length $L_x$ and width $L_z$ of the channel domain respectively. The processed surface has a cut-off radial wavenumber at $k=80$ (dotted line) compared to the original tile. The 2D power spectrum of the original surface is shown in figure \ref{fig:surface}$(c)$. The `cross-pattern' is caused by the aliasing effects at the non-periodic boundaries of the original rough tiles. Figure \ref{fig:surface_profile} shows a direct comparison of a portion of the surface profile between the original and processed surface. Note that the small scales are smoothed out but the profile remains a reasonable approximation.

\subsection{Problem description}\label{sec:problem description}

 Simulations are performed  at $Re_{\tau}=u_{\tau} \delta /\nu =400$ and $600$, where $u_{\tau}$ is the friction velocity, $\delta=(L_y-y_0)/2$ is the channel half-height and $y_0$ is the reference bottom plane, that is taken to be the arithmetic mean height of the roughness. The rough surface described in \S \ref{sec:surface data processing} is used as the bottom wall. No-slip boundary conditions are applied on both the top and bottom walls and periodicity is enforced in the streamwise ($x$) and spanwise ($z$) directions; non-uniform grids are used in the wall-normal ($y$) direction where the grid is clustered near the rough wall region. Domain sizes and relevant grid details are summarized in table \ref{tab:simdet}. The random-rough simulation at $Re_{\tau}=400$ is denoted by Case R400 and that at $Re_{\tau}=600$ is Case R600. $k_s^+$ is equal to 6.4 for Case R400, and 9.6 for Case R600. 
 

\begin{table}
\begin{center}
    \begin{tabular}{ lcccccccc| }
    Case & $Re_{\tau}$ & $N_x\times N_y\times N_z$ & $L_x\times L_y \times L_z$ & $\Delta x^{+}$ & $\Delta z^{+}$ & $\Delta y^{+}_{min}$ & $\Delta y^{+}_{max}$\\
    \hline
    R400 & 400 & $768\times 320\times 384$ & $2\pi\delta\times 2.03\delta\times \pi\delta$ & $3.27$ & $3.27$ & $0.85$ & $5.64$\\
    R600 & 600 & $1154\times 540\times 577$ & $2\pi\delta\times 2.03\delta\times \pi\delta$ & $3.27$ & $3.27$ & $0.43$ & $5.64$\\
    R400f & 400 & $667\times 540\times 400$ & $4\delta\times 2.03\delta\times 2.4\delta$ & $2.40$ & $2.40$ & $0.28$ & $3.80$\\
    R600f & 600 & $1572\times 540\times 787$ & $2\pi\delta\times 2.03\delta\times \pi\delta$ & $2.40$ & $2.40$ & $0.43$ & $5.64$\\
    \hline
    \end{tabular}
    \caption{Simulation parameters for the rough channel simulations.}
    \label{tab:simdet}
\end{center}
\end{table}


The computational time step $\Delta t$ is $5 \times 10^{-4} \delta / u_{\tau} $. In order to achieve statistical convergence, mean quantities and statistics were averaged over a period $T=50 \delta / u_{\tau}$. Since the roughness leads to spatial heterogeneity of the time-averaged variables, the double-averaging decomposition \citep{raupach1982averaging} is applied: 
\begin{equation}
    \theta(x,y,z,t) = \langle \overline{\theta} \rangle(y) + \Tilde{\theta}(x,y,z) + \theta'(x,y,z,t).
    \label{eqn:DA}
\end{equation}
Here, $\theta$ represents an instantaneous flow variable, and  $\overline{\theta}$ denotes its  time-average. The instantaneous turbulent fluctuation  $\theta'=\theta-\overline{\theta}$. The brackets denote the spatial-averaging operator,
\begin{equation}
    \langle \overline{\theta} \rangle(y) = 1/A_f\int \int_{A_f} \overline{\theta}(x,y,z)\mathrm{d}x \mathrm{d}z,
\end{equation}
where $A_f$ is the fluid-occupied area. This means that when calculating $\langle \overline{\theta} \rangle(y)$ in the rough regions, only the fluid cells are taken into account. The summation of $\overline{\theta}$ over the fluid cells at each $y$ location is then divided by the number of fluid cells at the corresponding $y$ location. The form-induced dispersive component, $\Tilde{\theta}$, is defined as $\Tilde{\theta}=\overline{\theta}-\langle \overline{\theta} \rangle$, which represents the spatial variation of the time-averaged flow quantities.

According to the decomposition in equation \ref{eqn:DA}, the Reynolds stress tensor is defined as \begin{equation}
   \overline{u_i'u_j'}  = \overline{(u_i-\overline{u_i})(u_j-\overline{u_j})},
\end{equation}
and the dispersive stress tensor is defined as 
\begin{equation}
   \Tilde{u_i}\Tilde{u_j}  =  (\overline{u_i}-\langle \overline{u_i} \rangle)(\overline{u_j}-\langle \overline{u_j} \rangle).
\end{equation}

For compactness, the streamwise mean velocity is denoted by $U$ where the overline (denoting temporal averaging) and angle brackets (denoting spatial averaging) are dropped,
\begin{equation}
 U(y) = \left< \overline{u} \right> = 1/A_f \int \int_{A_f} \overline{u}(x,y,z) \mathrm{d}x \mathrm{d}z.
\end{equation}
The bulk velocity is defined as
\begin{equation}
 U_b = (1/L_y) \int_{y_o}^{L_y} U(y) \mathrm{d}y.
\end{equation}
The spatial-averages of the Reynolds stress tensor are denoted by $\langle u_i'u_j' \rangle$ after dropping the overbar. The average wall shear stress is $\tau_w=\rho \delta K_1$. Since the top wall is smooth and the bottom wall is rough, the shear stress over the smooth wall, ${\tau}_{w}^{t}$, is calculated by averaging $\mu(\partial \overline u/\partial y)|_{y=L_y}$ over the wall. The rough-wall shear stress ${\tau}_{w}^{b}$ is then computed from the force balance between the drag of the walls and the constant pressure gradient. The bottom wall friction velocity is then calculated as $u_{\tau}^{b}=({\tau}_{w}^{b}/\rho)^{1/2}$.



\subsection{Grid convergence}

 \begin{figure}
\includegraphics[height=60mm]{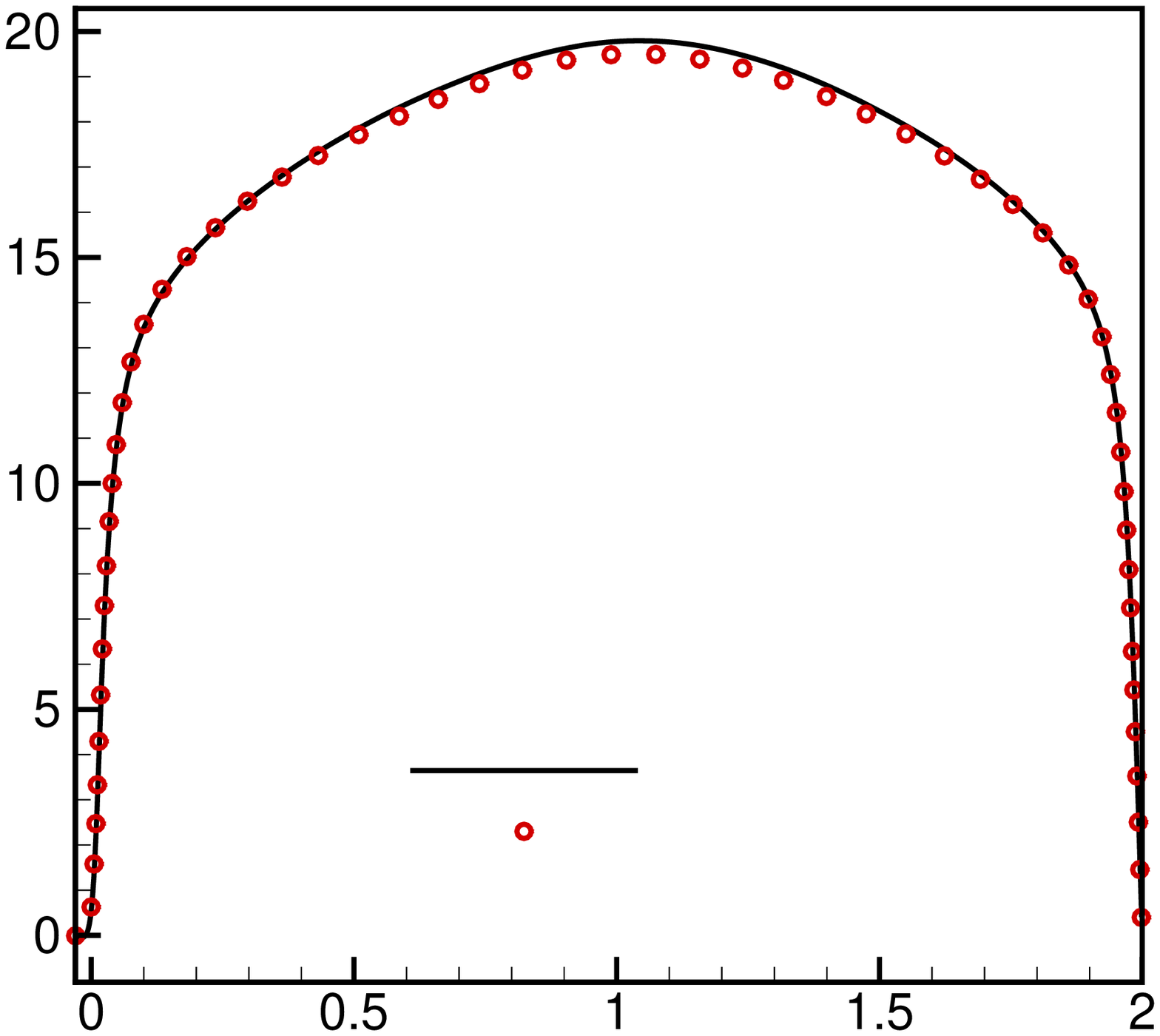}
\put(-100,160){$(a)$}
\put(-190,70){\rotatebox{90}{$U/u_{\tau}$}}
\put(-110,0){$(y-y_0)/\delta$}
\put(-85,45){R400}
\put(-85,35){R400f}
\includegraphics[height=60mm]{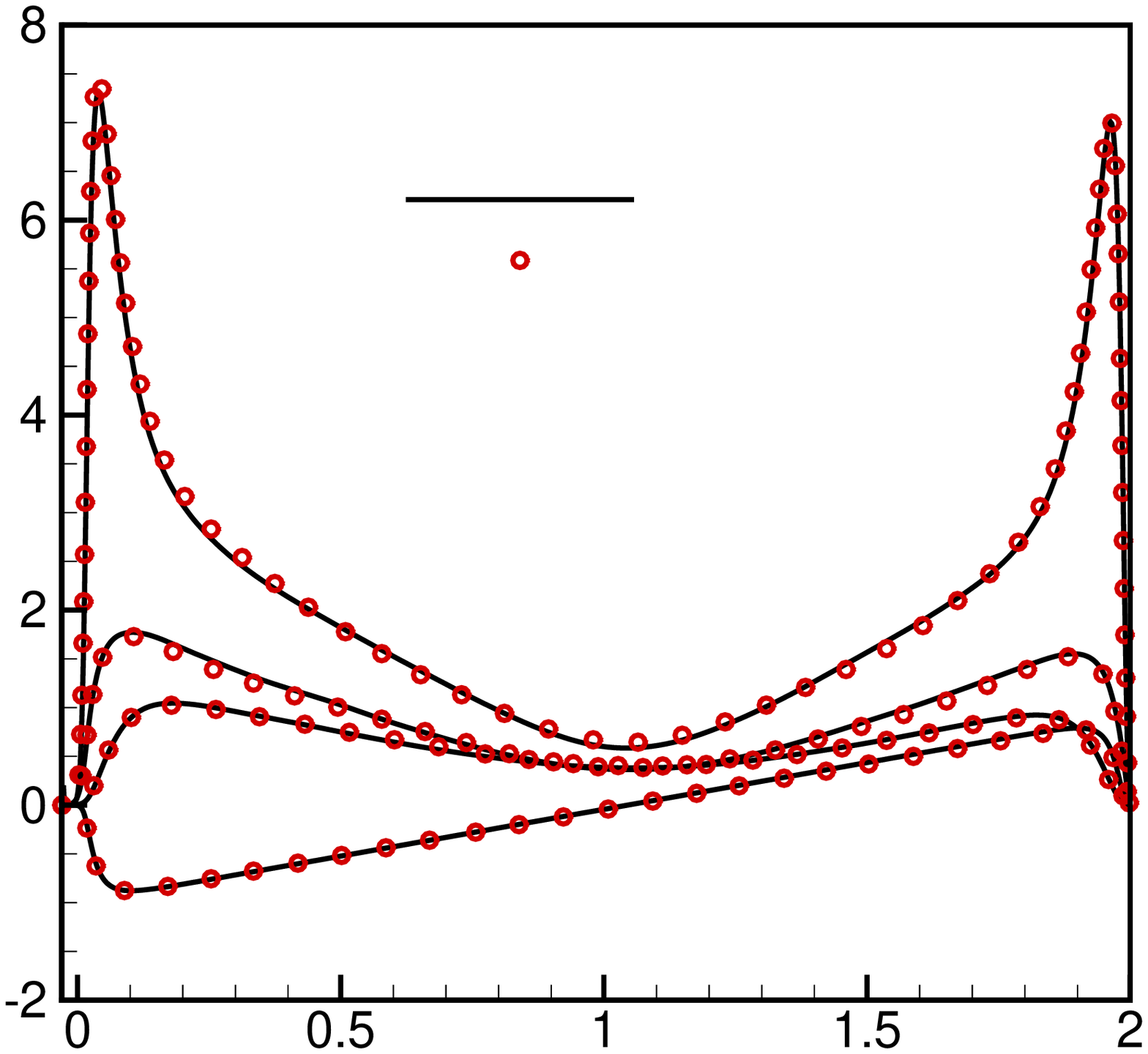}
\put(-100,160){$(b)$}
\put(-150,93){$\langle u'^{2}\rangle^+$}
\put(-165,73){$\langle w'^{2}\rangle^+$}
\put(-160,45){$\langle v'^{2}\rangle^+$}
\put(-130,28){$\langle u'v'\rangle^+$}
\put(-110,0){$(y-y_0)/\delta$}
\put(-85,125){R400}
\put(-85,115){R400f}
\hspace{5mm}
\includegraphics[height=60mm]{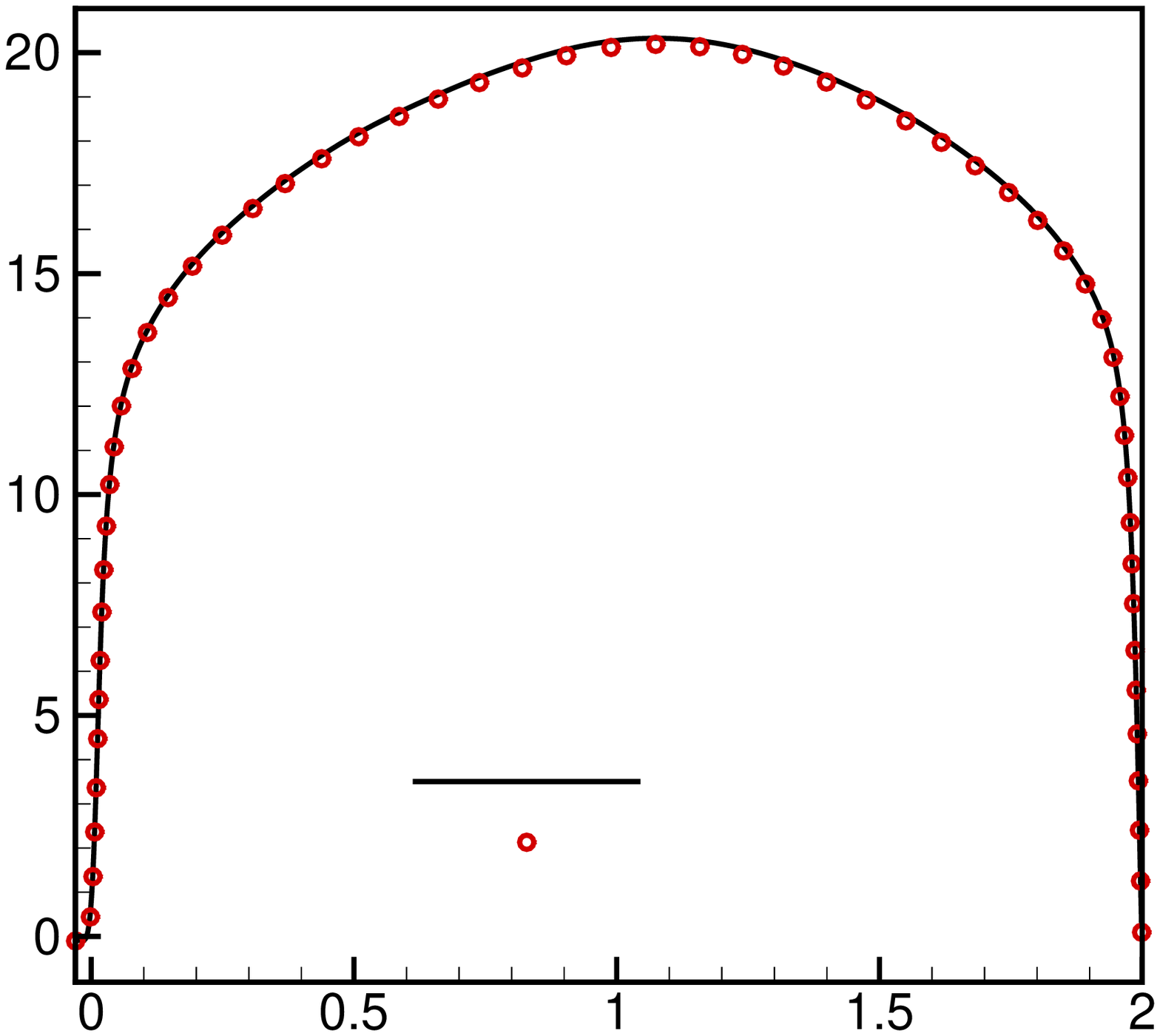}
\put(-100,160){$(c)$}
\put(-190,70){\rotatebox{90}{$U/u_{\tau}$}}
\put(-110,0){$(y-y_0)/\delta$}
\put(-85,45){R600}
\put(-85,35){R600f}
\includegraphics[height=60mm]{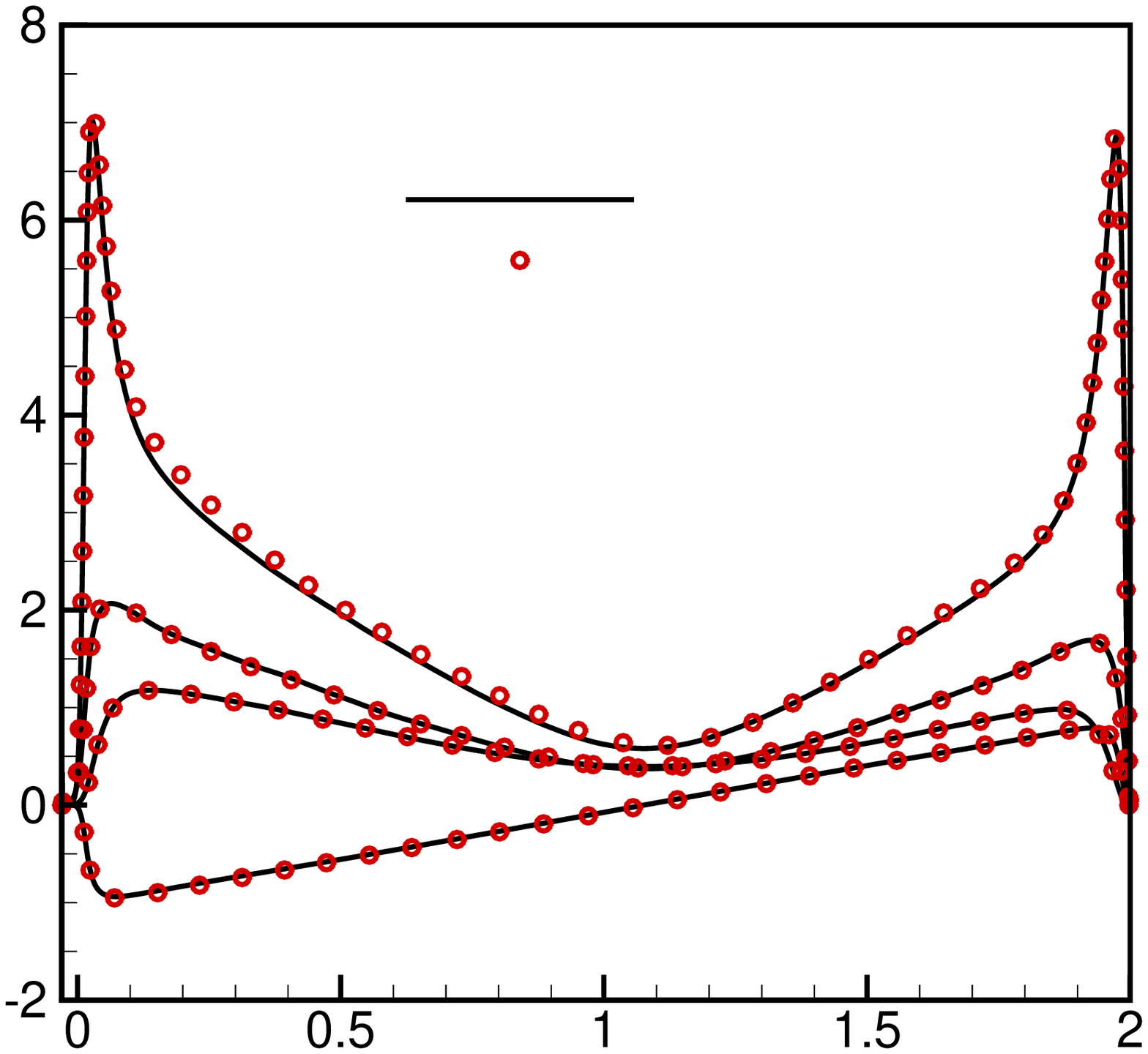}
\put(-100,160){$(d)$}
\put(-150,93){$\langle u'^{2}\rangle^+$}
\put(-165,73){$\langle w'^{2}\rangle^+$}
\put(-160,48){$\langle v'^{2}\rangle^+$}
\put(-130,28){$\langle u'v'\rangle^+$}
\put(-110,0){$(y-y_0)/\delta$}
\put(-85,125){R600}
\put(-85,115){R600f}
\caption{The grid-refined cases R400f and R600f compared to  R400 and R600: $(a)$ and $(c)$  mean velocity profile normalized with  average friction velocity $u_{\tau}$, $(b)$ and $(d)$ Reynolds stresses normalized with $u_{\tau}^2$.}
\label{fig:grid_convergence}
\end{figure}  

A grid convergence study is performed for the two $Re_{\tau}$, using finer grid simulations denoted by cases R400f and R600f. Details of the simulation are listed in table \ref{tab:simdet}. A smaller domain is used for Case R400f to reduce the computational cost, while the normal domain is used for Case R600f. Past work on smooth turbulent channel \citep{lozano2014effect} and turbulent channel with urban-like cubical obstacles \citep{coceal2006mean} suggest that a domain size of $L_x \times L_z=4\delta \times 2.4\delta$ is sufficient for obtaining mean and turbulent statistics without any significant confinement effects. 

The wall-normal grid resolution for DNS in rough-wall channel flows depends on the roughness topography and the strength of roughness effects on the flow.  \cite{busse2015direct} recommended $\Delta y^+ <1$ within the roughness layer and a maximum resolution of $\Delta y^+ \approx 5$ at the centerline of the channel for  transitionally rough cases. 
For the present simulations, 
the grid resolution is comparable to that used by \cite{busse2015direct} in the wall-normal direction. The streamwise and spanwise resolutions also adhere to the criteria suggested by \cite{busse2015direct}. The smallest roughness wavelength is resolved by $14$ grid points.

For Case R400f, two original rough tiles are interpolated onto grids of size $667\times 200$. The interpolated tiles are then randomly rotated and tiled in the spanwise direction to achieve a domain size of $4\delta\times 2.4\delta$. The length scales of the  rough surface are normalized by the channel half-height $12.5mm$. For Case R600f, since the wall-normal resolution of Case R600 is sufficiently fine, the grid is refined in the streamwise and spanwise directions. The mean velocity and Reynolds stresses profiles from the finer grids are compared to those from cases R400 and R600 in figure \ref{fig:grid_convergence}. Good agreement is observed between both profiles.

 \begin{figure}
 \centering{
\includegraphics[height=60mm]{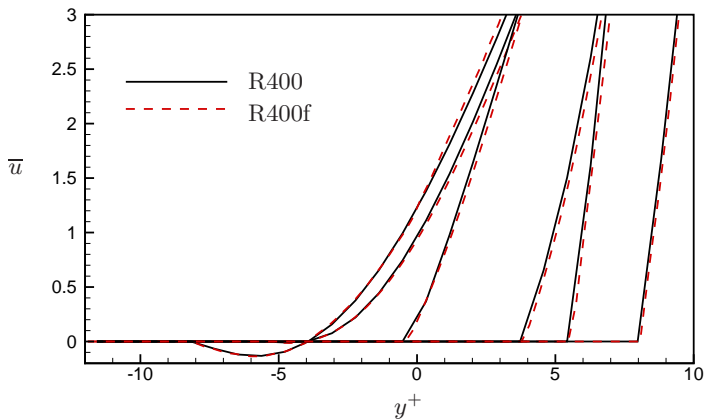}
\put(-150,0){$y^+$}
\put(-295,90){$\overline{u}$}
\put(-205,122){R400}
\put(-205,110){R400f}}
\caption{Time-averaged streamwise velocity profiles within the roughness layer at a few random locations on the rough surface.} 
\label{fig:nearwall_u}
\end{figure}

Figure \ref{fig:nearwall_u} shows the time-averaged velocity profiles in the roughness layer for a few randomly chosen points. Note that the velocity profiles drop to zero at the surface and that the profiles for coarse and fine grids are in acceptable agreement, indicating that the flow is adequately resolved.

\section{Results}\label{sec:results}

 
\subsection{Skin-friction coefficient}\label{subsec:Skin-friction coefficient}

\begin{figure}
  \hspace{32mm}
  \includegraphics[height=60mm,trim={0.2cm 0.2cm 0.2cm 0.2cm},clip]{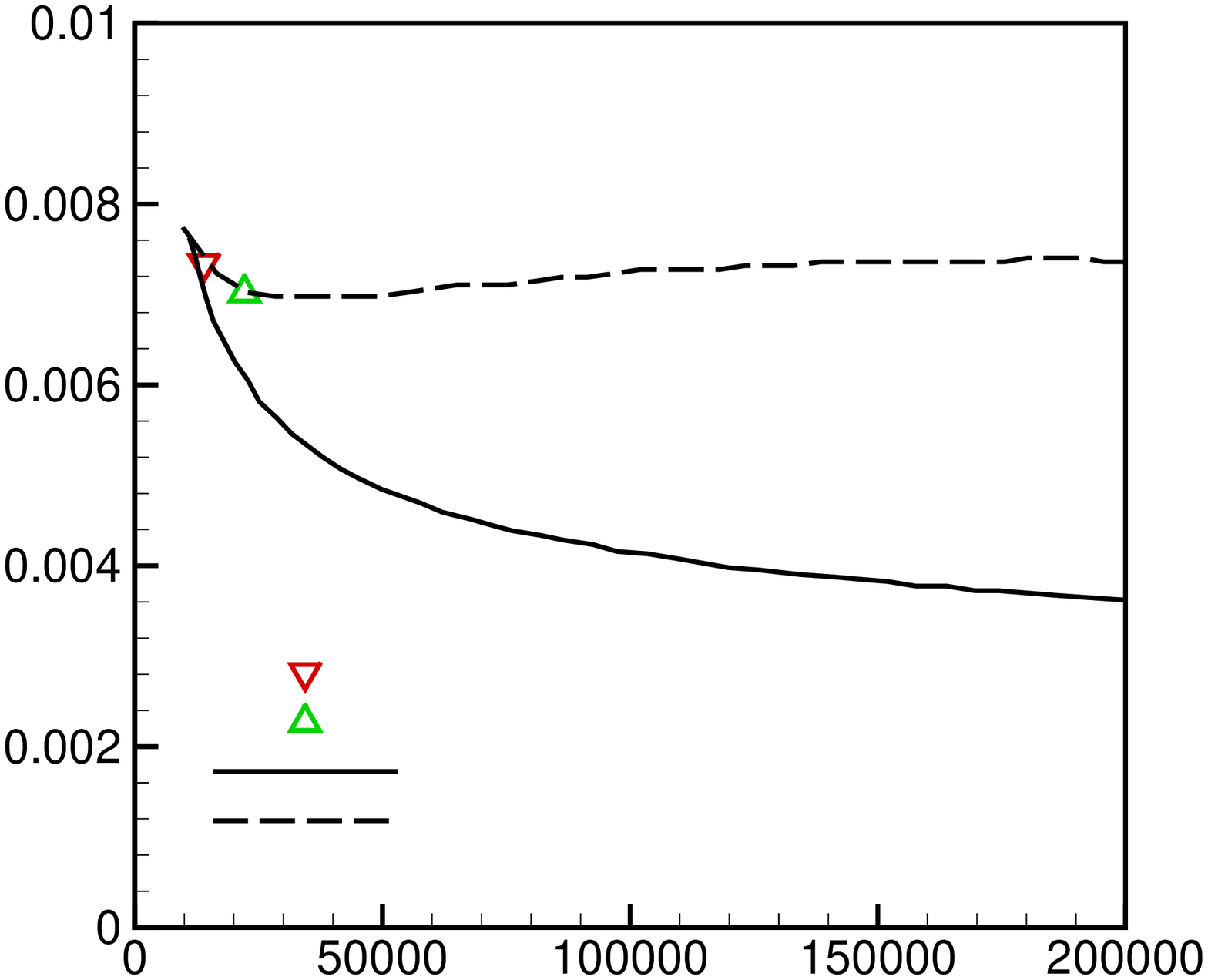}
  \put(-205,75){\rotatebox{90}{$C_f$}}
  \put(-100,0){$Re$}
    \put(-125,53){\scriptsize{R400}}
  \put(-125,46){\scriptsize{R600}}
  \put(-125,39){\scriptsize{Smooth, Flack et al.}}
  \put(-125,32){\scriptsize{$k_{rms}\approx 88 \mu m$, Flack et al.}}
  \caption{Skin-friction coefficient of cases R400 and R600, compared to  experimental results  \cite{flack2019skin}. The rough surfaces correspond to the surface with $k_{rms}\approx 88 \mu m$ and $Sk=-0.07$ in \cite{flack2019skin}.}
\label{fig:cf}
\end{figure}

The mean skin-friction coefficient $C_f={\tau}_{w}^{b}/(\frac{1}{2}{\rho} U_b^{2})$, where $U_b$ is the bulk velocity. Figure \ref{fig:cf} compares the computed values of $C_f$ at the two $Re_{\tau}$ to the experimental results of \cite{flack2019skin}. Note that the rough surface is nearly hydraulically smooth at the lower Reynolds number. The skin friction coefficient decreases as $Re$ increases and the flow becomes transitionally rough. Both the viscous drag and form drag contribute to the overall skin friction in this flow regime. As $Re$ further increases, the rough surface exhibits fully-rough behavior where the skin friction becomes independent of $Re$. The $C_f$ values of Case R400 and Case R600 are shown in table \ref{tab:cf}. The errors of the simulation relative to the experiment at the two $Re_{\tau}$ are $0.8\%$ and $0.3\%$ respectively, showing acceptable agreement.

\begin{table}
\begin{center}
    \begin{tabular}{ lcccc| }
    Case & Experiment & DNS & Error  \\
    \hline
    R400 & 0.00740 & 0.00734 & 0.8 \% \\
    R600 & 0.00704 & 0.00702 & 0.3 \% \\
    \hline
    \end{tabular}
    \caption{Skin-friction coefficient from experiments.}
    \label{tab:cf}
\end{center}
\end{table}


\subsection{Double-averaged velocity}
 \begin{figure}
\includegraphics[height=60mm]{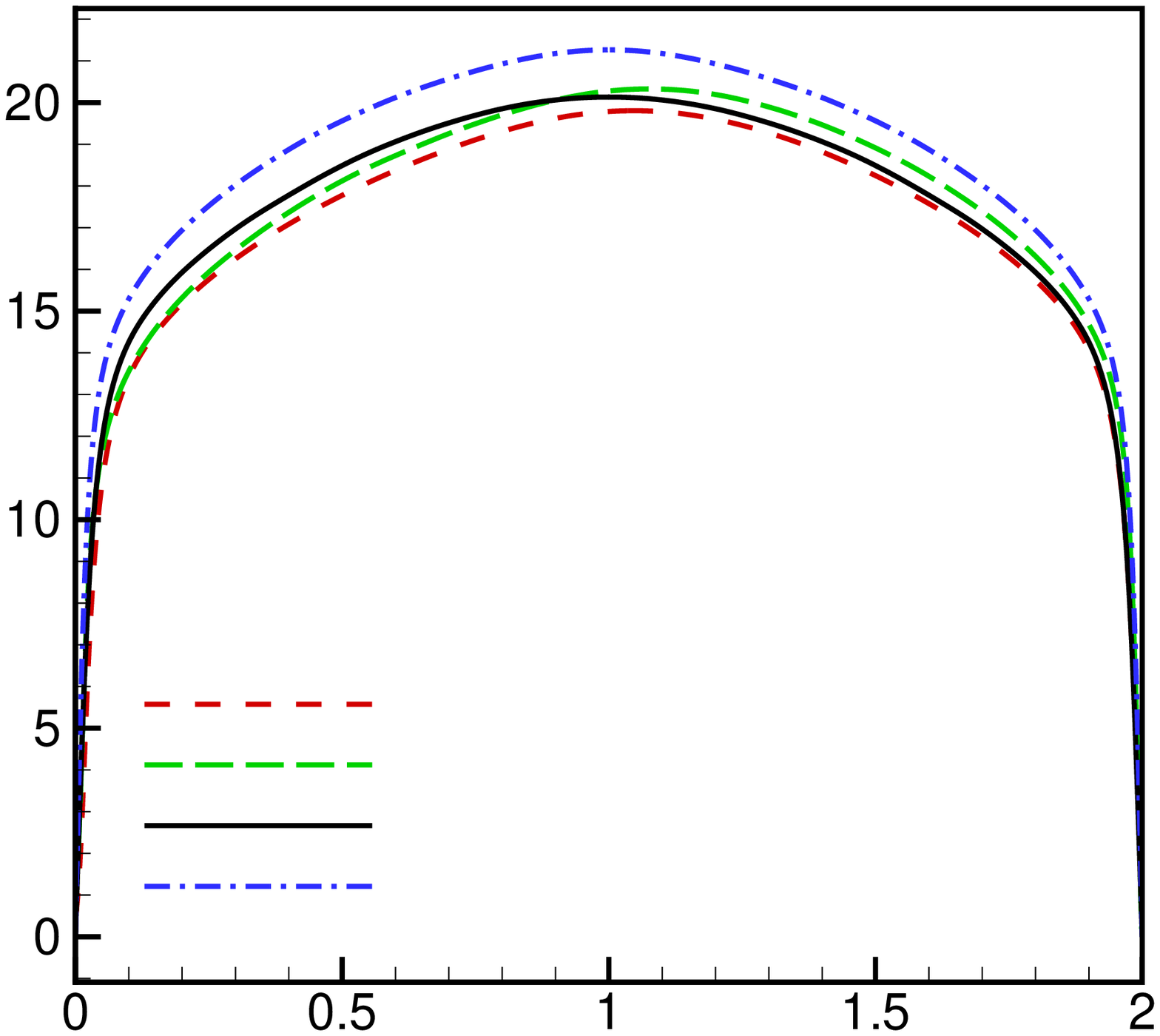}
\put(-100,160){$(a)$}
\put(-192,70){\rotatebox{90}{$U/u_{\tau}$}}
\put(-110,0){$(y-y_0)/\delta$}
  \put(-125,54){\scriptsize{R400}}
  \put(-125,46){\scriptsize{R600}}
  \put(-125,38){\scriptsize{Moser et al., $Re_{\tau}=395$}}
  \put(-125,30){\scriptsize{Moser et al., $Re_{\tau}=590$}}
\includegraphics[height=60mm]{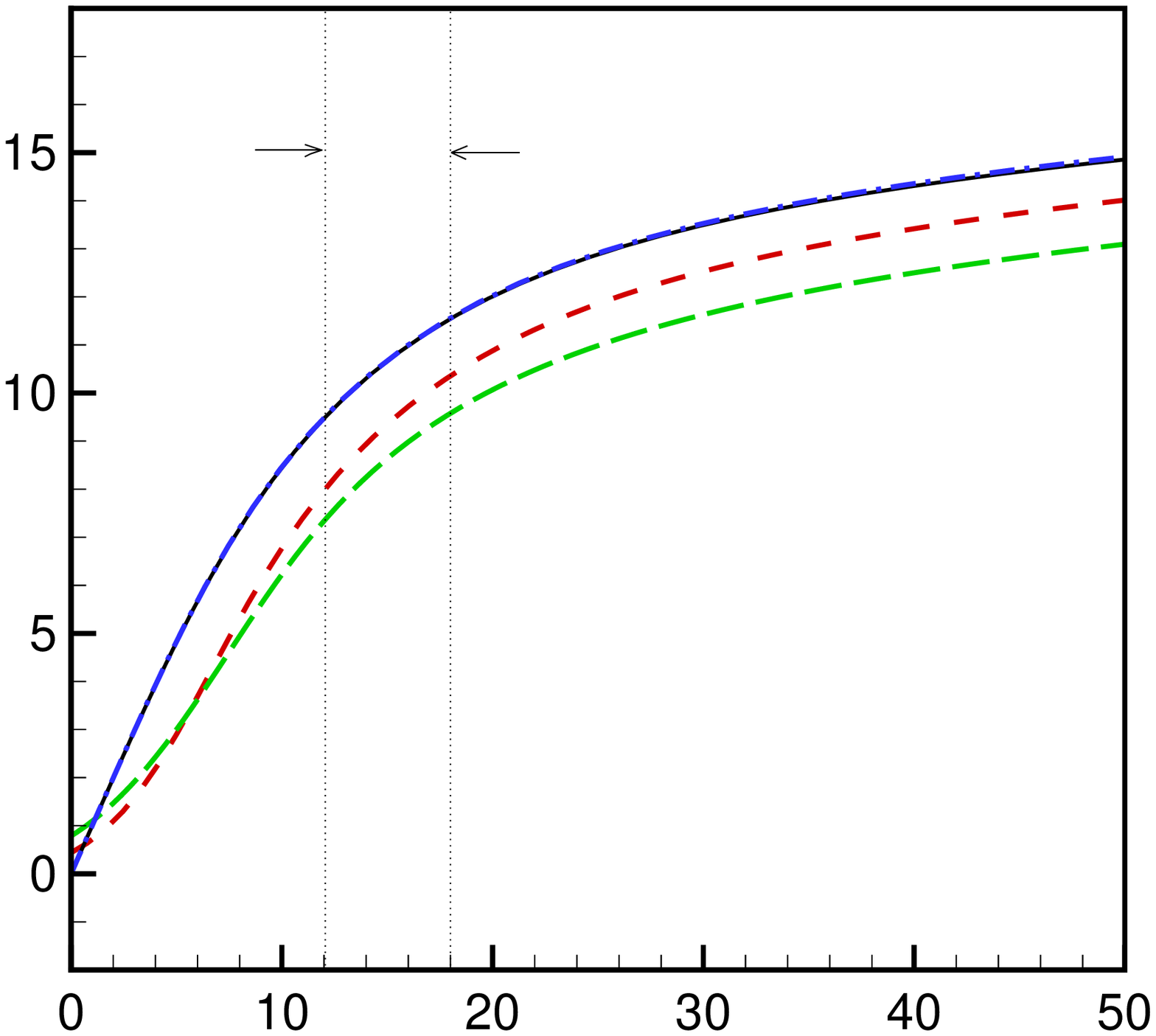}
\put(-100,160){$(b)$}
\put(-192,70){\rotatebox{90}{$U/u_{\tau}^b$}}
\put(-120,0){$(y-y_0)u_{\tau}^{b}/\nu$}
\put(-163,134){\scriptsize{$k_c^+$,R400}}
  \put(-113,134){\scriptsize{$k_c^+$,R600}}
\caption{Mean velocity profile  normalized by: $(a)$ average friction velocity $u_{\tau}$  $(b)$  rough-wall friction velocity $u_{\tau}^b$, compared to  DNS of  smooth channel flow by \cite{moser1999}.   The vertical dotted lines in $(b)$ denote the top of the roughness layer, $k_c^+$.}
\label{fig:umean_krms100}
\end{figure}


The double-averaged (DA) velocity profiles for cases R400 and R600 are discussed below.  Figure \ref{fig:umean_krms100}$(a)$ shows the streamwise mean velocity profile in outer   coordinates. $U$ is normalized by the average friction velocity $u_{\tau}$ and $y$ is shifted by subtracting the reference plane $y_0$, and then normalized by $\delta$. The smooth cases at $Re_{\tau}=395$ and $Re_{\tau}=590$ from \cite{moser1999} are also presented for comparison. A velocity deficit can be observed for the rough cases and the velocity decrease is more significant in the lower half-channel. Compared to the smooth case, the peak of the mean velocity profile at $Re_{\tau}=400$ decreases by $1.6\%$ and shifts away from the rough wall by $5\%$. For  the higher $Re_{\tau}$, a more significant velocity deficit is observed. The peak mean velocity is reduced by $4.4\%$ and shifted by $8.4\%$ compared to the smooth case at the same $Re_{\tau}$.

Figure \ref{fig:umean_krms100}$(b)$ provides a closer view of the viscous wall region for the rough-wall. The mean velocity and the wall-normal distance of the smooth cases are normalized by the average friction velocity $u_{\tau}$ and $\nu/u_{\tau}$ respectively. The smooth wall curves at different $Re_{\tau}$ collapse in  inner coordinates.  Scaling with the bottom-wall friction velocity $u_{\tau}^{b}$ provides a more accurate representation of slip and drag effects (e.g.  \cite{alame2019wall}). Hence, for the rough cases, the mean velocity profile of the lower half-channel is normalized by $u_{\tau}^{b}$. Case R400 shows a slip velocity at the wall and the profile is lowered by $6.7\%$ relative to the smooth case. The slip velocity is increased further for Case R600 and the profile is shifted downwards by $13.2\%$. The two profiles intersect at $y^{+}=5.7$. This indicates that a larger slip velocity exists at the wall and an overall increase of drag is exhibited in the viscous wall region as $Re_{\tau}$ increases. 

 \begin{figure}
\includegraphics[height=60mm]{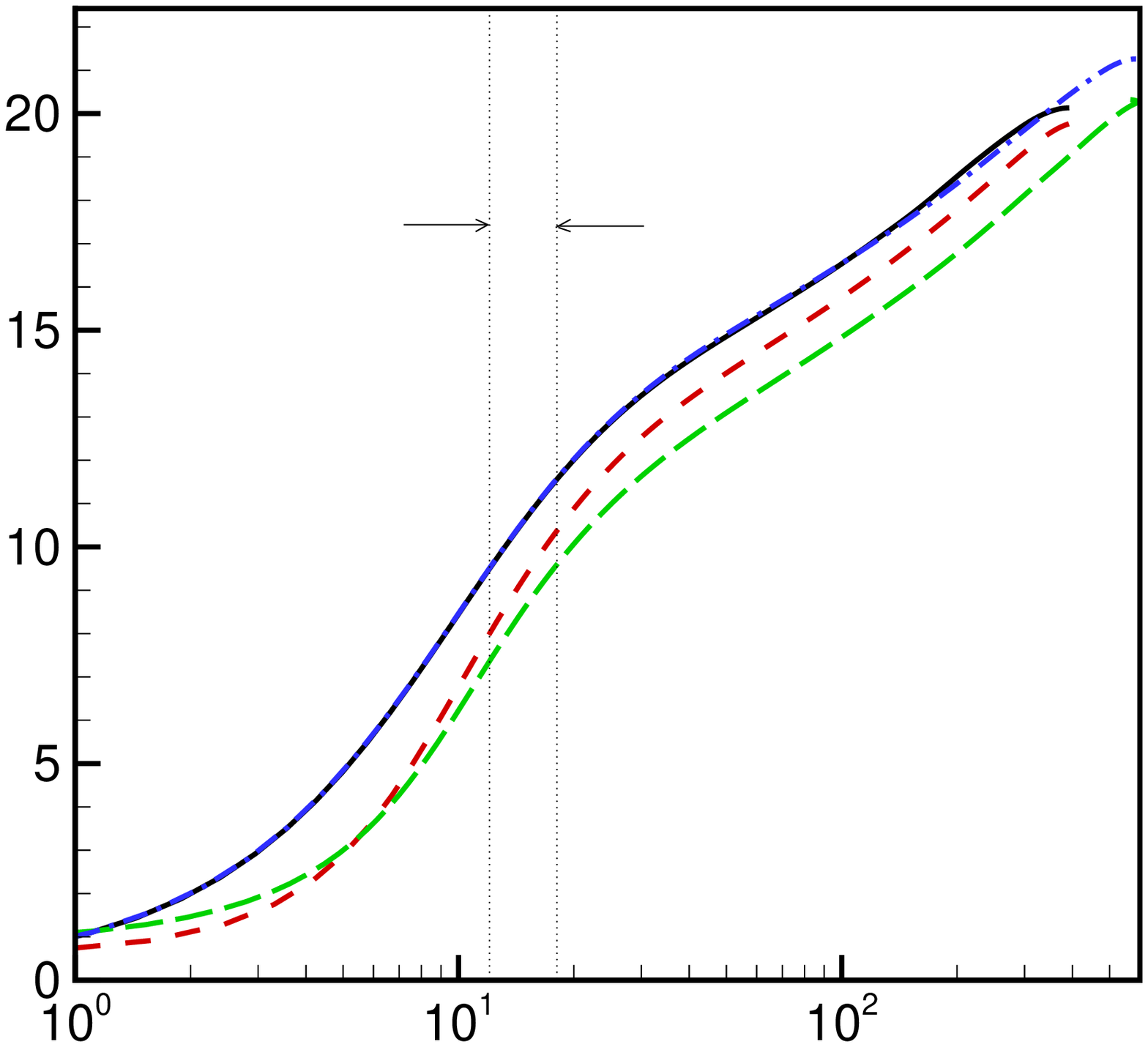}
\put(-100,160){$(a)$}
\put(-192,70){\rotatebox{90}{$U/u_{\tau}^b$}}
\put(-120,0){$(y-y_0)u_{\tau}^{b}/\nu$}
\put(-145,124){\scriptsize{$k_c^+$,R400}}
  \put(-98,124){\scriptsize{$k_c^+$,R600}}
\includegraphics[height=60mm]{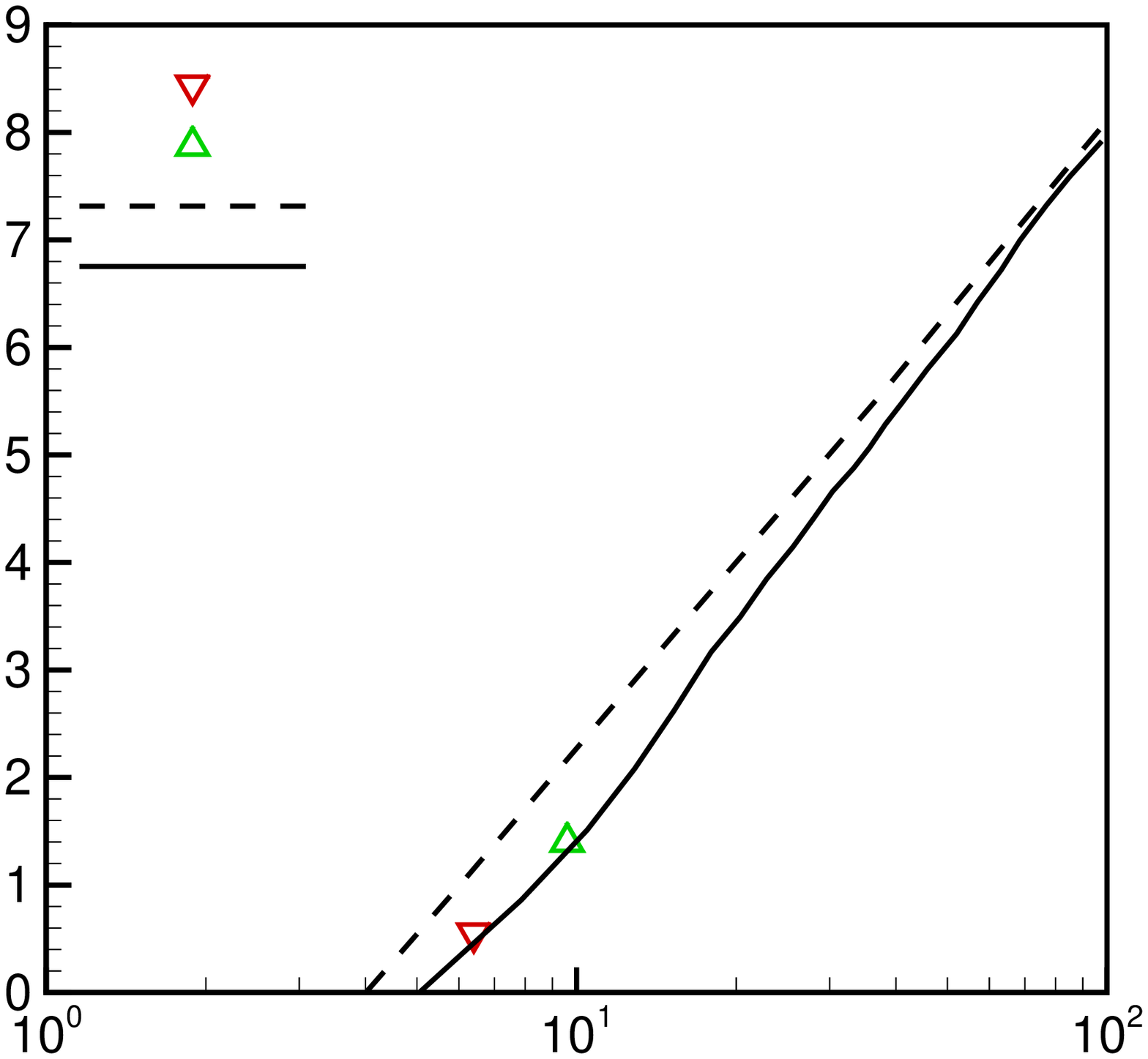}
\put(-100,160){$(b)$}
\put(-192,70){\rotatebox{90}{$\Delta U^{+}$}}
\put(-100,0){$k_{s}^{+}$}
  \put(-127,143){\scriptsize{R400}}
  \put(-127,133){\scriptsize{R600}}
  \put(-127,123){\scriptsize{Fully-rough asymptote}}
  \put(-127,113){\scriptsize{Flack et al.}}
\caption{$(a)$  Mean velocity profile in the bottom half of the channel in inner coordinates. The legend is the same as figure \ref{fig:umean_krms100}. $(b)$ Roughness function compared to the experimental results of  \cite{flack2019skin} and the fully-rough asymptote.}
\label{fig:umean_krms100_inner}
\end{figure}


The mean velocity profile is shown in figure \ref{fig:umean_krms100_inner}$(a)$ using the same scaling as figure \ref{fig:umean_krms100}$(b)$, but in  semi-log coordinates. Due to the slip effect of the roughness, the velocity profile shows a gradual increase away from the wall. This trend appears to end at $y^{+}=5$ which is the transition from the viscous sublayer to the buffer layer. In the buffer layer, the velocity for Case R600 increases more slowly and displays a more significant velocity deficit than Case R400. A similar trend was observed by \cite{busse2017reynolds}. In the log-law region, the smooth-wall profiles follow the logarithmic law:
\begin{equation}
 U_{s}^{+}=\frac{1}{\kappa} \ln{y^{+}}+ B,
 \label{eqn:low_smooth}
\end{equation}
where $\kappa$ is the Von Karman constant and $B$ is the intercept for a smooth wall. The rough-wall profiles conform to the log-law but display an offset from the smooth-wall profiles, where the roughness effect on mean velocity can be evaluated from this difference. \cite{nikuradse1933laws} found that the logarithmic velocity distribution for the mean velocity profile still held for rough walls, with the same value of $\kappa$ as
\begin{equation}
 U_{r}^{+}=\frac{1}{\kappa} \ln({y/k_{s}})+ 8.5
 \label{eqn:low_rough}
\end{equation}
The roughness function is obtained by taking the difference of mean velocities in wall units between smooth and rough walls within the logarithmic layer. 
\begin{equation}
 B-\Delta U^{+}+\frac{1}{\kappa} \ln{k_{s}^{+}}=8.5
 \label{eqn:low}
\end{equation}
\cite{flack2010review} show  good collapse in the fully rough regime for different roughness types  when using $k_{s}$ as the roughness scale. However, in the transitional regime, the roughness function depends on the roughness type, and the onset of the fully rough regime is unknown for most surfaces.  $\Delta U^{+}$ from the simulations is plotted in figure \ref{fig:umean_krms100_inner}$(b)$  as a function of the corresponding roughness Reynolds number $k_{s}^{+}$ obtained from experiment. The results show that cases R400 and R600 are in the transitionally rough regime,  and match the experimental results of \cite{flack2019skin}.




\subsection{Reynolds stresses}
\begin{figure}
 \includegraphics[height=45mm]{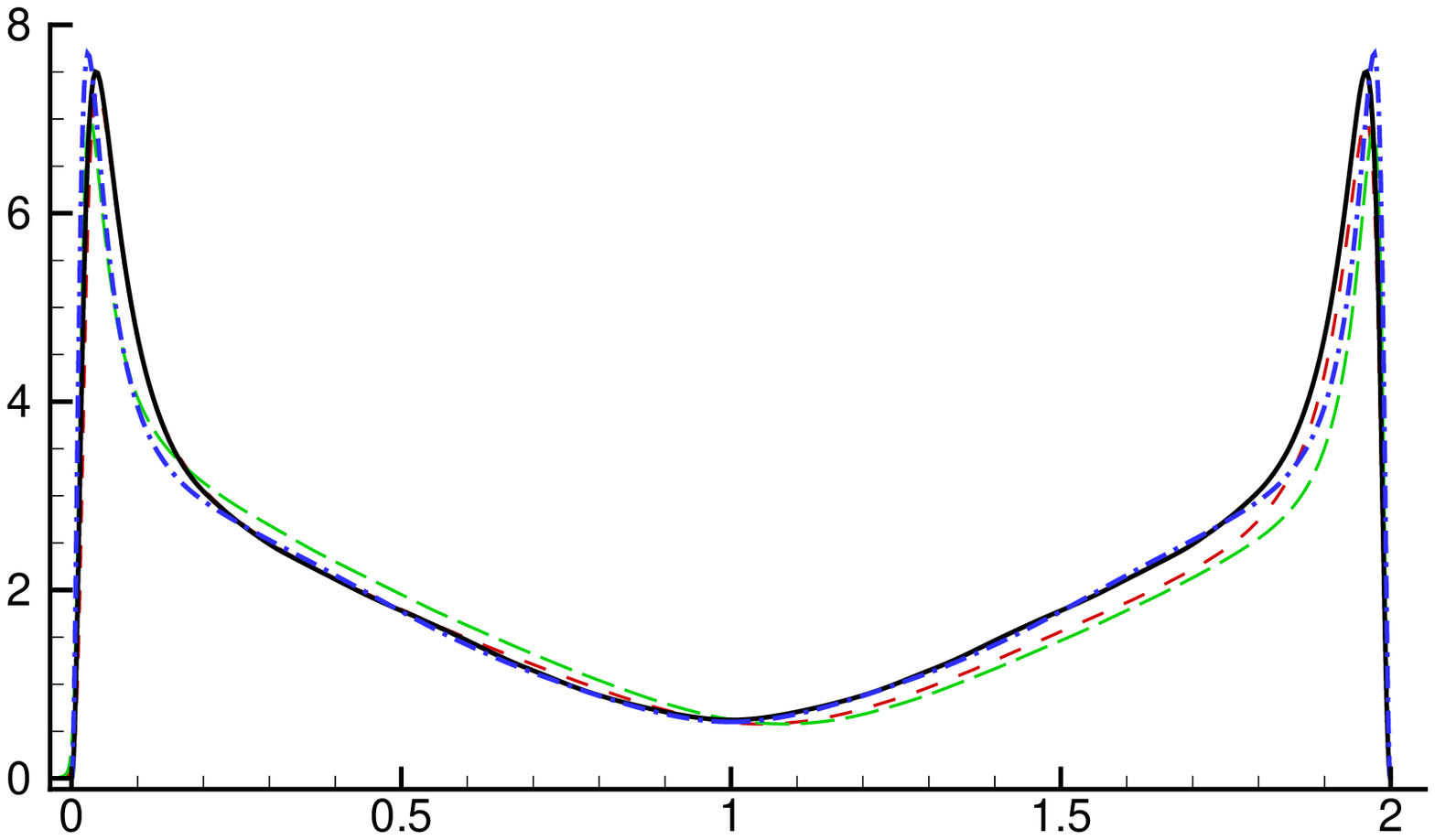}
 \put(-110,110){$(a)$}
 \put(-205,45){\rotatebox{90}{$\langle u'u' \rangle/(u_{\tau})^{2}$}}
 \put(-120,-5){$(y-y_0)/\delta$}
 \includegraphics[height=45mm]{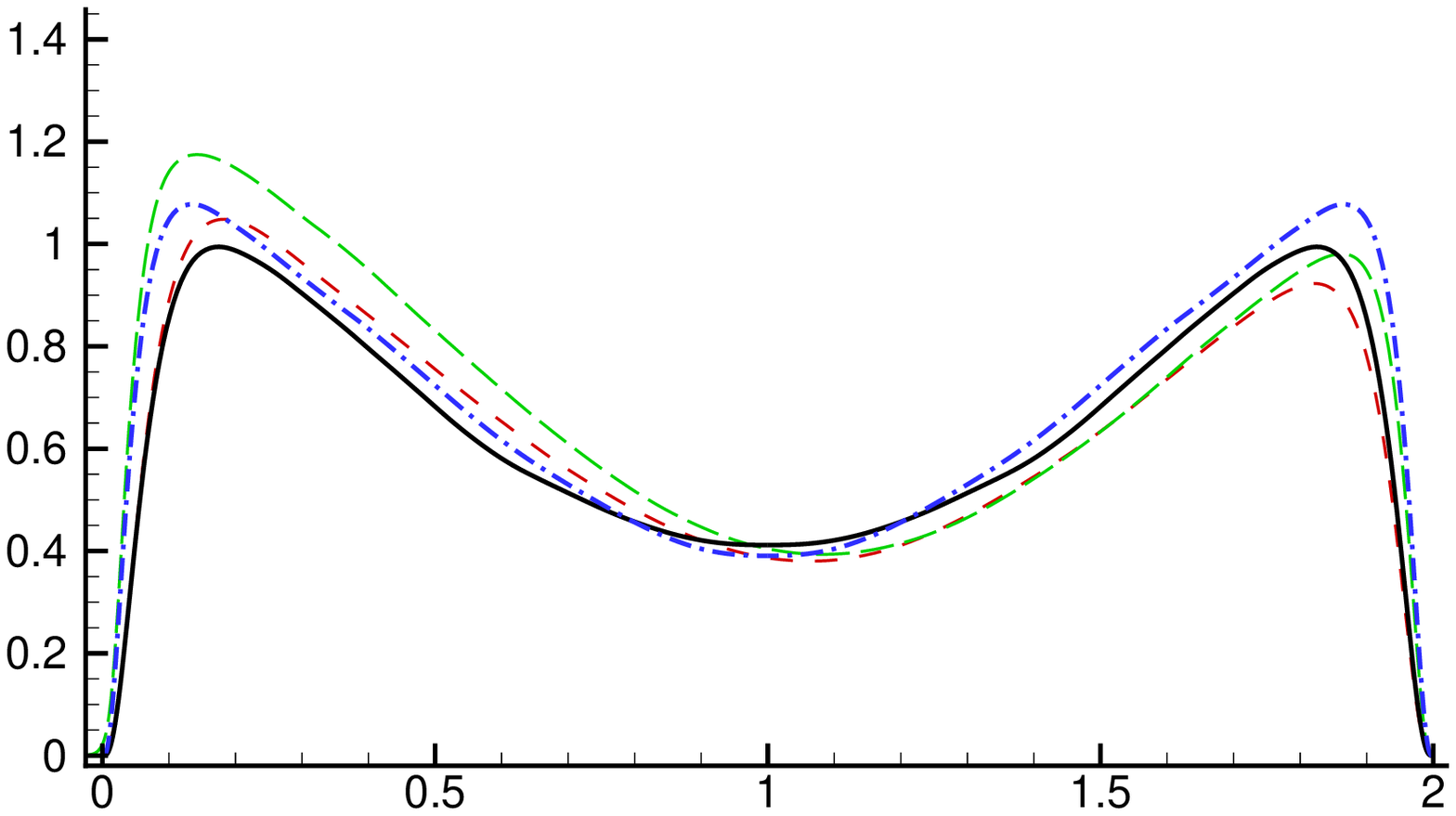}
  \put(-110,110){$(b)$}
 \put(-205,45){\rotatebox{90}{$\langle v'v' \rangle/(u_{\tau})^{2}$}}
 \put(-120,-5){$(y-y_0)/\delta$}
 \hspace{10mm}
 \includegraphics[height=45mm]{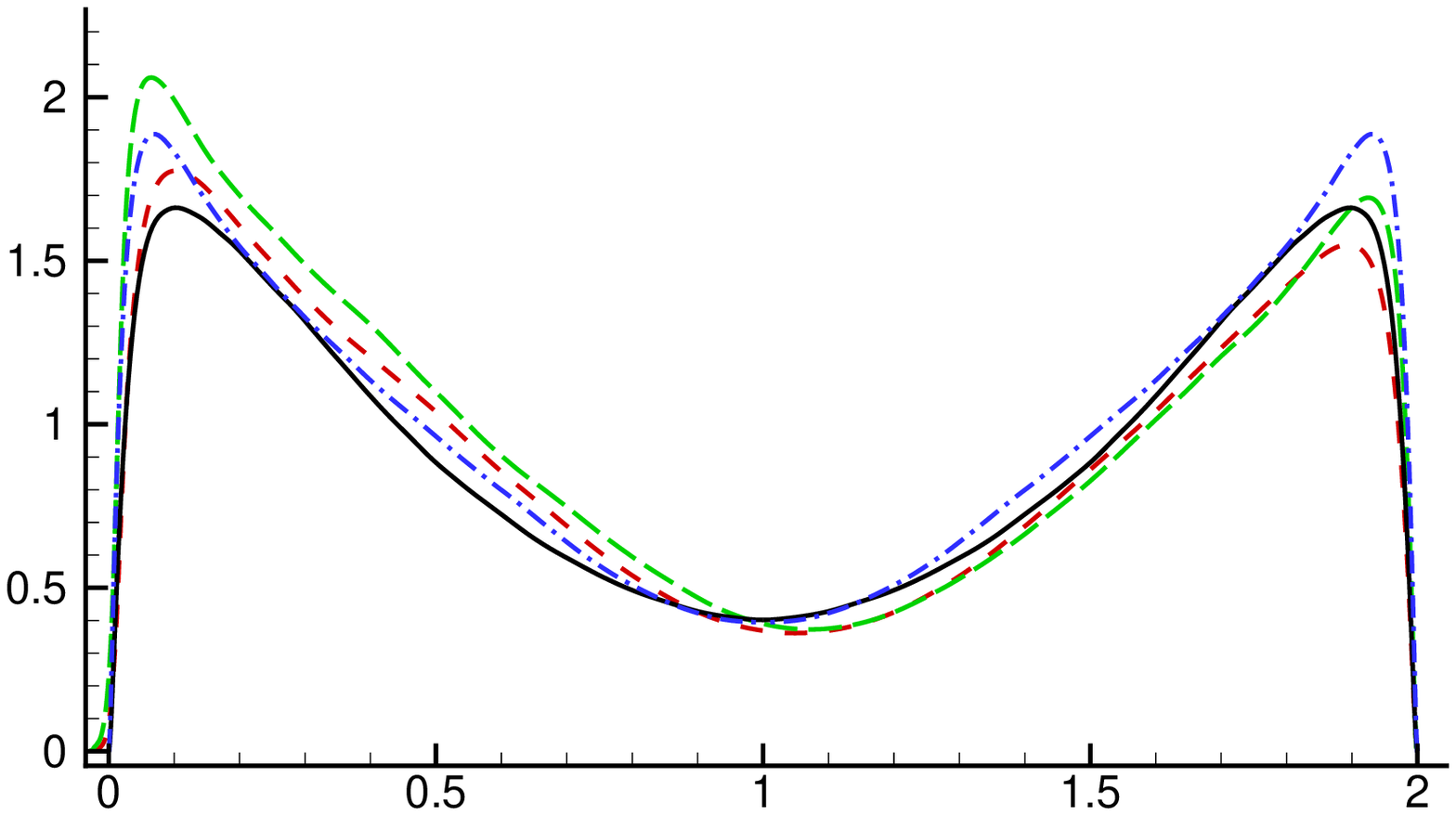}
  \put(-110,110){$(c)$}
 \put(-205,45){\rotatebox{90}{$\langle w'w' \rangle/(u_{\tau})^{2}$}}
 \put(-120,-5){$(y-y_0)/\delta$}
 \includegraphics[height=45mm]{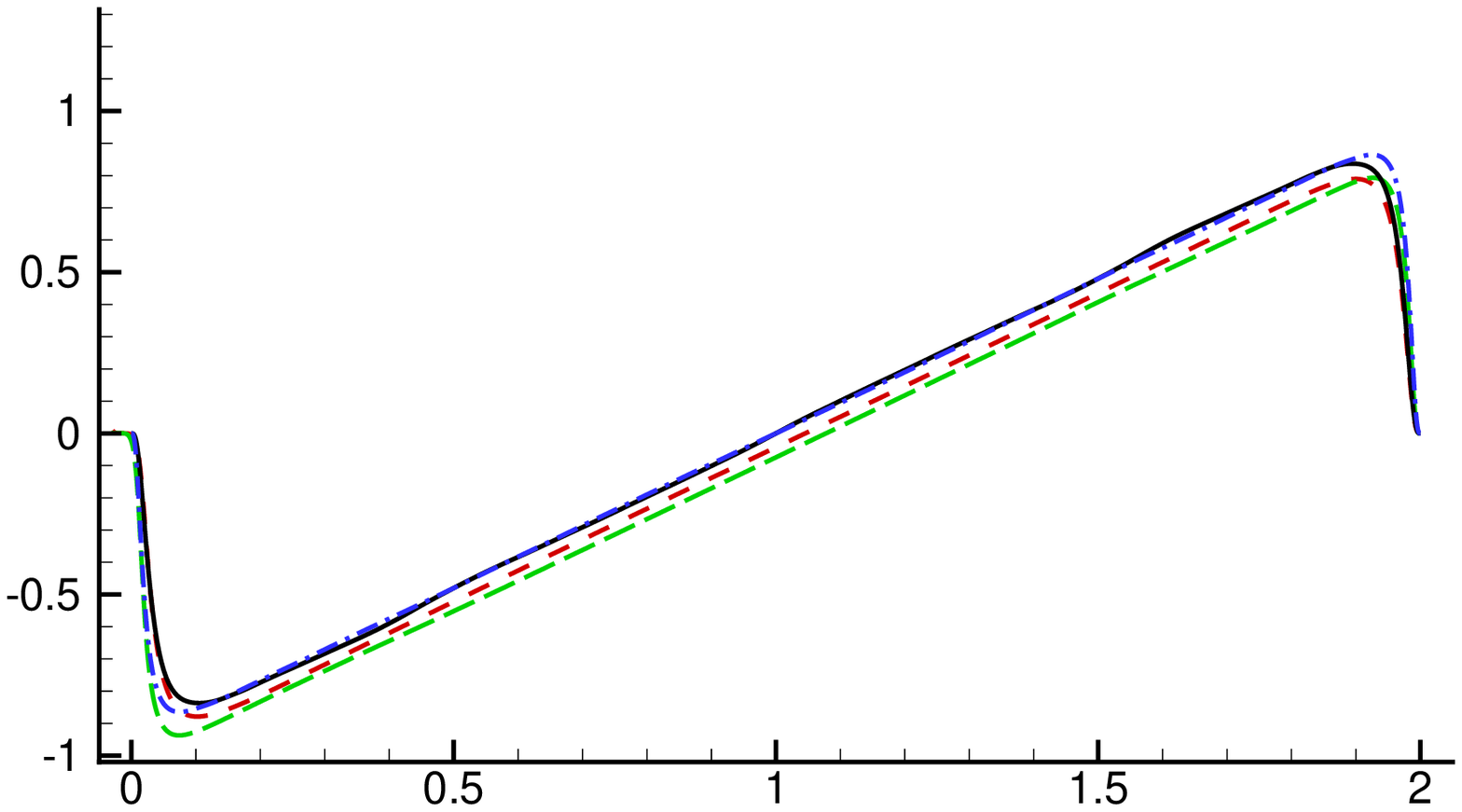}
  \put(-110,110){$(d)$}
 \put(-205,45){\rotatebox{90}{$\langle u'v' \rangle/(u_{\tau})^{2}$}}
 \put(-120,-5){$(y-y_0)/\delta$}
  \caption{Reynolds stress profiles normalized by $u_\tau$: $(a)$ streamwise, $(b)$ wall-normal, $(c)$ spanwise components, and $(d)$ Reynolds shear stress. The legend is the same as figure \ref{fig:umean_krms100}.}
\label{fig:intensity}
\end{figure}


The  Reynolds stresses scaled by $u_{\tau}^{2}$ are shown in  outer coordinates in figure \ref{fig:intensity}.  The location of the peak of $\langle u'u' \rangle^+$ is $Re$ dependent, as shown in figure \ref{fig:intensity}$(a)$, hence both smooth and rough cases at $Re_{\tau}=600$ are closer to the wall than  $Re_{\tau}=400$. The peak of $\langle u'u' \rangle^+$ in the lower half-channel for  
Case R400 is decreased by $2.8\%$ compared to the smooth wall of \cite{moser1999}. This behavior is consistent with \cite{busse2015direct}. The minimum value is shifted to the top wall by $7.9\%$ due to the asymmetry induced by the roughness. For $Re_{\tau}=600$, the peak value of $\langle u'u' \rangle^+$ is further decreased by $9.1\%$ and the minimum value is shifted to the top wall by $9.2\%$; i.e.  the roughness has a  more significant effect on the streamwise velocity fluctuations as $Re_{\tau}$ increases.

The  wall-normal Reynolds stress is shown in figure \ref{fig:intensity}$(b)$. Compared to the smooth channel,   $\langle v'v' \rangle^+$ in the lower half-channel is increased, while that in the upper half-channel is decreased. The peak value of $\langle v'v' \rangle^+$ in the lower half-channel is increased by $5.3\%$ at $Re_{\tau}=400$ and $8.9\%$ at $Re_{\tau}=600$. The spanwise velocity fluctuations also increase in the lower half-channel (figure \ref{fig:intensity}$(c)$), and they decrease in the upper half-channel compared to the smooth channel. The  Reynolds shear stress increases in magnitude at the rough wall, (figure \ref{fig:intensity}$(d)$), consistent with the increase in wall-normal velocity fluctuations.  The profile of $\langle u'v' \rangle^+$ shifts downwards by $4.9 \%$ at $Re_{\tau}=400$ and $8.4 \%$ at $Re_{\tau}=600$. 

\begin{figure}
\includegraphics[height=60mm]{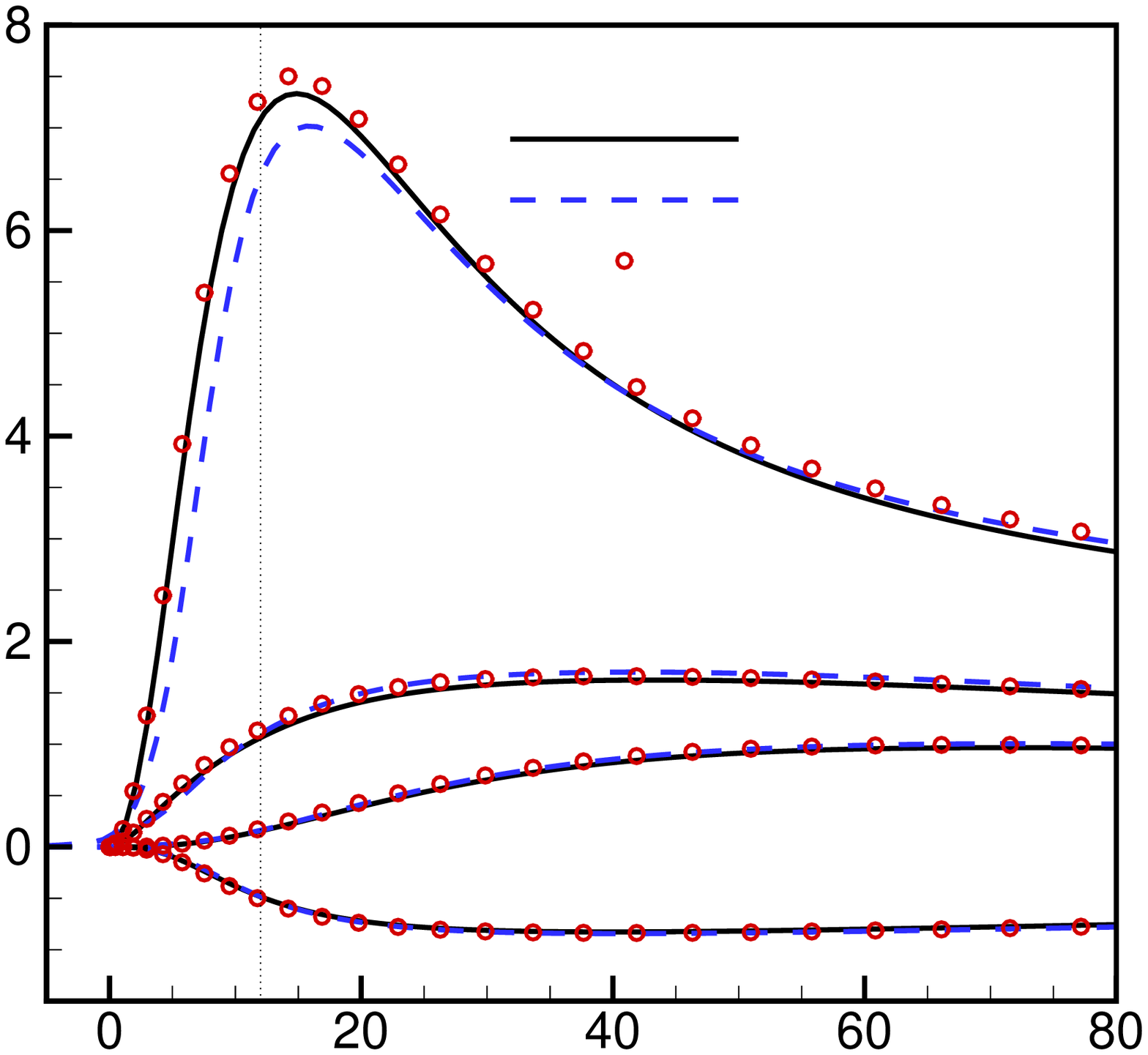}
\put(-100,160){$(a)$}
\put(-190,65){\rotatebox{90}{$\langle u_{i}'u_{j}'\rangle/(u_{\tau}^l)^{2}$}}
\put(-110,0){$(y-y_0)u_{\tau}^{l}/\nu$}
  \put(-122,100){$\langle u'^{2}\rangle^+$}
  \put(-125,67){$\langle w'^{2}\rangle^+$}
  \put(-115,40){$\langle v'^{2}\rangle^+$}
  \put(-95,30){$\langle u'v'\rangle^+$}
  \put(-72,134){\scriptsize{R400, top}}
  \put(-72,124){\scriptsize{R400, bottom}}
  \put(-72,114){\scriptsize{Moser et al.}}
\includegraphics[height=60mm]{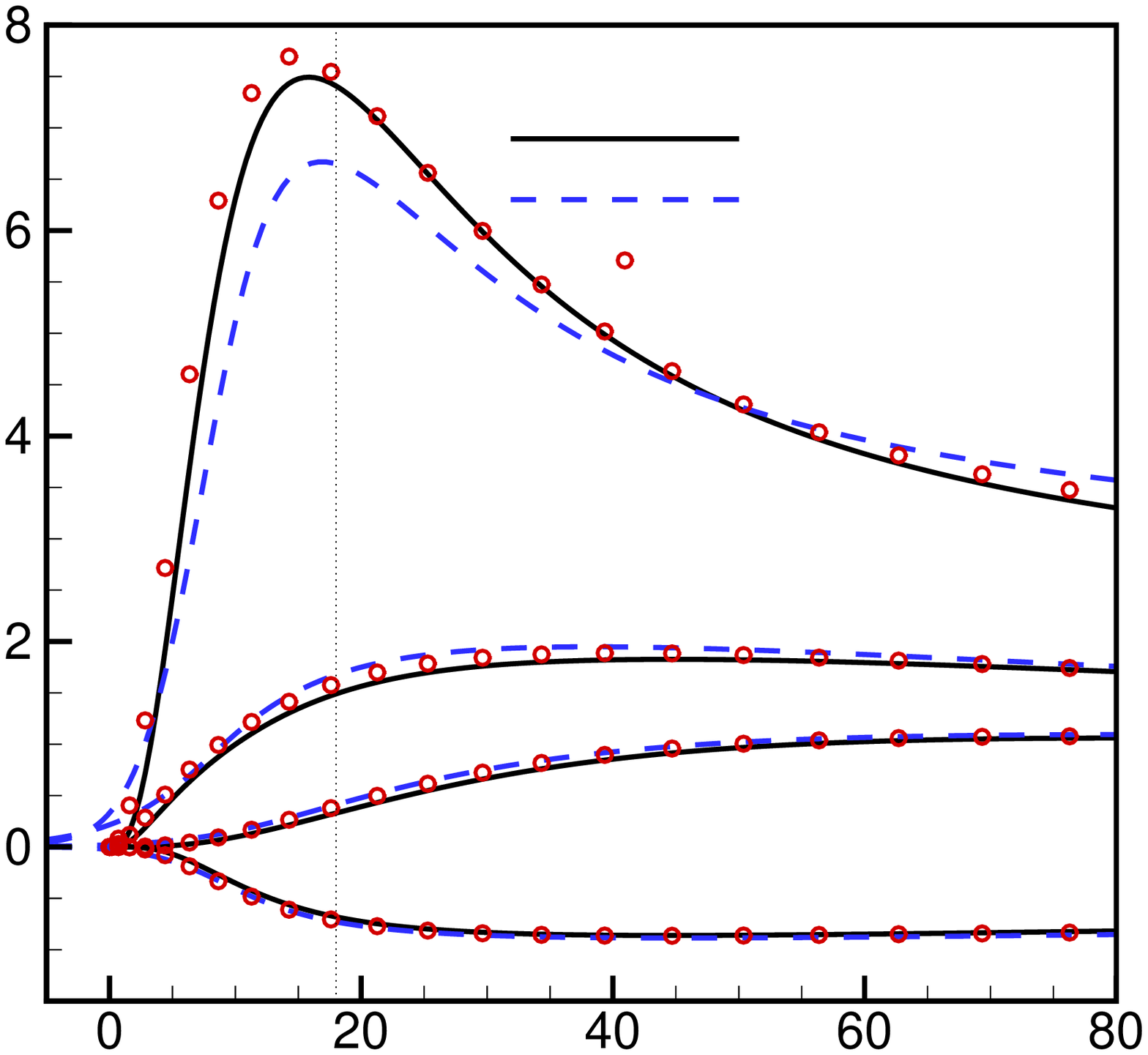}
\put(-100,160){$(b)$}
\put(-190,65){\rotatebox{90}{$\langle u_{i}'u_{j}'\rangle/(u_{\tau}^l)^{2}$}}
\put(-110,0){$(y-y_0)u_{\tau}^{l}/\nu$}
  \put(-122,100){$\langle u'^{2}\rangle^+$}
  \put(-125,70){$\langle w'^{2}\rangle^+$}
  \put(-115,40){$\langle v'^{2}\rangle^+$}
  \put(-95,30){$\langle u'v'\rangle^+$}
  \put(-72,134){\scriptsize{R600, top}}
  \put(-72,124){\scriptsize{R600, bottom}}
  \put(-72,114){\scriptsize{Moser et al.}}
\caption{Near-wall Reynolds stresses scaled by local friction velocity, $u_{\tau}^l=u_{\tau}^b$ for the bottom wall and $u_{\tau}^t$ for the top wall. $(a)$ Case R400 and $(b)$ Case R600. The vertical dashed line denotes the top of the roughness layer $k_c^+$.}
\label{fig:intensity_inner_nearpeak}
\end{figure}


The Reynolds stresses are normalized by the local friction velocity at each wall and plotted separately in figure \ref{fig:intensity_inner_nearpeak}. This scaling yields  better collapse in the logarithmic regions of the smooth and rough walls. In the near-wall region of the rough wall, the peak value of the streamwise component of the rough-wall side decreases. The spanwise component shows a higher peak value while the wall-normal component and Reynolds shear stress show relatively weaker variation in their peak values. These trends become more pronounced with  increasing $Re_{\tau}$. 

\cite{busse2015direct} studied a series of surfaces with different levels of filtering (a low-pass filter was used to remove the high-wavenumber contributions to the surface data). For their  smallest roughness height ($k_{rms}^+=4.6$) and largest skewness ($Sk=1.15$, where $Sk>0$ means the surface is peak-dominant), they found an increase in the peak of $\langle v'v'\rangle^+$, and a slight increase in both $\langle v'v'\rangle^+$ and $\langle u'v'\rangle^+$ in the roughness layer. Compared to their surfaces, the present rough surfaces have smaller roughness heights in wall units for cases R400 and R600, and zero skewness. That is, our surfaces have fewer, and smaller asperities to increase $\langle v'v'\rangle^+$ and $\langle u'v'\rangle^+$. 

\subsection{Dispersive flux and correlations} \label{subsec:dispersive flux}

\begin{figure}
\includegraphics[height=60mm]{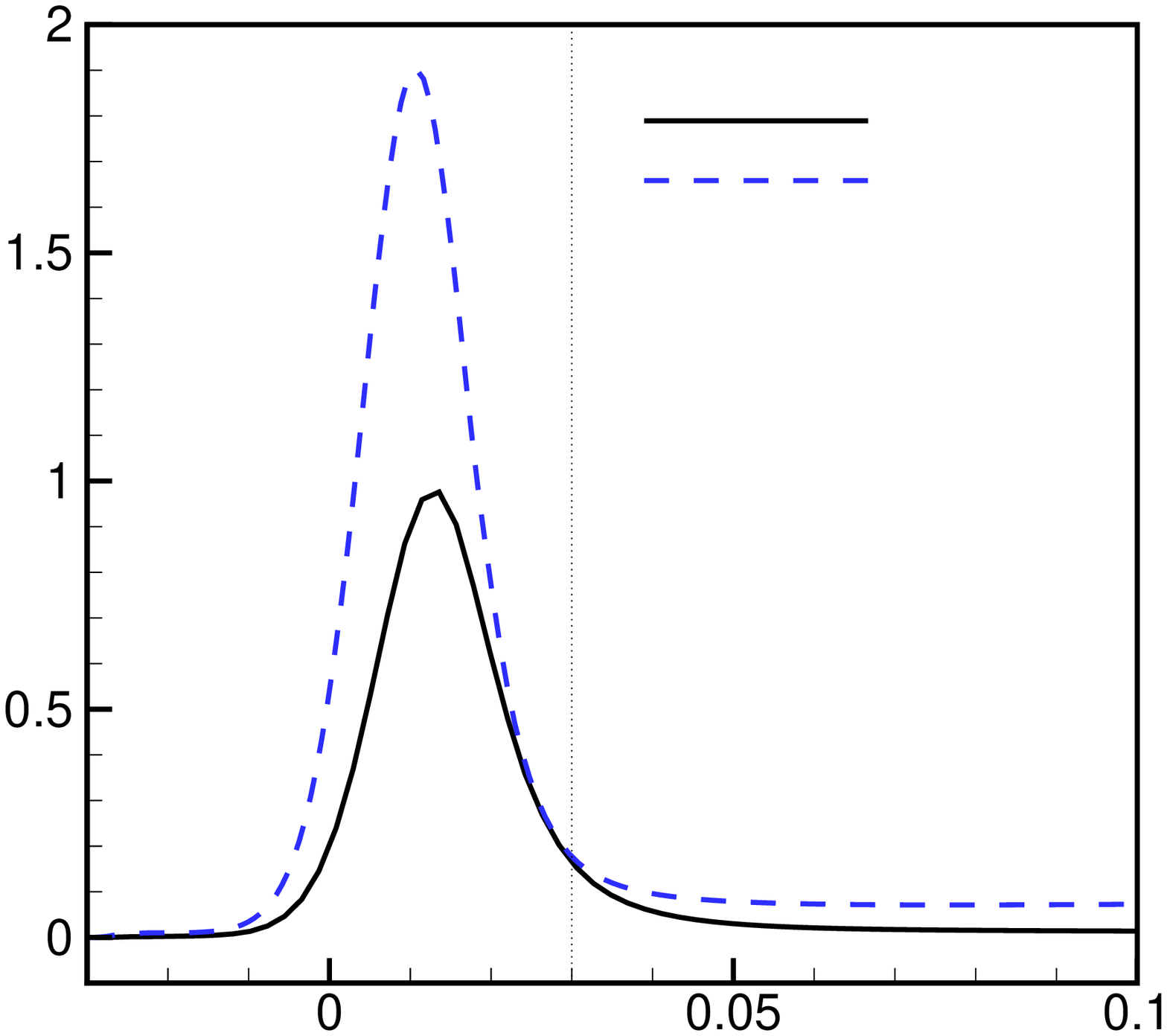}
\put(-100,160){$(a)$}
\put(-190,70){\rotatebox{90}{$\langle \Tilde{u_i} \Tilde{u_j} \rangle^+ $}}
\put(-115,0){$(y-y_0)/\delta$}
  \put(-55,135){R400}
  \put(-55,125){R600}
  \put(-155,85){$\langle \Tilde{u} \Tilde{u} \rangle^+ $}
\includegraphics[height=60mm]{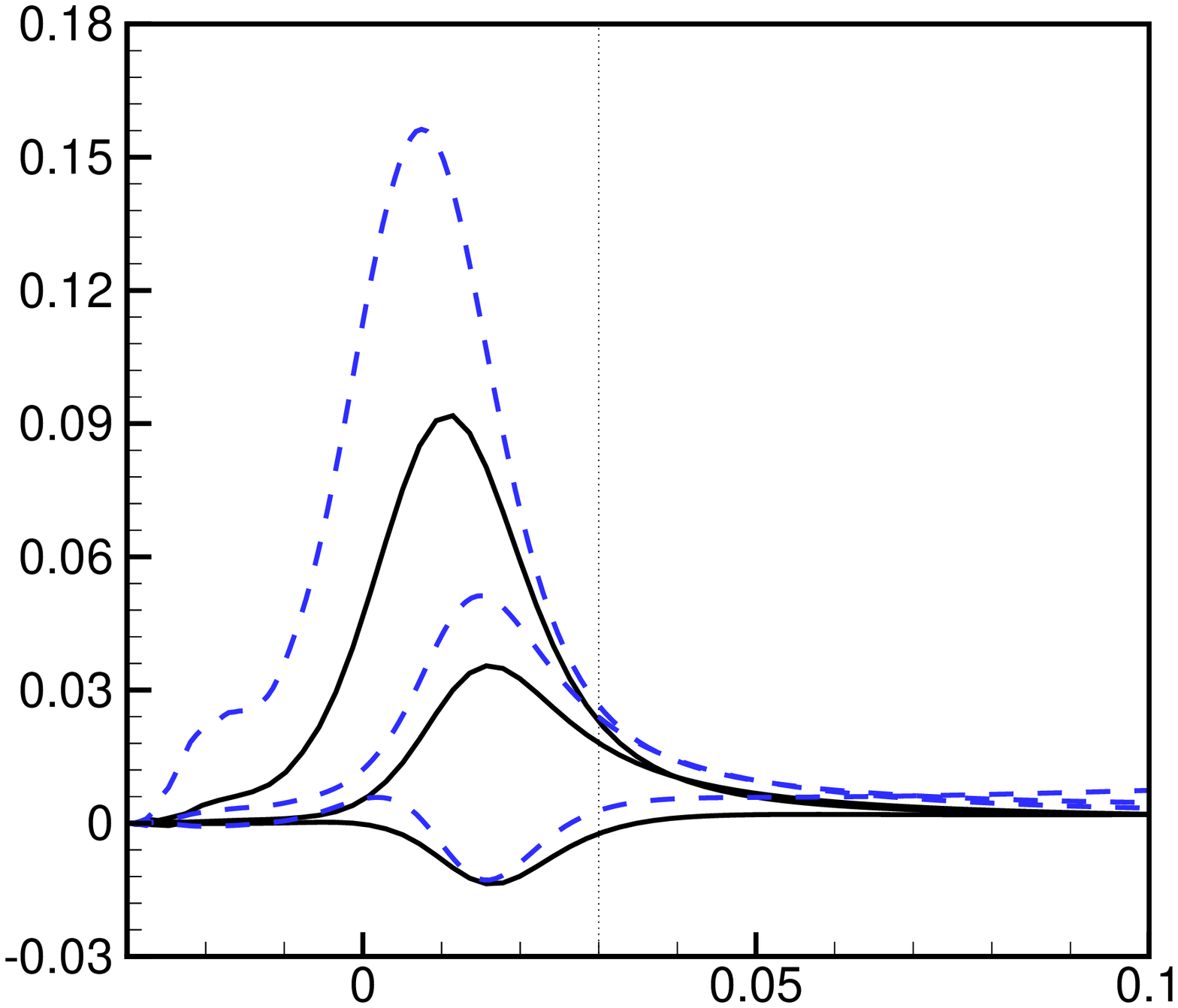}
\put(-100,160){$(b)$}
\put(-115,0){$(y-y_0)/\delta$}
\put(-95,50){$\langle \Tilde{v} \Tilde{v} \rangle^+ $}
\put(-160,100){$\langle \Tilde{w} \Tilde{w} \rangle^+ $}
\put(-100,25){$\langle \Tilde{u} \Tilde{v} \rangle^+ $}
\caption{Dispersive stress profiles  scaled by 
$u_{\tau}^b$ over the bottom half of the channel. The vertical dotted line denotes the top of the roughness layer $k_c$. The legend of $(b)$ is same as $(a)$.}
\label{fig:dispersive stress}
\end{figure}

The dispersive stresses obtained from double-averaging are examined in the near-wall region of the rough wall in figure \ref{fig:dispersive stress}. Compared to the Reynolds stresses, the dispersive stresses are mostly confined to the roughness layer,  consistent with the observations of  \cite{yuan2018topographical} and \cite{jelly2019reynolds}. For Case R400, $\langle \Tilde{u} \Tilde{u} \rangle^+$ is  maximum at $(y-y_0)/\delta=0.015$ and drops to zero by $(y-y_0)/\delta=0.05$, underlining the importance of spatial inhomogeneity in the roughness layer.  The profiles of $\langle \Tilde{v} \Tilde{v} \rangle^+$ and $\langle \Tilde{w} \Tilde{w} \rangle^+$  peak at $(y-y_0)/\delta=0.011$ and $0.016$, respectively. The dispersive shear stress, $\langle \Tilde{u} \Tilde{v} \rangle^+$, shows the same peak location as $\langle \Tilde{v} \Tilde{v} \rangle^+$ and is comparatively small. The level of dispersive stresses is enhanced for higher $Re_{\tau}$, especially in the roughness layer, and the peak values occur closer to the wall. 



\begin{figure}
\includegraphics[width=70mm]{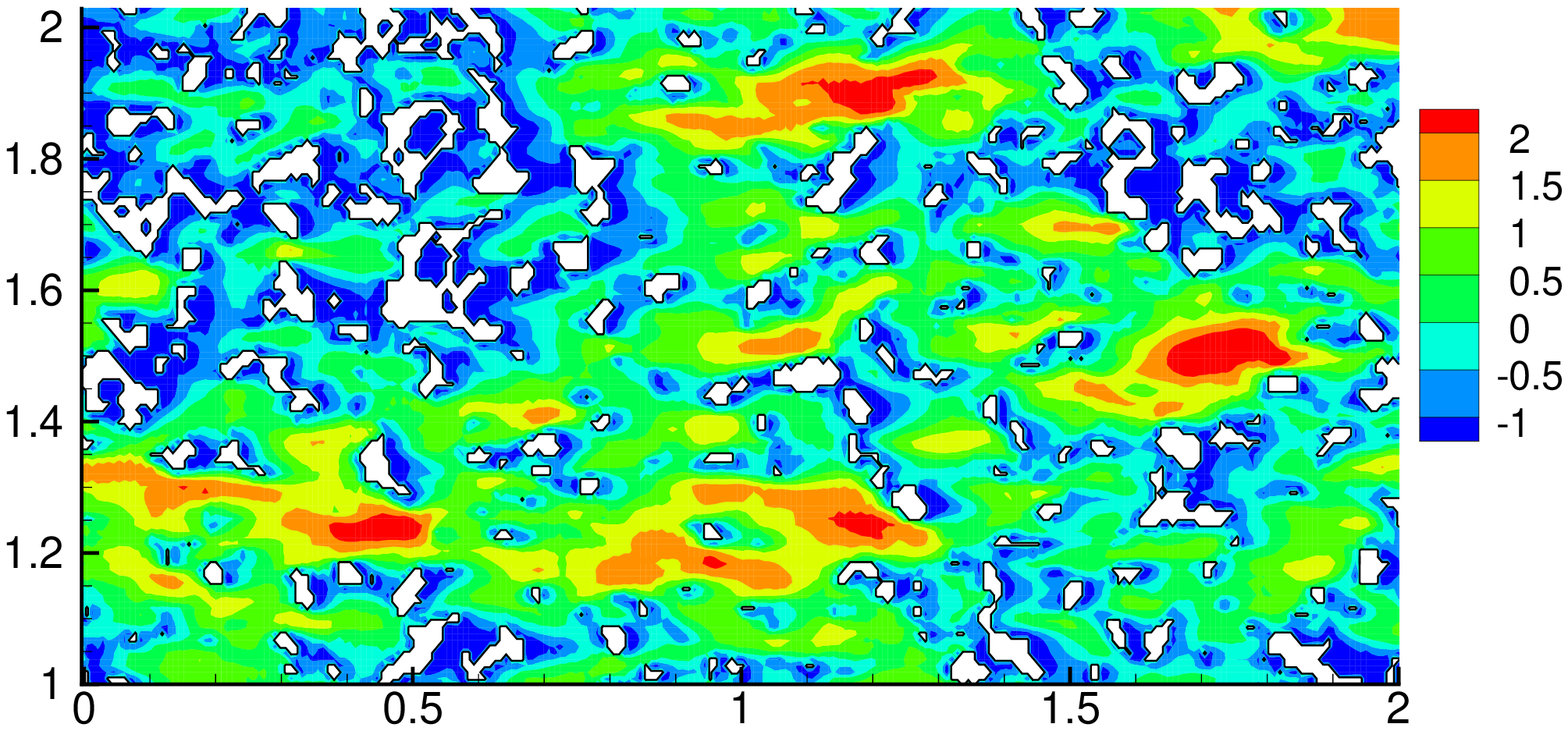}
\put(-195,43){\rotatebox{90}{$z/\delta$}}
\put(-105,-3){$x/\delta$}
\put(-20,80){$ \Tilde {u} ^+$}
 \put(-100,90){$(a)$}
\includegraphics[width=70mm]{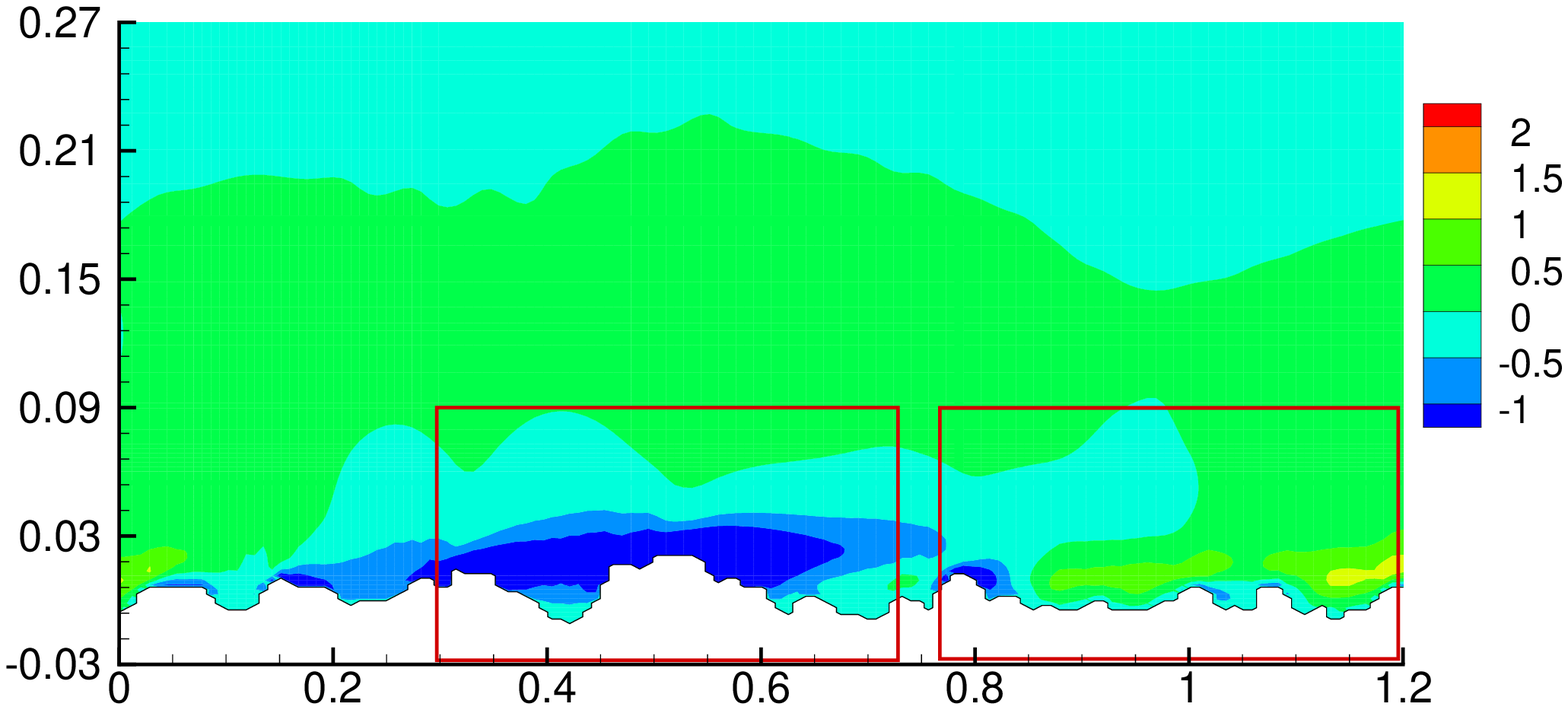}
\put(-198,30){\rotatebox{90}{$(y-y_0)/\delta$}}
\put(-105,-3){$x/\delta$}
\put(-115,30){$A$}
\put(-55,30){$B$}
\put(-20,80){$ \Tilde {u} ^+$}
 \put(-100,90){$(b)$}
\hspace{10mm}
\includegraphics[width=70mm]{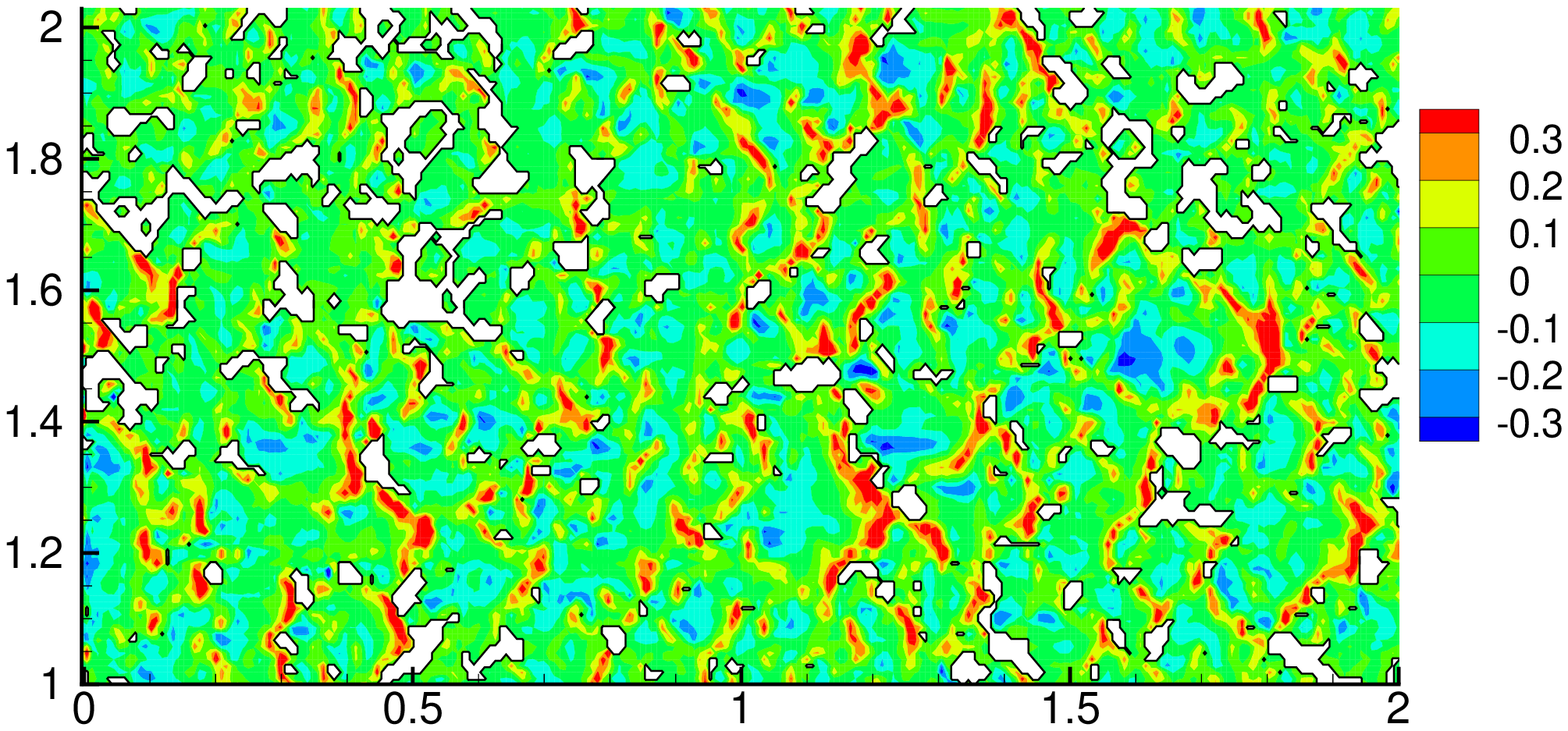}
\put(-195,43){\rotatebox{90}{$z/\delta$}}
\put(-105,-3){$x/\delta$}
\put(-20,80){$ \Tilde {v} ^+$}
 \put(-100,90){$(c)$}
 \includegraphics[width=70mm]{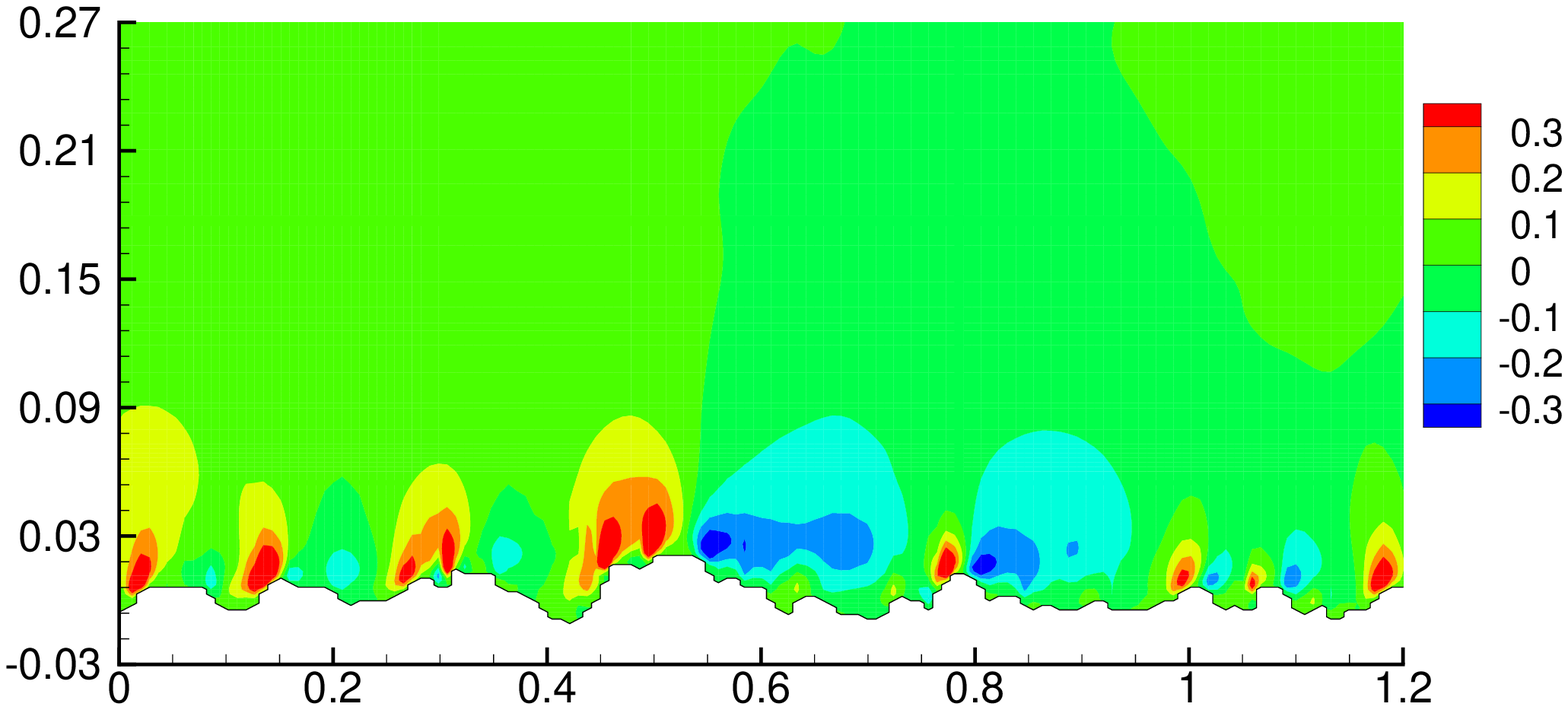}
\put(-198,30){\rotatebox{90}{$(y-y_0)/\delta$}}
\put(-105,-3){$x/\delta$}
\put(-20,80){$ \Tilde {v} ^+$}
 \put(-100,90){$(d)$}
 \hspace{10mm}
\includegraphics[width=70mm]{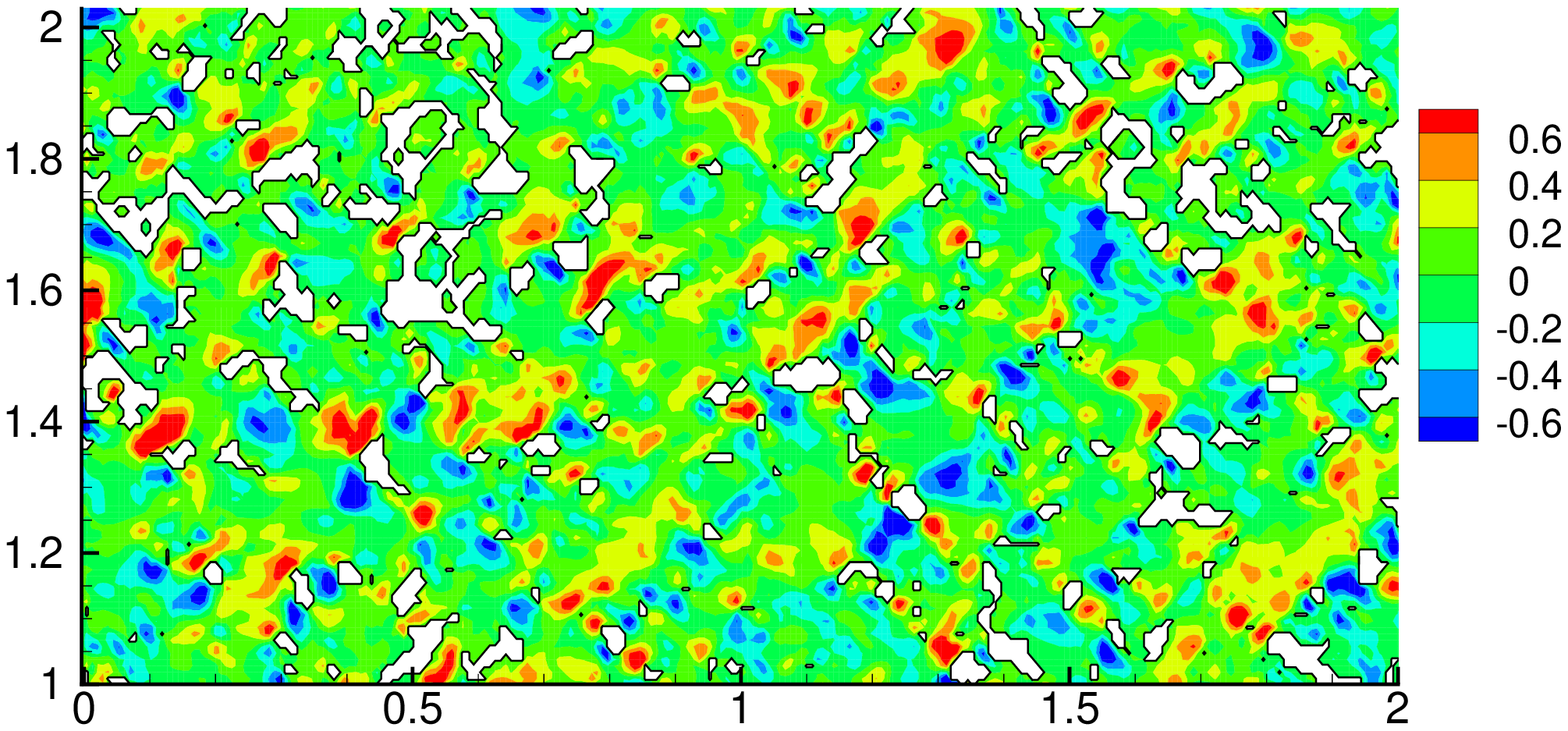}
\put(-195,43){\rotatebox{90}{$z/\delta$}}
\put(-105,-3){$x/\delta$}
\put(-20,80){$ \Tilde {w} ^+$}
 \put(-100,90){$(e)$}
 \includegraphics[width=70mm]{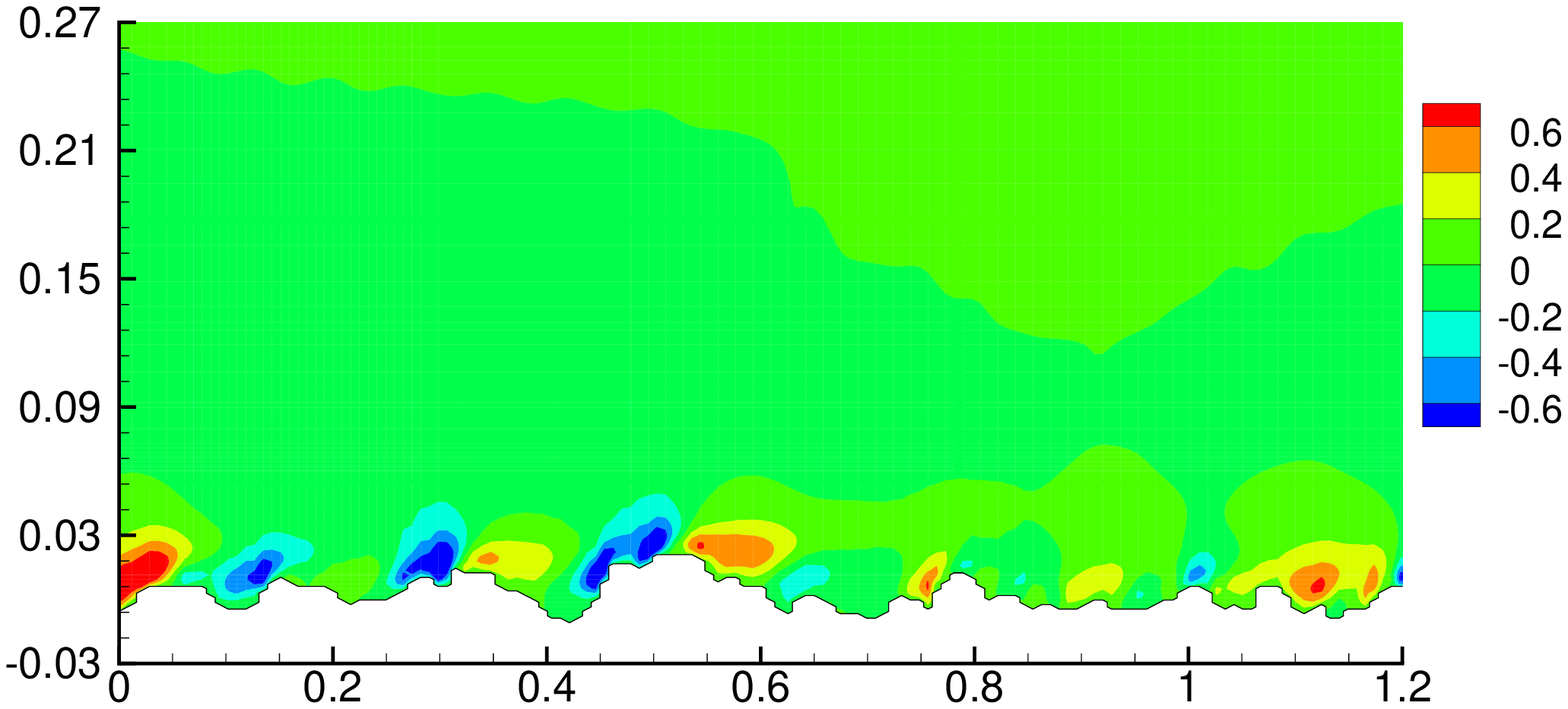}
\put(-198,30){\rotatebox{90}{$(y-y_0)/\delta$}}
\put(-105,-3){$x/\delta$}
\put(-20,80){$ \Tilde {w} ^+$}
 \put(-100,90){$(f)$}
  \hspace{10mm}
\includegraphics[width=70mm]{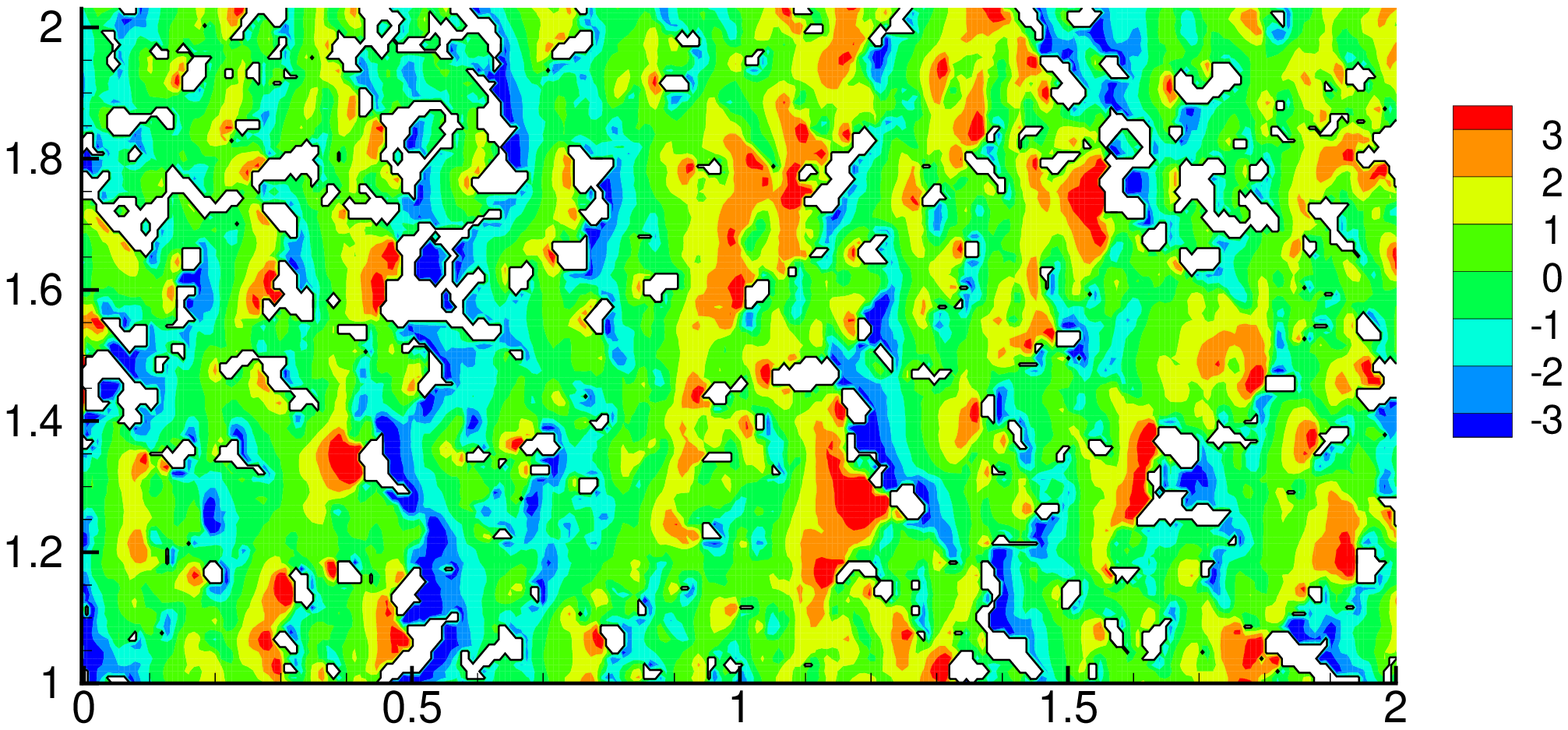}
\put(-195,43){\rotatebox{90}{$z/\delta$}}
\put(-105,-3){$x/\delta$}
\put(-20,80){$ \Tilde {p} ^+$}
 \put(-100,90){$(g)$}
 \includegraphics[width=70mm]{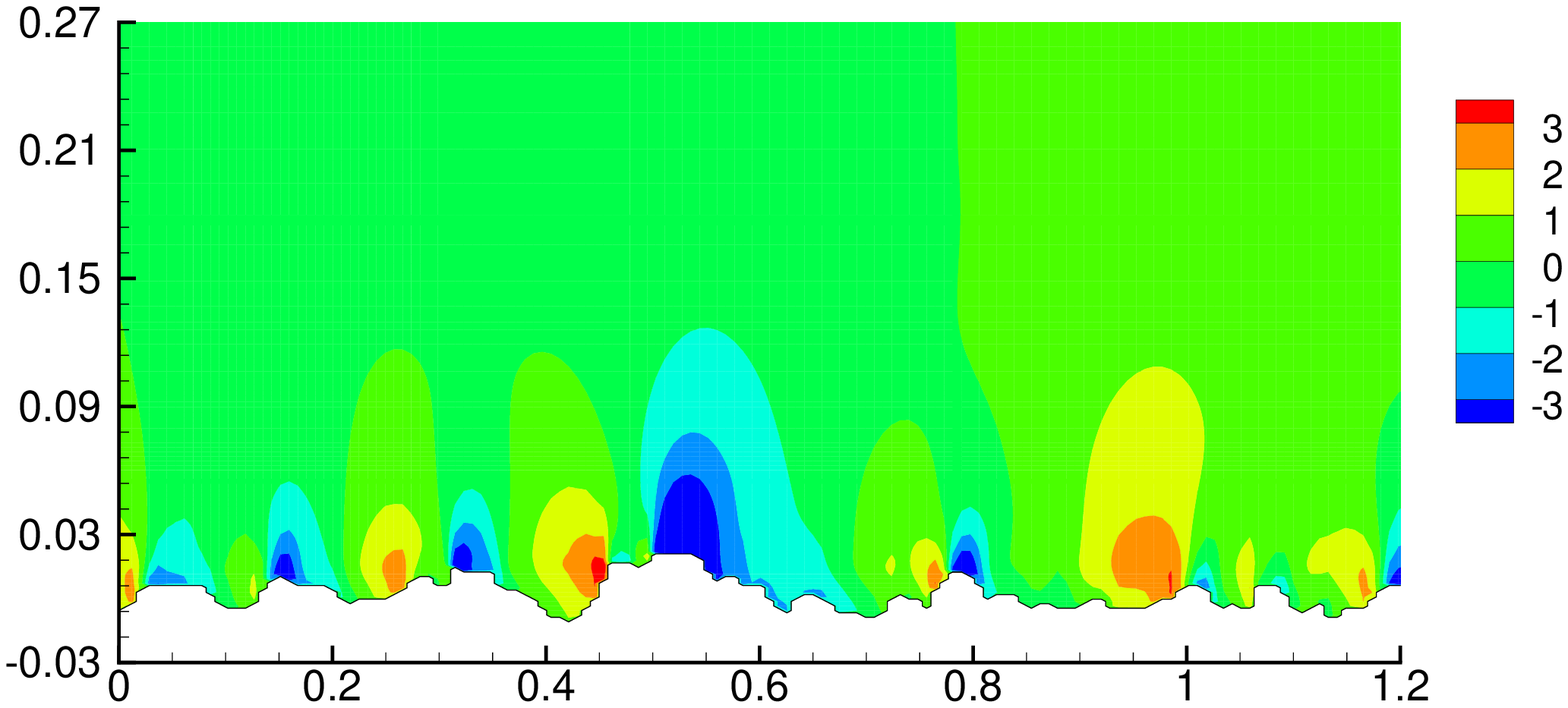}
\put(-198,30){\rotatebox{90}{$(y-y_0)/\delta$}}
\put(-105,-3){$x/\delta$}
\put(-20,80){$ \Tilde {p} ^+$}
 \put(-100,90){$(h)$}
\caption{Form-induced quantities for Case R400 at $(y-y_0)/\delta=0.0075$ ($y^+=3$): ($a$) $ \Tilde {u} ^+$, ($c$) $ \Tilde {v} ^+$, ($e$) $ \Tilde {w} ^+$, ($g$) $ \Tilde {p} ^+$. A slice at $z/\delta=1.57$: ($b$) $ \Tilde {u} ^+$, ($d$) $ \Tilde {v} ^+$, ($f$) $ \Tilde {w} ^+$, ($h$) $ \Tilde {p} ^+$. The velocities  are normalized by $u_{\tau}^b$ and pressure is normalized by $(u_{\tau}^b)^2$. The rectangular regions labeled by $A$ and $B$ are two representative roughness geometries for the following discussion. } \label{fig:tilt_R400}
\end{figure}

 \begin{figure}
\includegraphics[height=60mm]{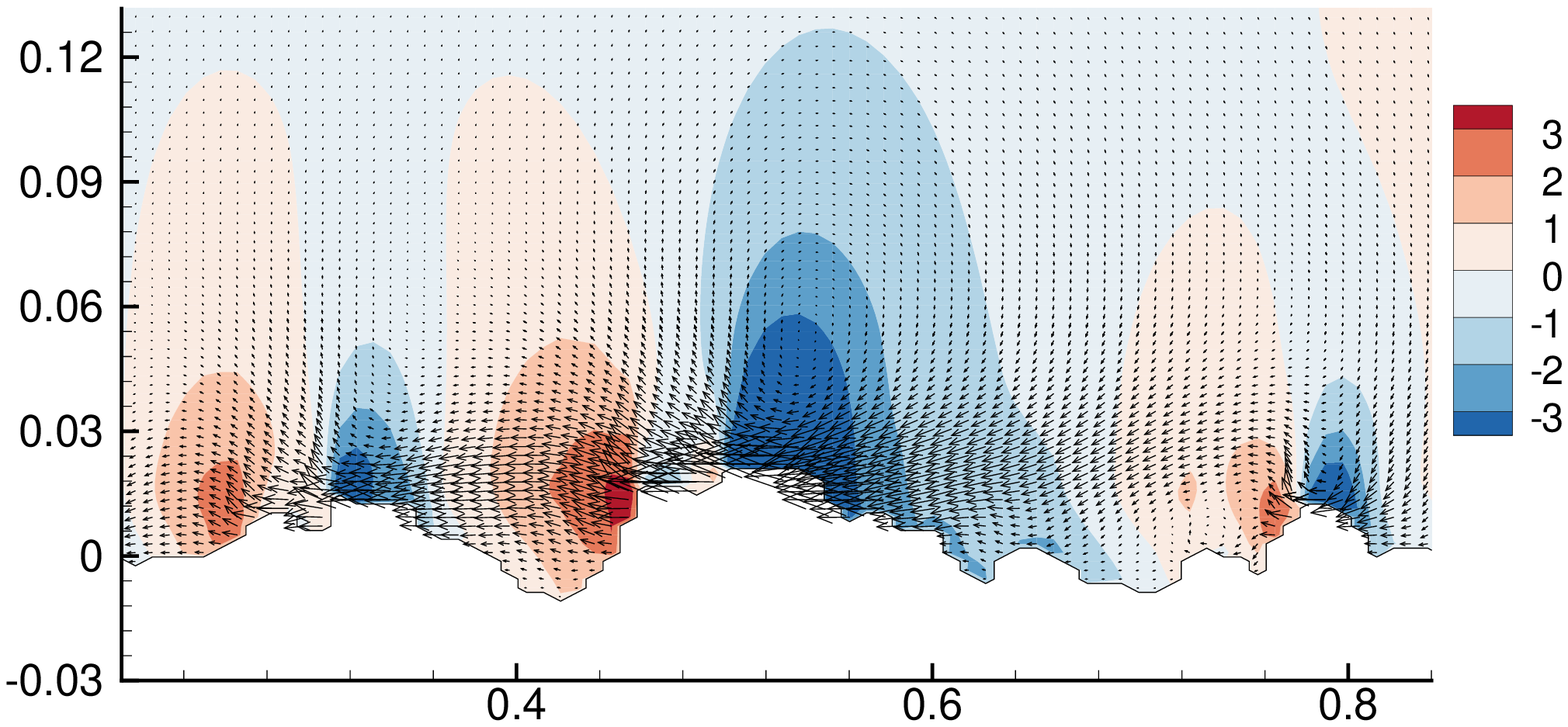}
\put(-170,155){$(a)$}
\put(-170,3){$x/\delta$}
\put(-330,60){\rotatebox{90}{$(y-y_0)/\delta$}}
\put(-30,140){$ \Tilde {p} ^+$}
\put(-285,130){$A$}
 \hspace{5mm}
\includegraphics[height=60mm]{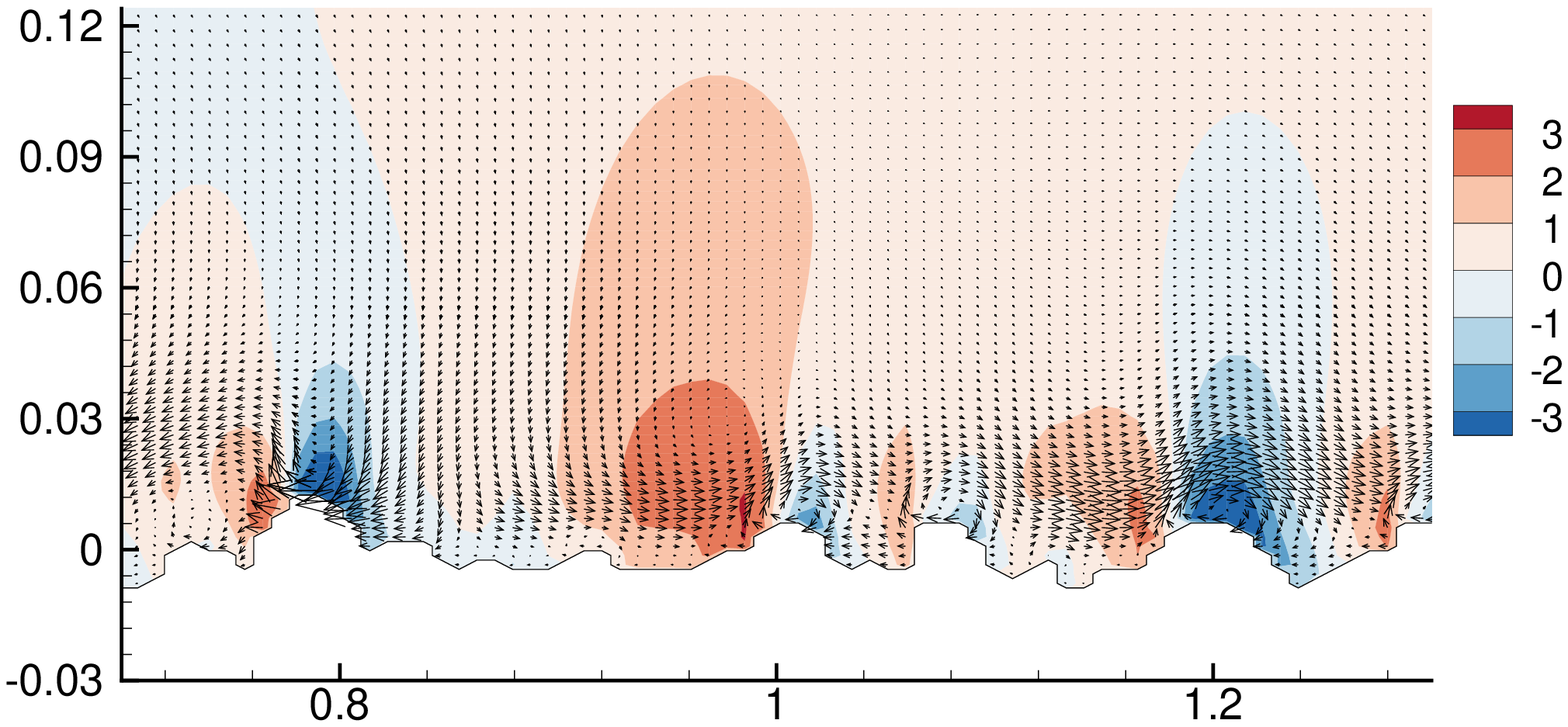}
\put(-170,155){$(b)$}
\put(-170,3){$x/\delta$}
\put(-330,60){\rotatebox{90}{$(y-y_0)/\delta$}}
\put(-30,140){$ \Tilde {p} ^+$}
\put(-285,130){$B$}
\caption{Vectors of ($\Tilde{u}$,$\Tilde{v}$) with  background contours of $\Tilde{p}^+$ for representative roughness sections. $(a)$ and $(b)$ correspond to the labeled region $A$ and $B$ in figure \ref{fig:tilt_R400}$(b)$.} 
\label{fig:vector}
\end{figure}  

Dispersive stresses arise from regions where the local temporal average is different from the average computed over time and the entire rough wall. The form-induced velocity and pressure are examined at $(y-y_0)/\delta=0.0075$ and $z/\delta=1.57$ in figure \ref{fig:tilt_R400} to illustrate how the roughness topography contributes to the local inhomogeneity in the roughness layer. Figure \ref{fig:tilt_R400}$(a)$ shows that large positive $\Tilde{u}$ (red region) occurs in the large trough regions between roughness asperities, while large negative $\Tilde{u}$ (blue region) occurs in the wake regions behind the roughness elements. Both these high and low-momentum regions contribute to the streamwise dispersive stress. In the areas where the roughness protrusions are relatively peaky and dense, corresponding to the $x$-$y$ plane probed in figure \ref{fig:tilt_R400}$(b)$,  regions with negative $\Tilde{u}$ are observed near the roughness protrusions. 

Figures \ref{fig:tilt_R400}$(c)$ and \ref{fig:tilt_R400}$(d)$ show that impulsive upward velocities, shown by the high-magnitude positive $\Tilde{v}$ (red regions), occur mostly in front of the roughness protrusions.  Downward velocities, shown by the negative $\Tilde{v}$ (blue regions), occur in the wake of the protrusions. The strength of these upward and downward motions is positively correlated with the roughness height. 

Figures \ref{fig:tilt_R400}$(e)$ and \ref{fig:tilt_R400}$(f)$ show pairs of  large positive $\Tilde{w}$ (red region) and negative $\Tilde{w}$ (blue region)  in front of the roughness crests. This suggests that the impulsive upward velocity produces a pair of streamwise vortices in front of the roughness elements. Similar behavior was observed by \cite{muppidi2012direct} in their investigation of an idealized rough-wall supersonic boundary layer. 
Also,  $\Tilde{v}$ displays larger wall-normal length scales than  $\Tilde{w}$, which explains why  the peak location of $\langle \Tilde{v} \Tilde{v} \rangle^+$ is higher than that of $\langle \Tilde{w} \Tilde{w} \rangle^+$ in figure \ref{fig:dispersive stress}$(b)$. 

The pressure perturbations at the same locations are shown in figures \ref{fig:tilt_R400}$(g)$ and \ref{fig:tilt_R400}$(h)$. High $\Tilde{p}$ (red region) is found to occur in front of the roughness asperities where the flow stagnates, while low $\Tilde{p}$ (blue region) occurs at the crests and behind the roughness elements. The form drag is therefore mainly produced by the peaks of the rough surface.

The form-induced quantities for Case R600 were also examined but are not shown here in the interests of brevity. The main features remain the same, while their magnitudes increase at the higher  $Re_{\tau}$.

Figure \ref{fig:vector} takes a closer look at how roughness protrusions contribute to mean wall-normal fluxes. Two representative regions labeled  $A$ and $B$ in figure \ref{fig:tilt_R400}$(b)$ are considered. Region A represents a dense and peaky region where the flow is dominated by negative $\Tilde{u}$, while B is a relatively flat and smooth region. Both figure \ref{fig:vector}$(a)$ and \ref{fig:vector}$(b)$ show that  negative $\Tilde{u}$ is induced at the back of the protrusions, and the biggest crest yields the strongest negative $\Tilde{u}$. The upward $\Tilde{v}$ induced in front of the crests accompanies the negative $\Tilde{u}$, resulting in  "ejection" motions, borrowing terminology from the boundary layer literature \citep{bailey2013turbulence}. Similar strong "ejection" motions are present in front of the crests while "sweeping" motions occur within the troughs. Both "ejection" and "sweeping" motions are responsible for the negative dispersive shear stress shown in figure \ref{fig:dispersive stress}$(b)$. A clockwise "roll-up" motion is shown at the roughness crests in figure \ref{fig:vector}$(b)$; such motions occur when the surface topography is less rough and the flow is not dominated by negative $\Tilde{u}$. Thus roughness geometry influences wall-normal momentum transfer by influencing  the "ejection" and "roll-up" motions. 




 \begin{figure}
\centering
\includegraphics[height=40mm]{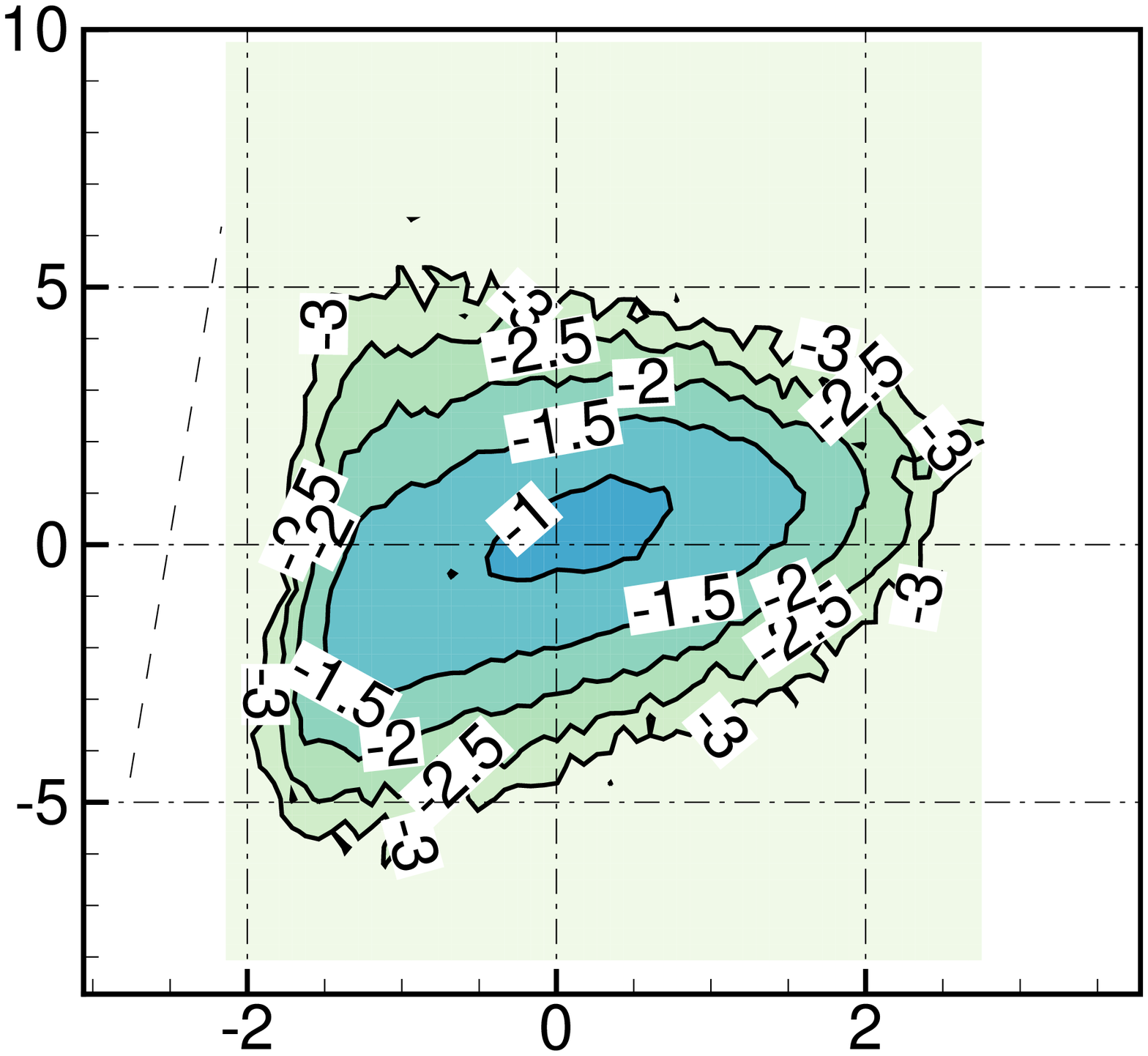}
\put(-70,110){$(a)$}
\put(-130,55){\rotatebox{90}{$\Tilde{p}^+$}}
\put(-70,-3){$\Tilde{u}^+$}
\put(-45,90){\scriptsize{$y^+=3$}}
\hspace{5mm}
\includegraphics[height=40mm]{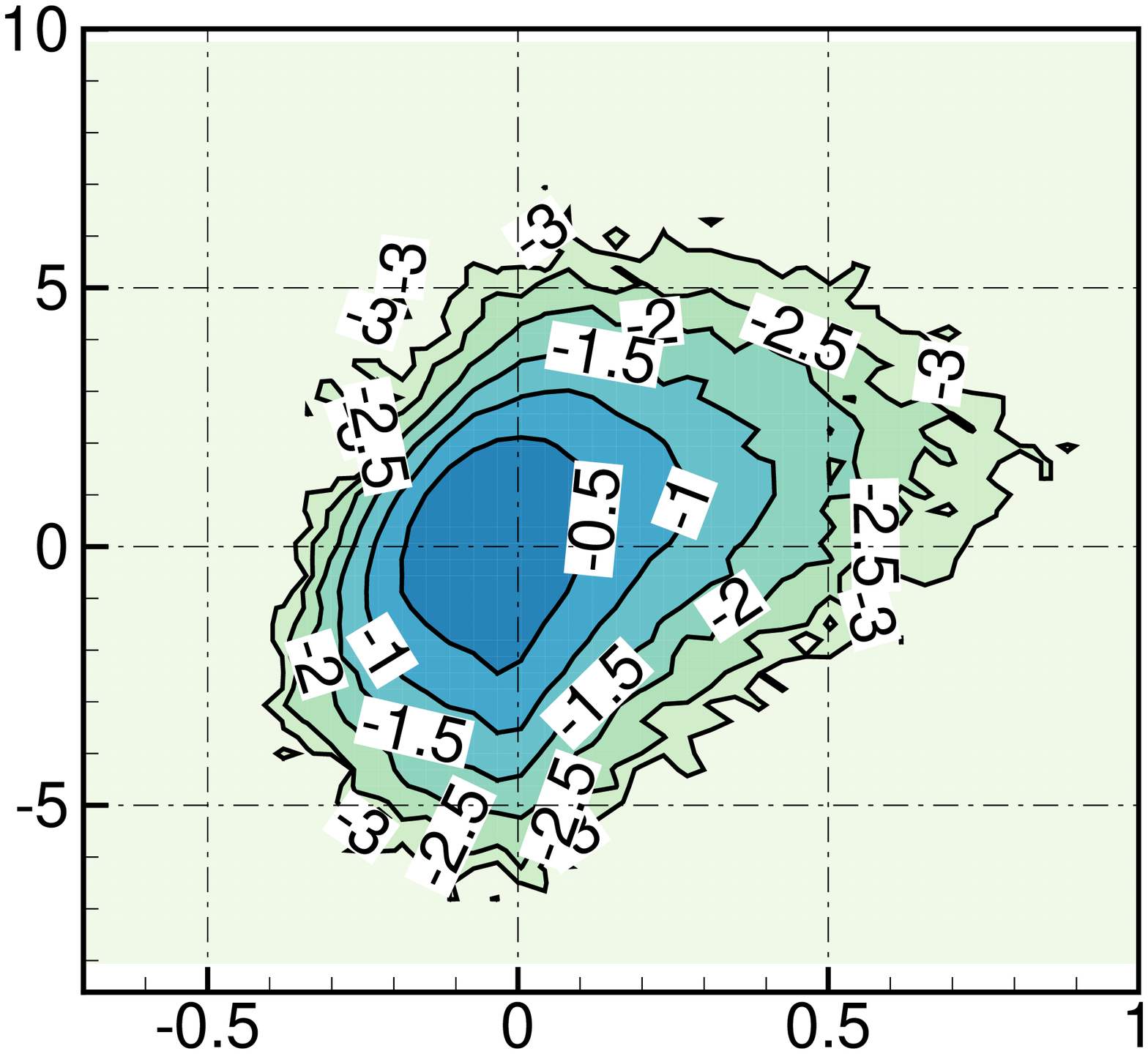}
\put(-70,110){$(b)$}
\put(-130,55){\rotatebox{90}{$\Tilde{p}^+$}}
\put(-70,-3){$\Tilde{v}^+$}
\put(-45,90){\scriptsize{$y^+=3$}}
\caption{Joint p.d.f of $(a)$ ($\Tilde{u}^+$, $\Tilde{p}^+$), and $(b)$ ($\Tilde{v}^+$, $\Tilde{p}^+$) for Case R400 at the wall-normal location $(y-y_0)/\delta=0.0075$ ($y^+=3$). Contour levels are defined by the common logarithm of joint p.d.f and the interval between successive contour levels is 0.5.}
\label{fig:jpdf_tilt_R400}
\end{figure}
Figure \ref{fig:jpdf_tilt_R400} shows the joint p.d.f of  ($\Tilde{u}$, $\Tilde{p}$) and ($\Tilde{v}$, $\Tilde{p}$) near the roughness mean height at $y^+=3$ for Case R400. The results for Case R600 are  not shown since the main conclusions remain the same. 
 Stronger correlations between $\Tilde{u}$ and $\Tilde{p}$ are observed in the third quadrant $Q3$ compared to the smooth channel \citep{lenaers2012rare}, which corresponds to the "roll-up" motions induced by the roughness crests. The regions with these motions account for the negative $\Tilde{u}$ and correlate with the negative $\Tilde{p}$, as shown in figure \ref{fig:vector}$(b)$.
 \cite{lenaers2012rare} found a correlation between negative $v$ and positive $p'$ near smooth walls, indicating  splatting events. In the rough case, however, the correlation in the second quadrant $Q2$ is weak since positive $p'$ is more due to stagnation  on the roughness elements. Meanwhile, the correlation in $Q1$ and $Q3$ is relatively stronger since the upward $\Tilde{v}$ is always induced in front of the crests where positive $\Tilde{p}$ occurs, while the downward $\Tilde{v}$ appears in the wake regions where negative $\Tilde{p}$ occurs. \\

\subsection{Mean momentum balance}

The mean momentum balance (MMB)  reveals the different contributions to momentum conservation in the roughness layer. The streamwise DA momentum equation for a fully-developed channel flow \citep{raupach1982averaging,yuan2018topographical} is 



\begin{equation}
 \underbrace{\frac{\partial \langle \Tilde{u} \Tilde{v} \rangle}{\partial y}}_{\text{A}} = -\underbrace{\frac{\partial \langle \overline{p} \rangle}{\partial x}}_{\text{B}} + \underbrace{\nu \frac{\partial^2  \langle \overline{u} \rangle}{\partial y^2}}_{\text{C}} -\underbrace{ \big \langle \frac{\partial \Tilde{p} }{\partial x} \big \rangle}_{\text{D}} + \underbrace{\nu \big \langle \frac{\partial^2 \Tilde{u} }{\partial x_j^2}\big \rangle}_{\text{E}} -
 \underbrace{\frac{\partial \langle \overline{u'v'} \rangle}{\partial y}}_{\text{F}} +  \langle K \rangle,
 \label{eqn:mmb}
\end{equation}
where the subscript $j$ denotes streamwise, wall-normal and spanwise components for position vectors and $K$ is the constant pressure gradient (divided by density). In this equation, term A is the dispersive shear stress gradient, term B represents the mean pressure gradient which is zero in our simulations, term C shows the viscous force (viscous stress gradient), term D and E account for the pressure drag and viscous drag caused by the roughness respectively, and term F is the net effect of turbulent inertia (Reynolds stress gradient).

 \begin{figure}
 \centering
\includegraphics[height=60mm]{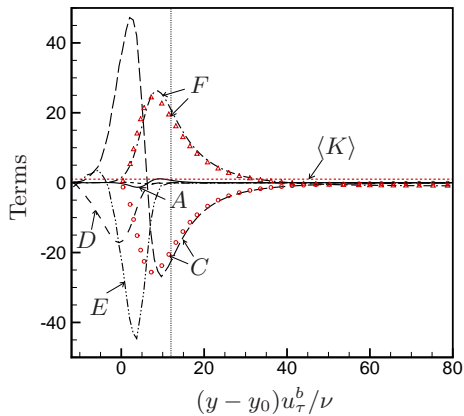}
\put(-190,70){\rotatebox{90}{Terms}}
\put(-120,0){$(y-y_0)u_{\tau}^b/\nu$}
\put(-130,75){$A$}
\put(-122,50){$C$}
\put(-165,60){$D$}
\put(-160,35){$E$}
\put(-122,118){$F$}
\put(-75,95){$\langle K \rangle$}
\caption{Budget of the streamwise DA momentum equation for Case SW (symbols) and R400 (lines). The vertical dotted line denotes the top of the roughness layer $k_c^+$. The summation of all terms is denoted by the short dashed line.}
\label{fig:DA_mmb}
\end{figure}


Figure \ref{fig:DA_mmb} shows the budget of terms in the mean momentum equation for the bottom half channel. Since terms A, D, and E are zero for Case SW, the balance is among the constant pressure gradient, the viscous and Reynolds stress gradients. For Case R400, the Reynolds stress gradient is slightly increased at the wall and the peak but follows a similar variation as Case SW in general. As $y^+$ increases, the viscous stress gradient increases and reaches a maximum value at $y^+=3$, then decreases to negative values and collapses with Case SW above the roughness layer. The pressure and viscous drag are induced by the roughness and only exist in the roughness layer. Compared to other terms, the contribution of the dispersive term is quite small. Overall, the roughness transfers  mean momentum downward through the viscous and Reynolds stress gradients to balance the pressure and viscous drag. The observation for Case R600 is the same except that the magnitude of each term is larger than Case R400. 

\subsection{Pressure fluctuations}\label{subsec:mean pressure}
\subsubsection{Mean-square pressure fluctuations}

\begin{figure}
\centering
 \includegraphics[height=50mm]{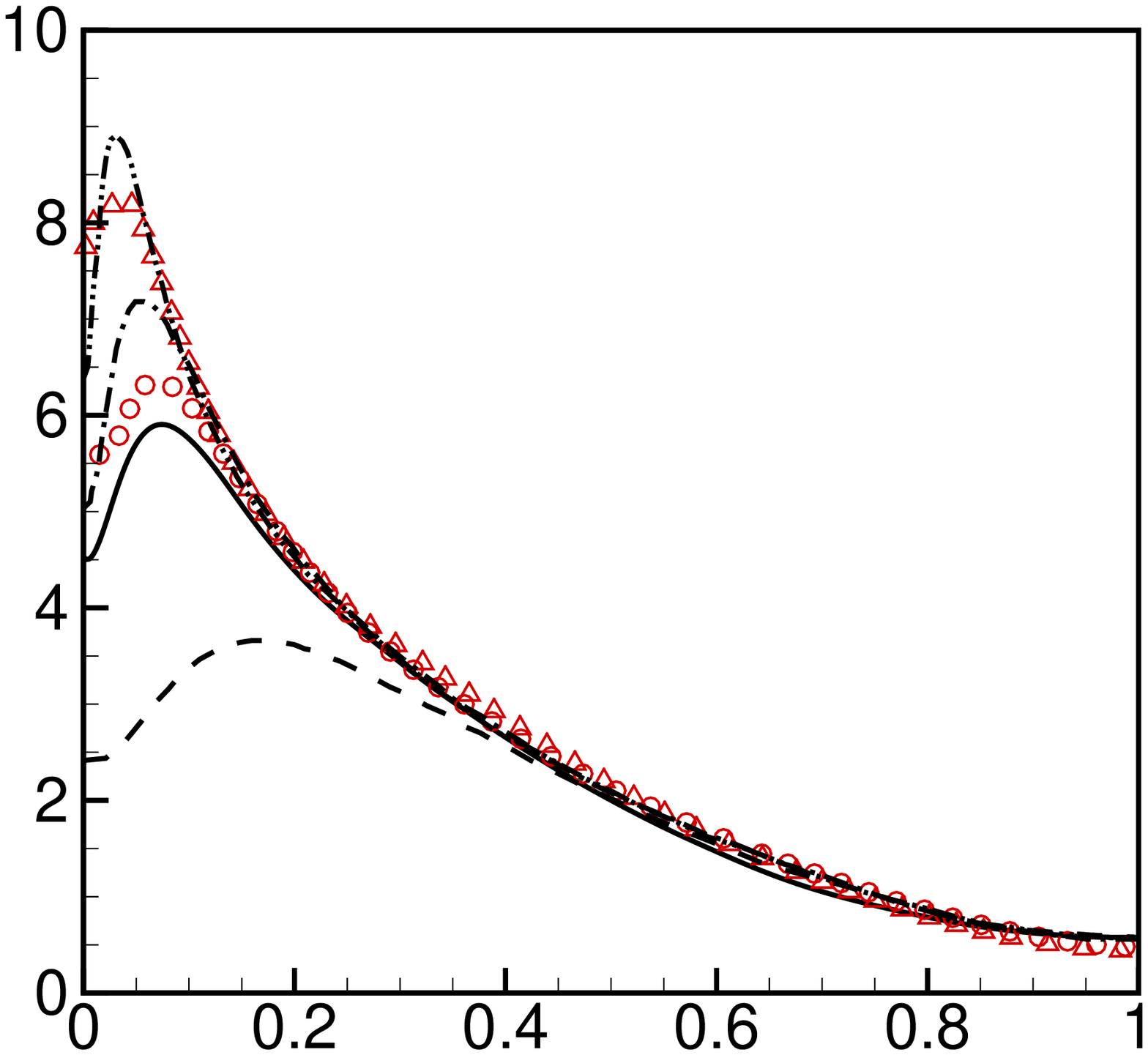}
 \put(-95,-3){$(y-y_0)/\delta$}
 \put(-160,55){\rotatebox{90}{$\langle \overline{p'p'} \rangle^+$}}
 \put(-85,130){$(a)$}
\hspace{5mm}
\includegraphics[height=50mm]{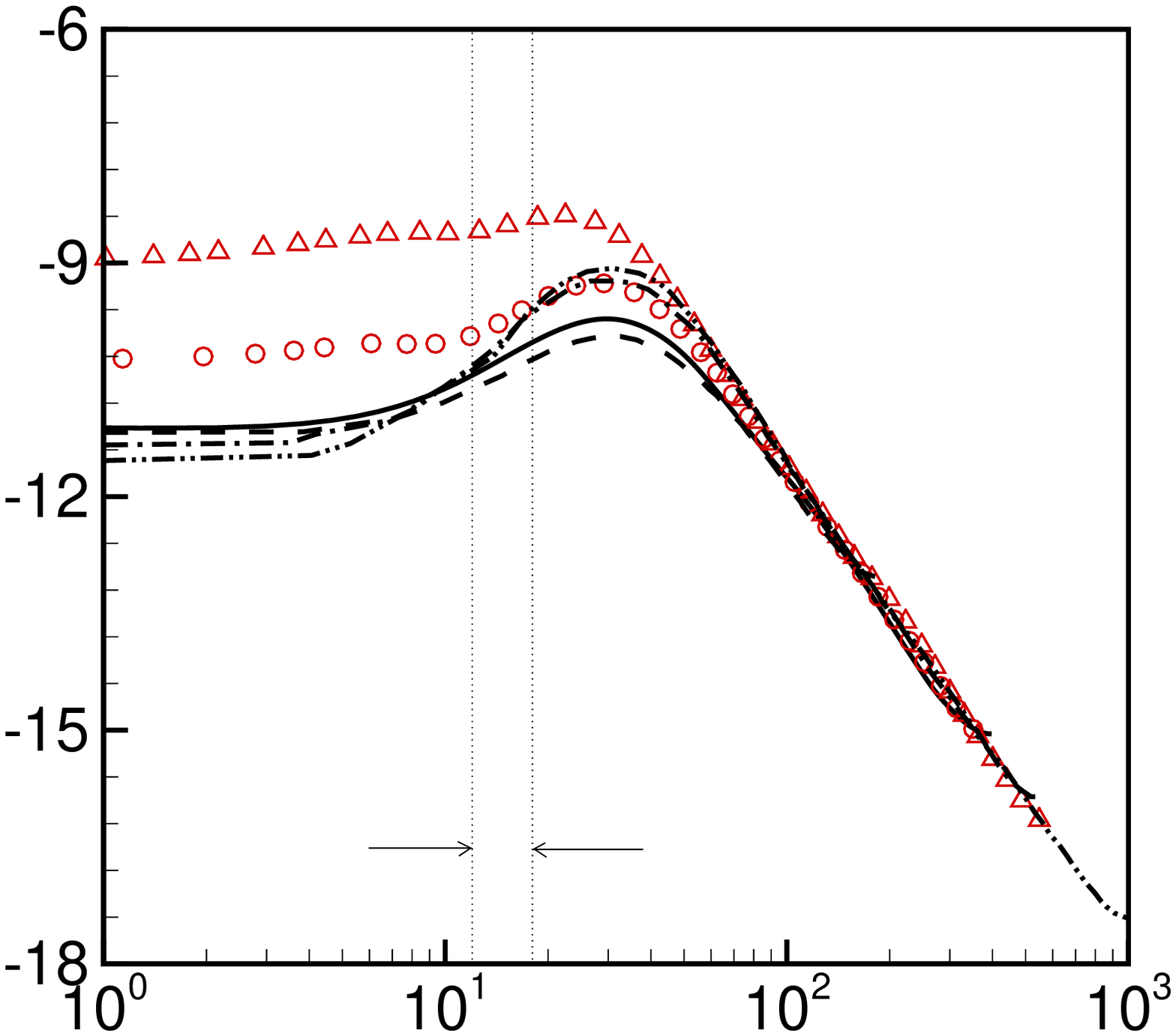}
\put(-97,-3){$(y-y_0)u_{\tau}^b/\nu$}
\put(-165,55){\rotatebox{90}{$\langle \phi \rangle(y^+)$}}
\put(-130,33){\scriptsize{$k_c^+$,R400}}
\put(-85,33){\scriptsize{$k_c^+$,R600}}
\put(-85,130){$(b)$}
\caption{$(a)$ Profiles of mean-square pressure fluctuations $\langle \overline{p'p'} \rangle$ normalized by $u_{\tau}^b$ and $(b)$ inner correlation  $\langle \phi \rangle(y^+)$. Symbols denote: Case SW (solid), Case R400 (circle), Case R600 (delta), \cite{panton2017correlation}, $Re_{\tau}=182$ (dashed), $Re_{\tau}=544$ (dash dot), $Re_{\tau}=1001$ (dash dot dot). The vertical dashed lines denote the roughness layers.}
\label{fig:prms}
\end{figure}

 Figure \ref{fig:prms}$(a)$ shows the mean-square pressure fluctuations  $ \langle \overline{p'p'} \rangle^+$ in outer coordinates for cases R400 and R600 over the bottom half of the channel. Also shown are  Case SW and other smooth channel flows from  \cite{panton2017correlation}. Note that away from the wall, both smooth and rough wall curves collapse on to the logarithmic correlation  described by \cite{panton2017correlation},
  \begin{equation}
     \langle \overline{p'p'} \rangle^+_{cp}(y/\delta) = -2.5625\ln(y/\delta) + 0.2703.
    \label{eqn:pp_cp}
  \end{equation}
As $Re_{\tau}$ increases,  $ \langle \overline{p'p'} \rangle^+$ increases near the wall, peaks closer to the wall, and collapses onto the logarithmic correlation closer to the wall. Roughness increases the peak levels of $ \langle \overline{p'p'} \rangle^+$ near the wall.

  \cite{panton2017correlation} derived an inner correlation to account for the $Re_\tau$ dependence for smooth channel flows,  
   \begin{equation}
    \langle \phi \rangle(y^+) =  \langle \overline{p'p'} \rangle^+(y^+,Re_{\tau}) - 2.5625\ln(Re_{\tau}) - 0.2703.
    \label{eqn:phi}
  \end{equation}
Figure \ref{fig:prms}$(b)$ shows $\langle \phi \rangle(y^+)$ computed for both smooth and rough cases. The smooth  curves show acceptable collapse both near, and away from the wall for all $Re_\tau$. The rough curves however collapse only beyond $y^+=50$. The increased level of pressure fluctuations below $y^+=50$ indicate that the roughness increases the pressure fluctuations in the inner layer and the enhancement is much stronger for higher $Re_{\tau}$. 

\begin{figure}
\includegraphics[width=70mm]{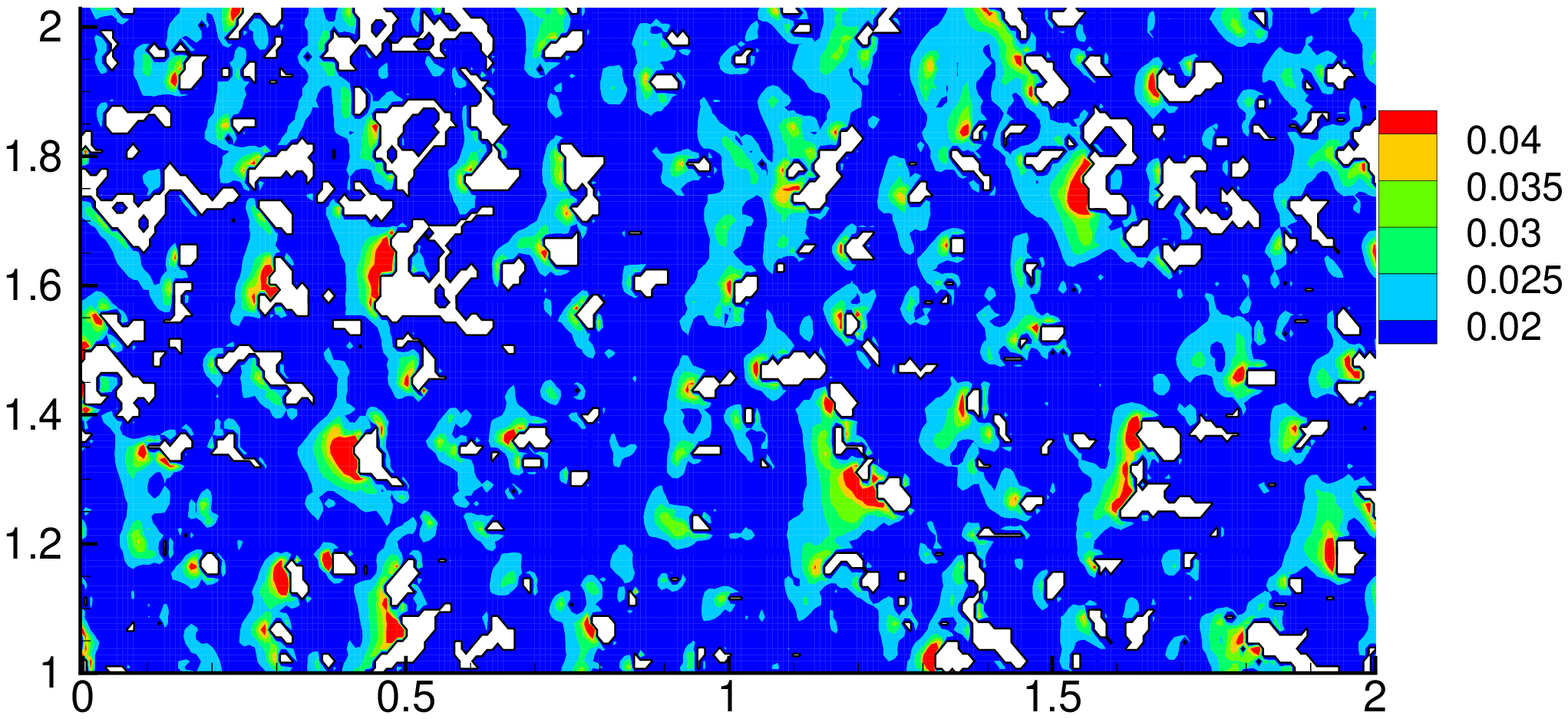}
\put(-195,47){\rotatebox{90}{$z/\delta$}}
\put(-100,-3){$x/\delta$}
\put(-22,80){$\overline {p'^2}/U_b^2$}
 \put(-195,92){$(a)$}
 \hspace{2mm}
\includegraphics[width=70mm]{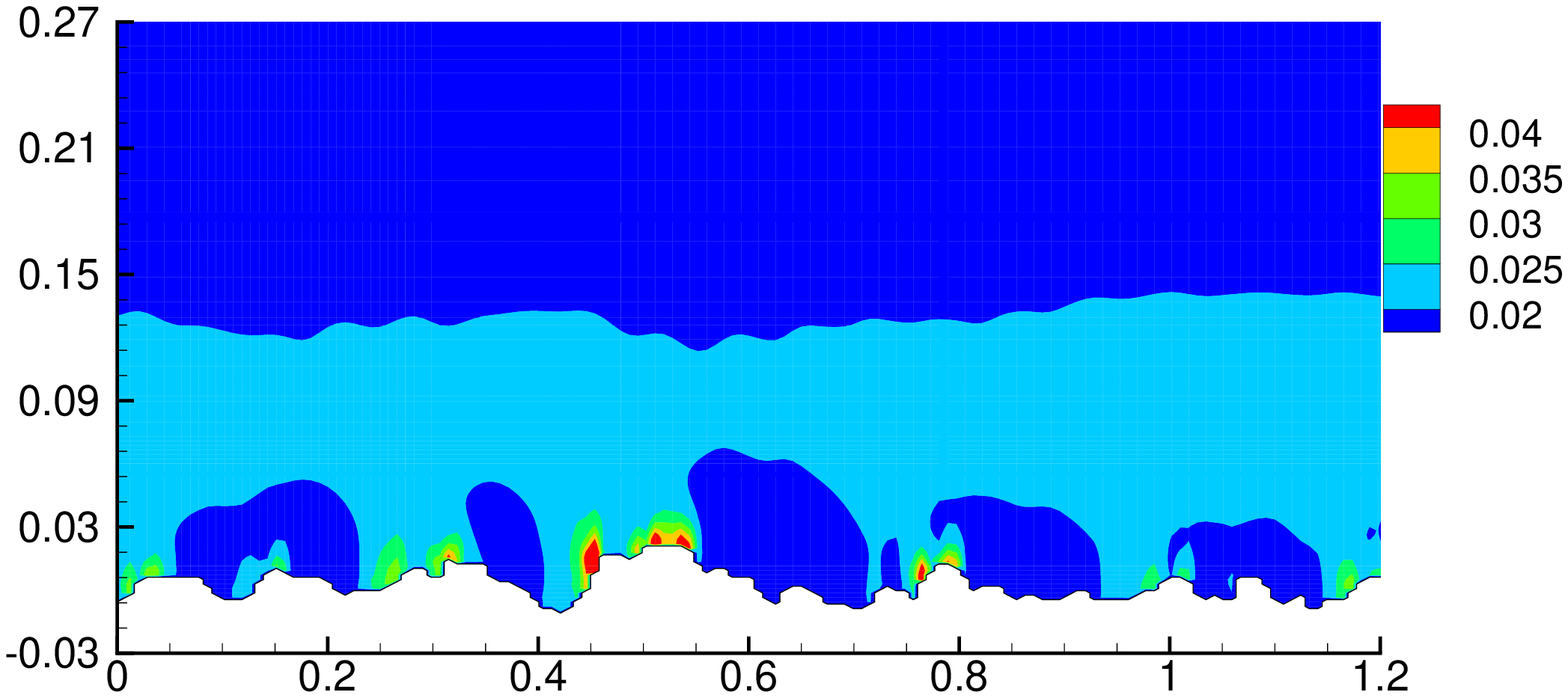}
\put(-195,47){\rotatebox{90}{$y/\delta$}}
\put(-100,-3){$x/\delta$}
\put(-22,80){$\overline {p'^2}/U_b^2$}
 \put(-195,92){$(b)$}
 \hspace{5mm}
 \includegraphics[width=70mm]{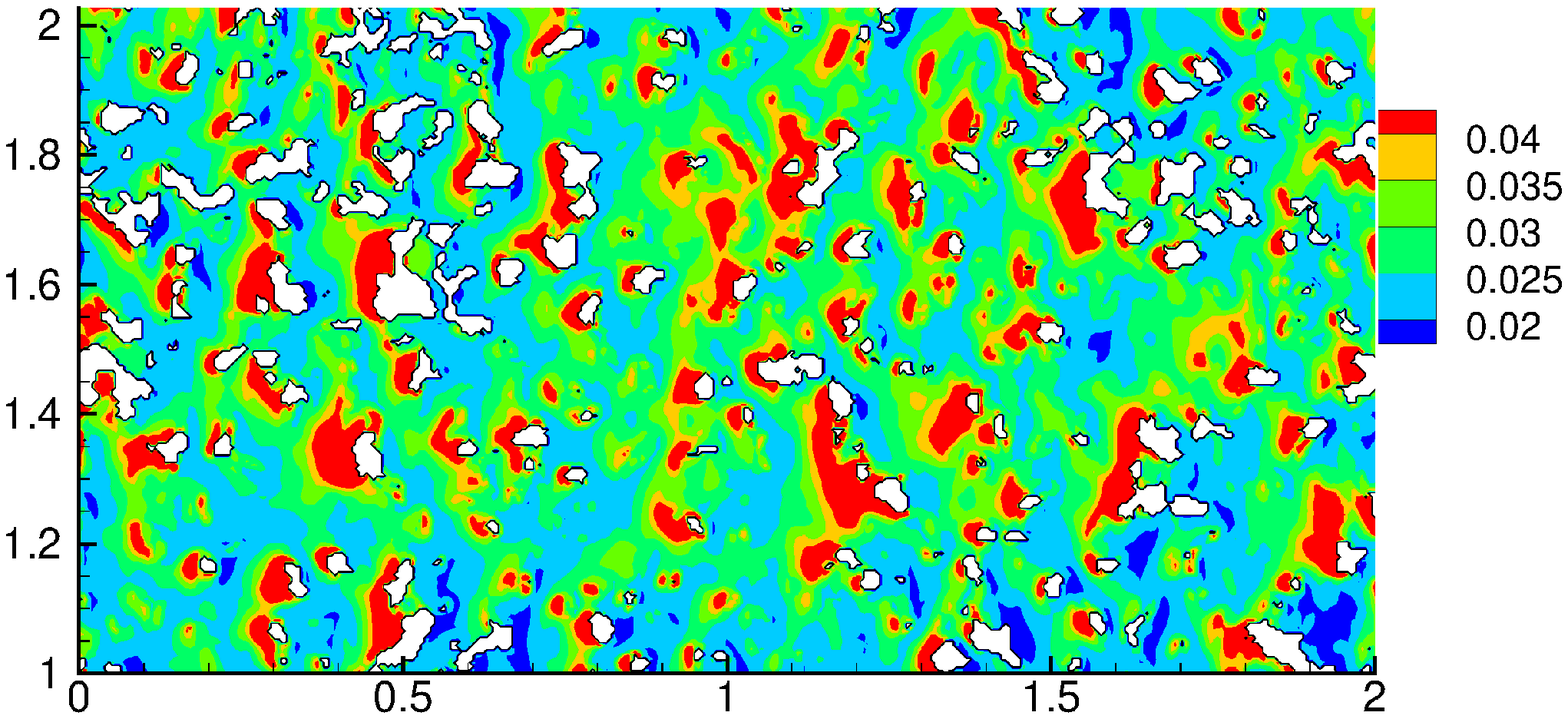}
\put(-195,47){\rotatebox{90}{$z/\delta$}}
\put(-100,-3){$x/\delta$}
\put(-22,80){$\overline {p'^2}/U_b^2$}
 \put(-195,92){$(c)$}
 \hspace{2mm}
\includegraphics[width=70mm]{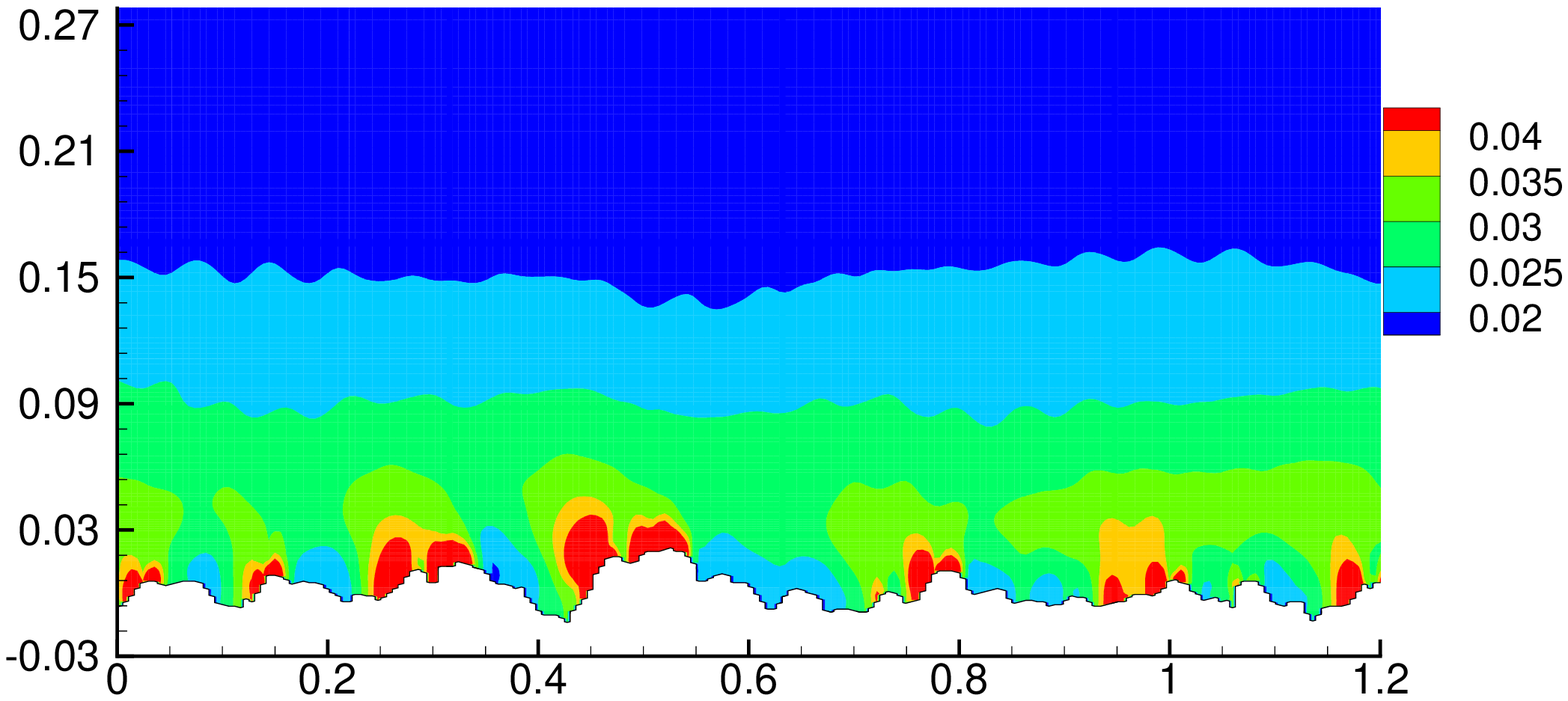}
\put(-195,47){\rotatebox{90}{$y/\delta$}}
\put(-100,-3){$x/\delta$}
\put(-22,80){$\overline {p'^2}/U_b^2$}
 \put(-195,92){$(d)$}
\caption{Mean-square pressure fluctuations $\overline {p'p'}$ normalized by $U_b$ for Case R400 at ($a$) $(y-y_0)/\delta=0.0075$ (corresponding to $y^+=3$), ($b$) $z/\delta=1.57$, and Case R600 at ($c$) $(y-y_0)/\delta=0.0075$ (corresponding to $y^+=4.5$), ($d$) $z/\delta=1.57$.} \label{fig:pp}
\end{figure}

The vertical lines in the figure illustrate the top of the roughness layer. Note that $\langle \overline{p'p'} \rangle^+$ peaks outside the roughness layer. Comparing Case R400 to Case SW, the level of pressure fluctuations  within the roughness layer  increases by $20.2 \%$, and the peak value by $ 7.8 \%$. For Case R600, the enhancement of pressure fluctuations in the roughness layer is even higher. 

The spatial variation  of $\overline {p'p'}$ is examined at a representative location, $(y-y_0)/\delta=0.0075$ and $z/\delta=1.57$ in figure \ref{fig:pp}. At both Reynolds numbers, high pressure fluctuations are observed in front of the roughness asperities, while the regions behind the roughness elements have lower levels.  The unsteady stagnation and wake regions induced by the roughness elements result in higher pressure fluctuations. Smooth wall flows do not display this behavior, and hence have lower levels of pressure fluctuations for the same $Re_\tau$.

\subsubsection{Poisson equation source terms}

The  source terms in the pressure Poisson equation provide further insight into how roughness influences pressure fluctuations.  The Poisson equation for the fluctuating pressure is given by, 
\begin{equation}
    \frac{\partial^2 p'}{\partial x_i \partial x_i} = - 2 \frac{\partial {\langle \overline{u_i} \rangle}}{\partial {x_j}}\frac{\partial u_j'}{\partial x_i} - \frac{\partial^2}{\partial x_i \partial x_j}(u_i'u_j' - \overline{u_i'u_j'}).
    \label{eqn:Pois}
\end{equation}
The first term on the right-hand side is called the mean-shear or rapid source term (MS) and the second term is known as the turbulence-turbulence or slow source term (TT).  The dominant component of the MS term is 
\begin{equation}
    MS_{12} = - 2 \frac{\partial {\langle \overline{u} \rangle}}{\partial {y}}\frac{\partial v'}{\partial x}. 
\end{equation}
Since this term involves the streamwise mean velocity gradient, it is large near the wall  \citep{johansson1987generation}. For the TT term, it has been shown that the component
\begin{equation}
    TT_{23}=\frac{\partial {v'}}{\partial {z}}\frac{\partial w'}{\partial y} - \frac{\partial ^2 \overline{v'w'}}{\partial y \partial z}
\end{equation}
is the dominant term, located in the buffer region whose peak location is correlated with the average position of quasi-streamwise vortices \citep{chang1999relationship}. The dominant terms $MS_{12}$ and $TT_{23}$  for the random-rough channel flows are discussed below.

\begin{figure}
\includegraphics[height=45mm]{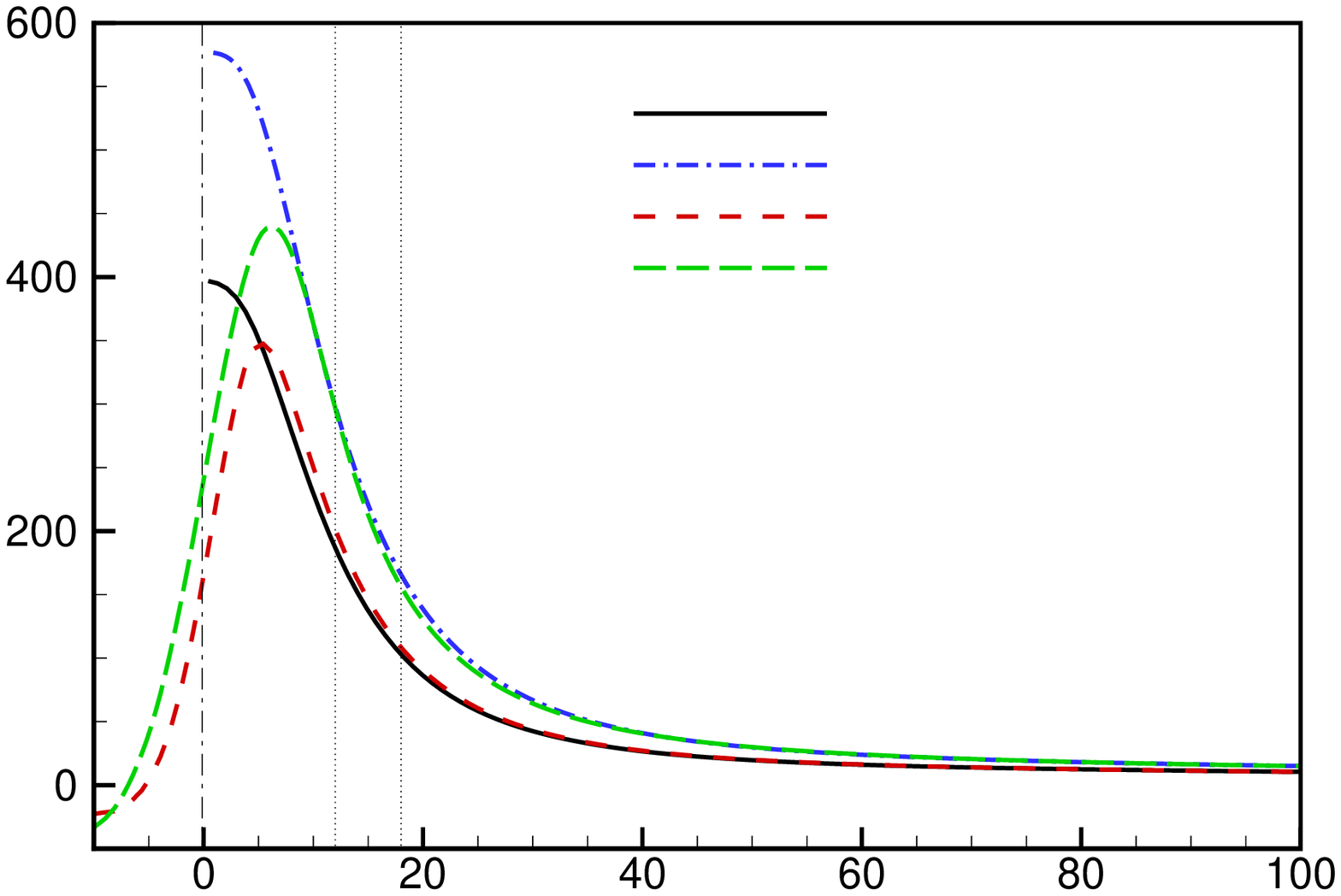}
\put(-120,0){$(y-y_0)u_{\tau}^l/\nu$}
\put(-195,50){\rotatebox{90}{$\partial \langle \overline{u}\rangle / \partial y $}}
\put(-78,100){\scriptsize{R400, top}}
\put(-78,93){\scriptsize{R600, top}}
\put(-78,86){\scriptsize{R400, bottom}}
\put(-78,79){\scriptsize{R600, bottom}}
\put(-100,118){$(a)$}
\includegraphics[height=45mm]{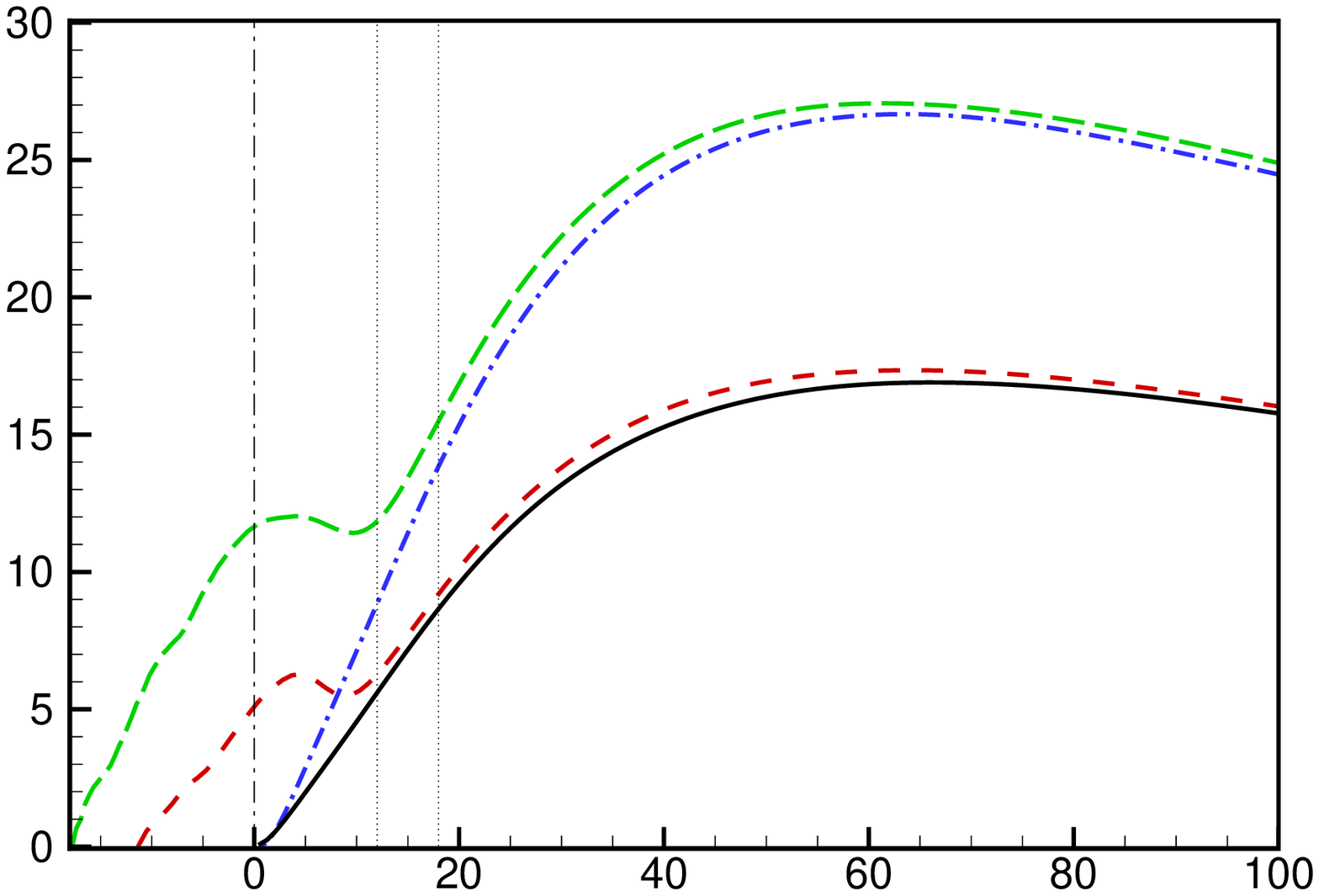}
\put(-120,0){$(y-y_0)u_{\tau}^l/\nu$}
\put(-195,40){\rotatebox{90}{rms of $\partial {v'} /\partial x $}}
\put(-100,118){$(b)$}
\hspace{5mm}
\includegraphics[height=45mm]{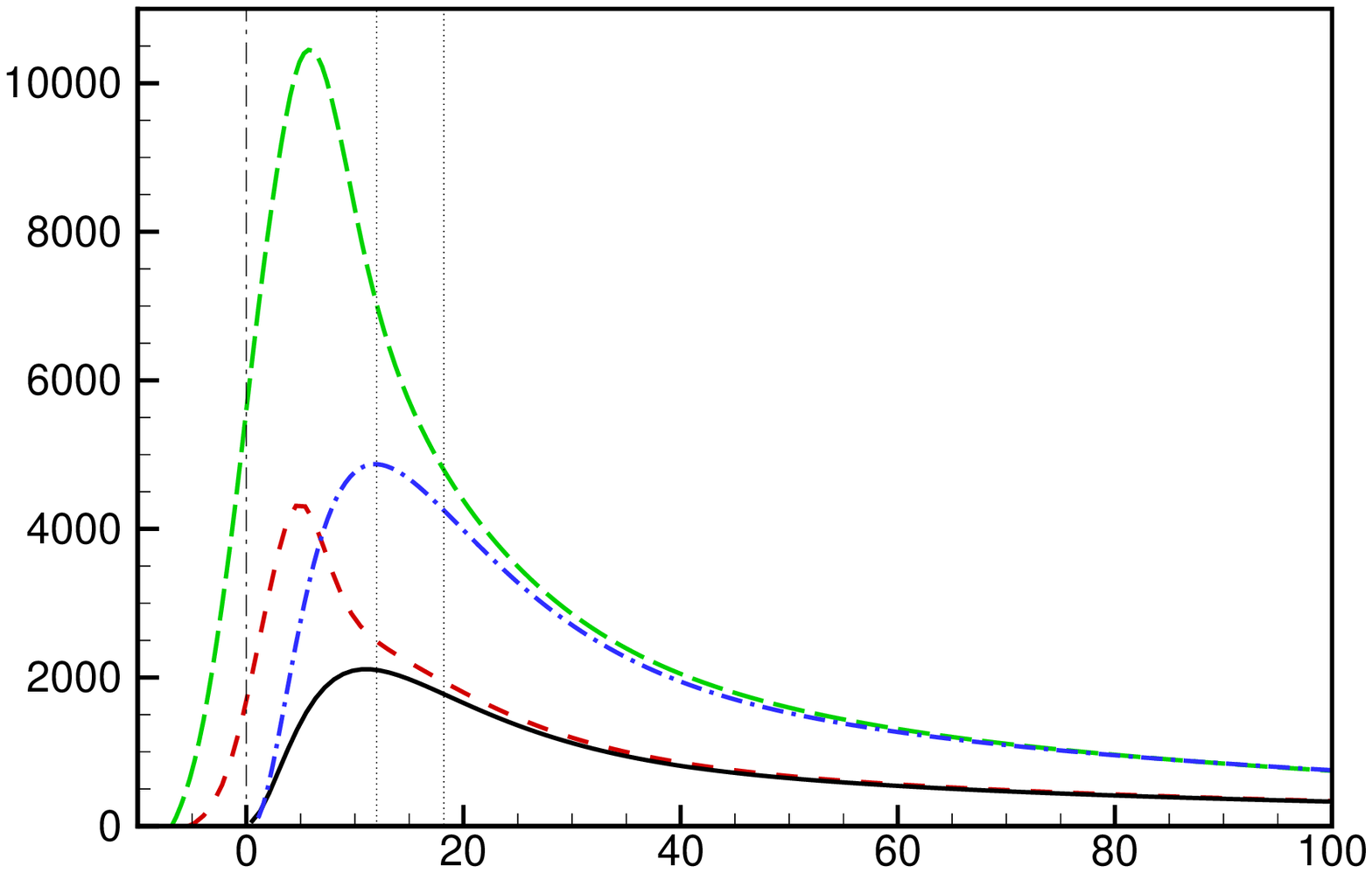}
\put(-120,0){$(y-y_0)u_{\tau}^l/\nu$}
\put(-195,40){\rotatebox{90}{rms of $MS_{12}$}}
\put(-100,118){$(c)$}
\includegraphics[height=45mm]{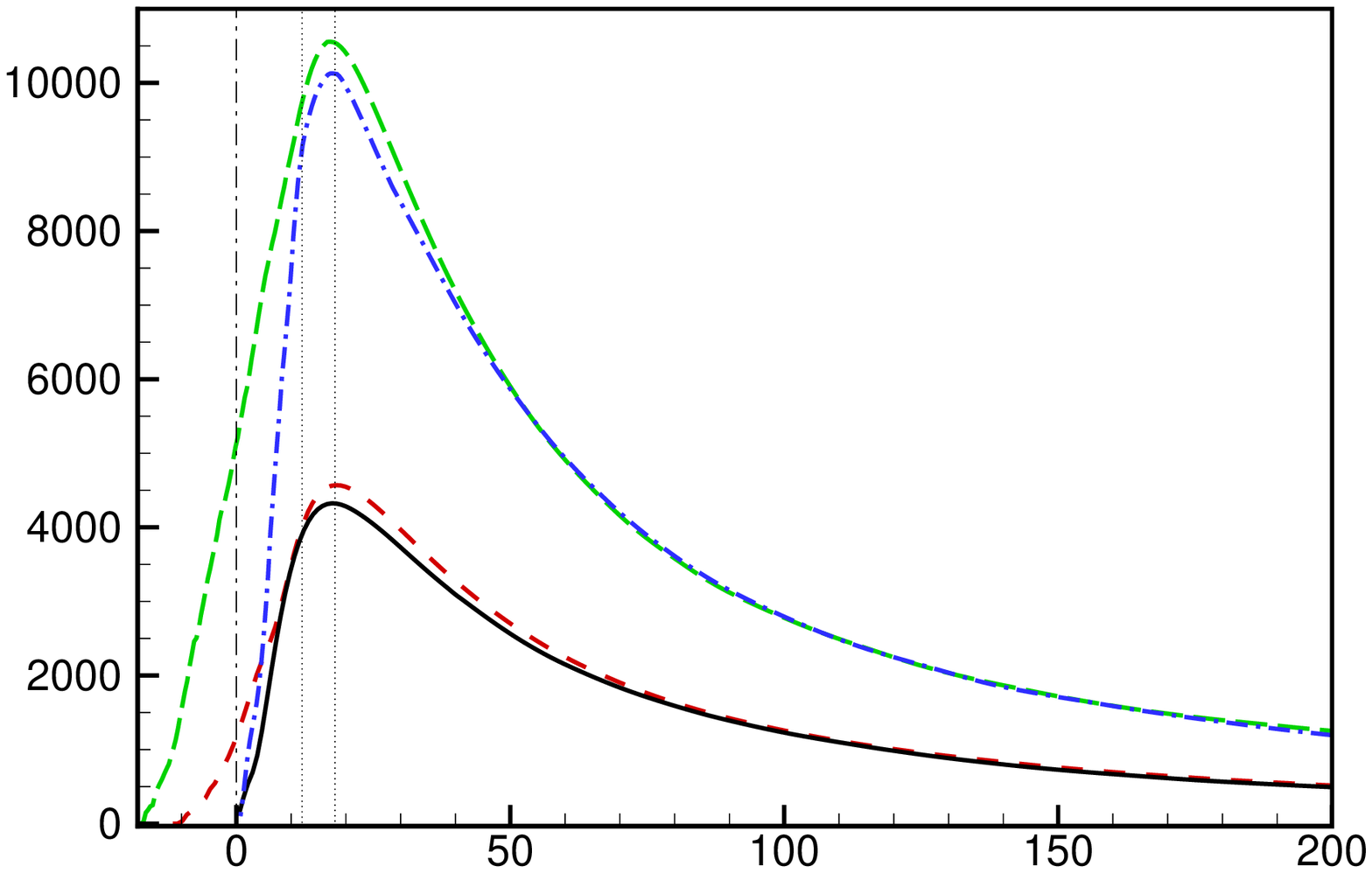}
\put(-120,0){$(y-y_0)u_{\tau}^l/\nu$}
\put(-195,40){\rotatebox{90}{rms of $TT_{23}$}}
\put(-100,118){$(d)$}
\caption{$(a)$ Streamwise mean velocity gradient, $(b)$  $\partial {v'} /\partial x $, $(c)$  $MS_{12}$, and $(d)$  $TT_{23}$. $u_\tau^l$ and $\delta$ are used for normalization. The vertical dash dot line denotes the mean roughness height $y=y_0$, and the vertical dashed lines denote the top of the roughness layer $k_c^+$.} 
\label{fig:source term}
\end{figure}



Figure \ref{fig:source term} shows how roughness impacts ${\partial {\langle \overline{u} \rangle}}/{\partial {y}}, {\partial v'}/{\partial x}$ and $MS_{12}$ individually. The profiles are normalized by the local friction velocity at the two walls. Roughness is seen to affect both the mean and fluctuating velocity gradients. Figure \ref{fig:source term}$(a)$ shows that for the smooth half of the channel, the mean velocity gradient, is largest at the wall and gradually decreases with increasing  $y$. On the other hand, at the rough wall, ${\partial {\langle \overline{u} \rangle}}/{\partial {y}}$ peaks at $y^+=6$ which lies within the roughness layer, and collapses onto the smooth-wall profile for larger $y$. 
Figure \ref{fig:source term}$(b)$ shows  $\partial {v'} /\partial x$. Note that the rms of $\partial {v'} /\partial x$ significantly increases within the roughness layer, and collapses with the smooth-wall profile above the roughness. Both terms combine to increase $MS_{12}$ as shown in figure \ref{fig:source term}$(c)$. The  peak values  are within the roughness layer at $y^+=6$, consistent with the peak locations observed in figure \ref{fig:source term}$(a)$. Both Reynolds numbers show the same behavior except that  Case R600 has higher magnitudes. 

\begin{figure}
 \includegraphics[width=70mm]{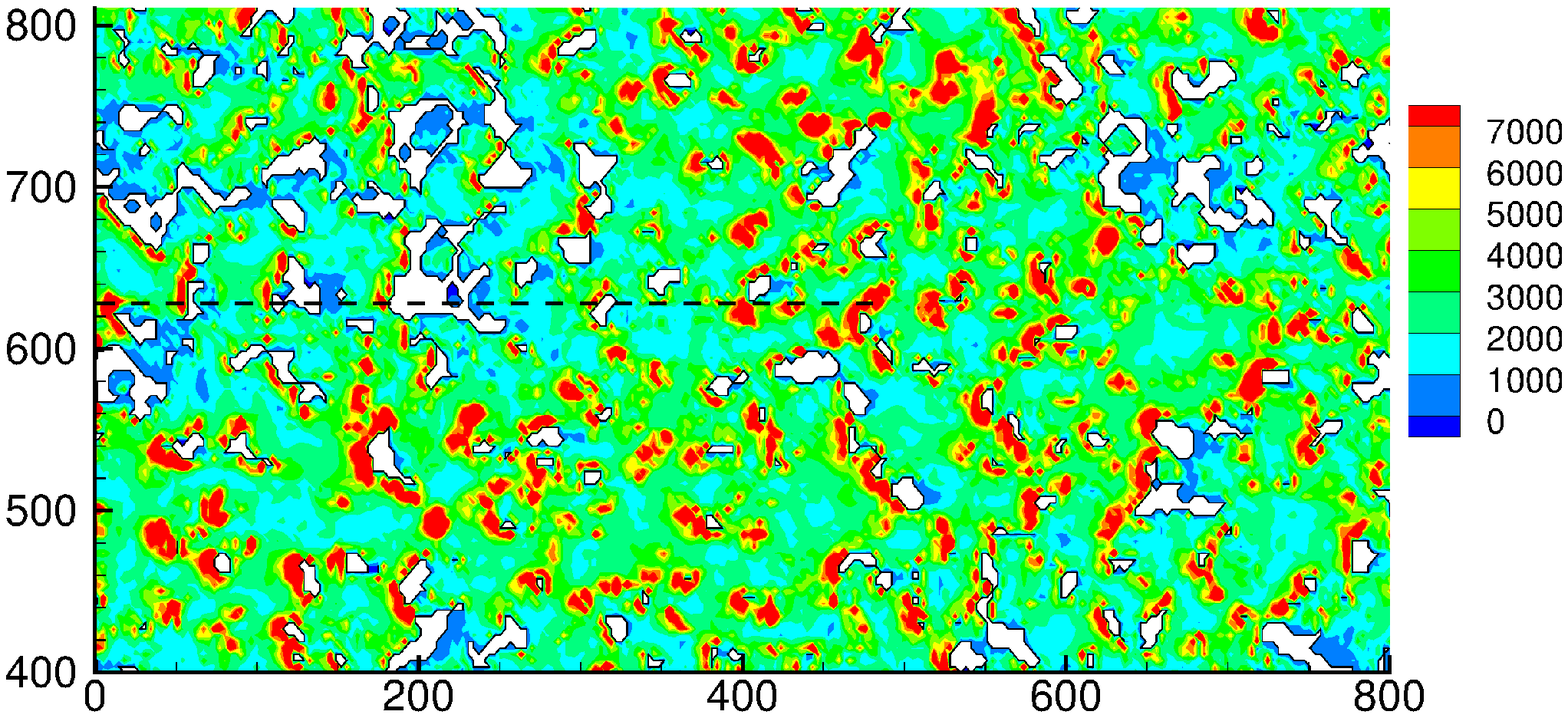}
\put(-195,47){\rotatebox{90}{$z^+$}}
\put(-100,-3){$x^+$}
\put(-25,80){\scriptsize{$MS_{12,rms}^+$}}
 \put(-190,90){$(a)$}
 \hspace{3mm}
\includegraphics[width=70mm]{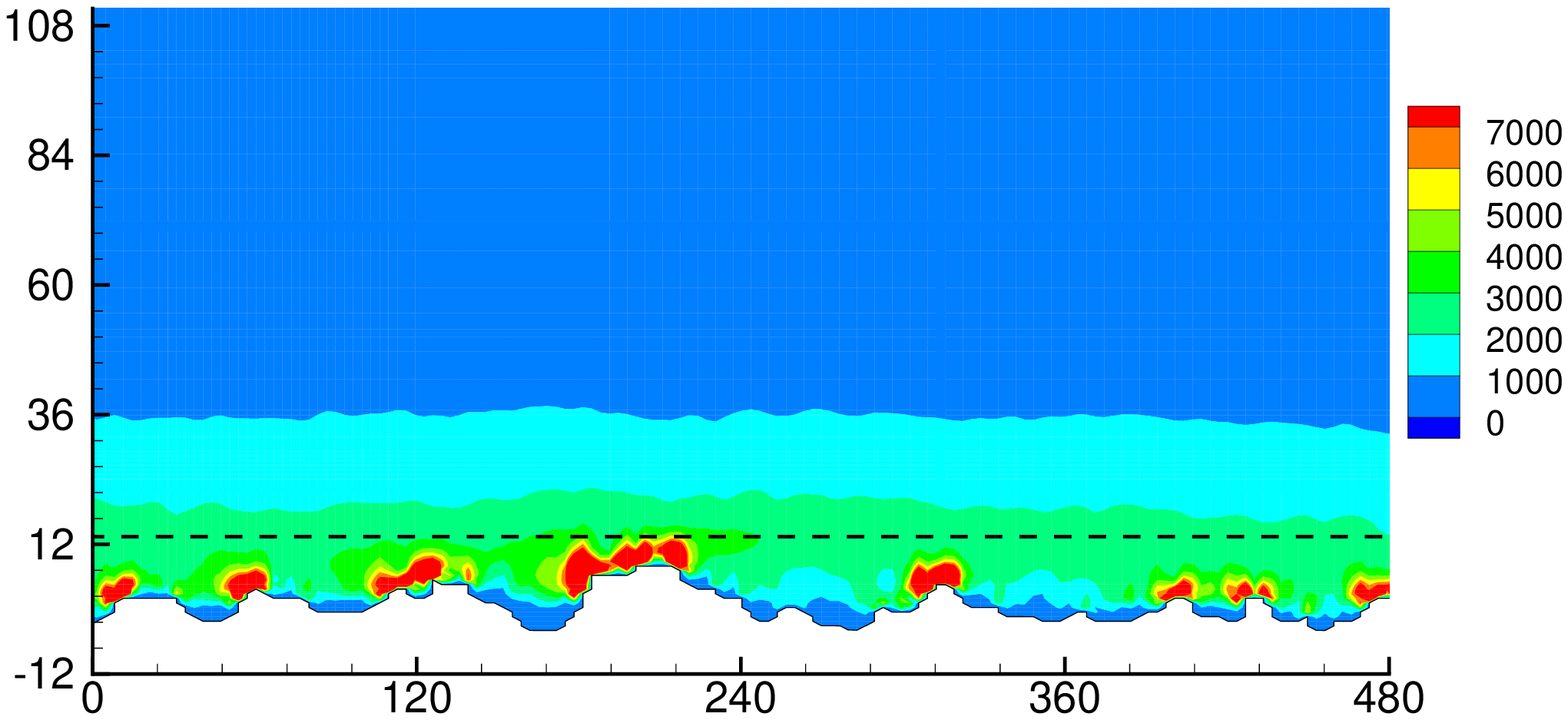}
\put(-195,47){\rotatebox{90}{$y^+$}}
\put(-100,-3){$x^+$}
\put(-25,80){\scriptsize{$MS_{12,rms}^+$}}
 \put(-190,90){$(b)$}
 \hspace{5mm}
  \includegraphics[width=70mm]{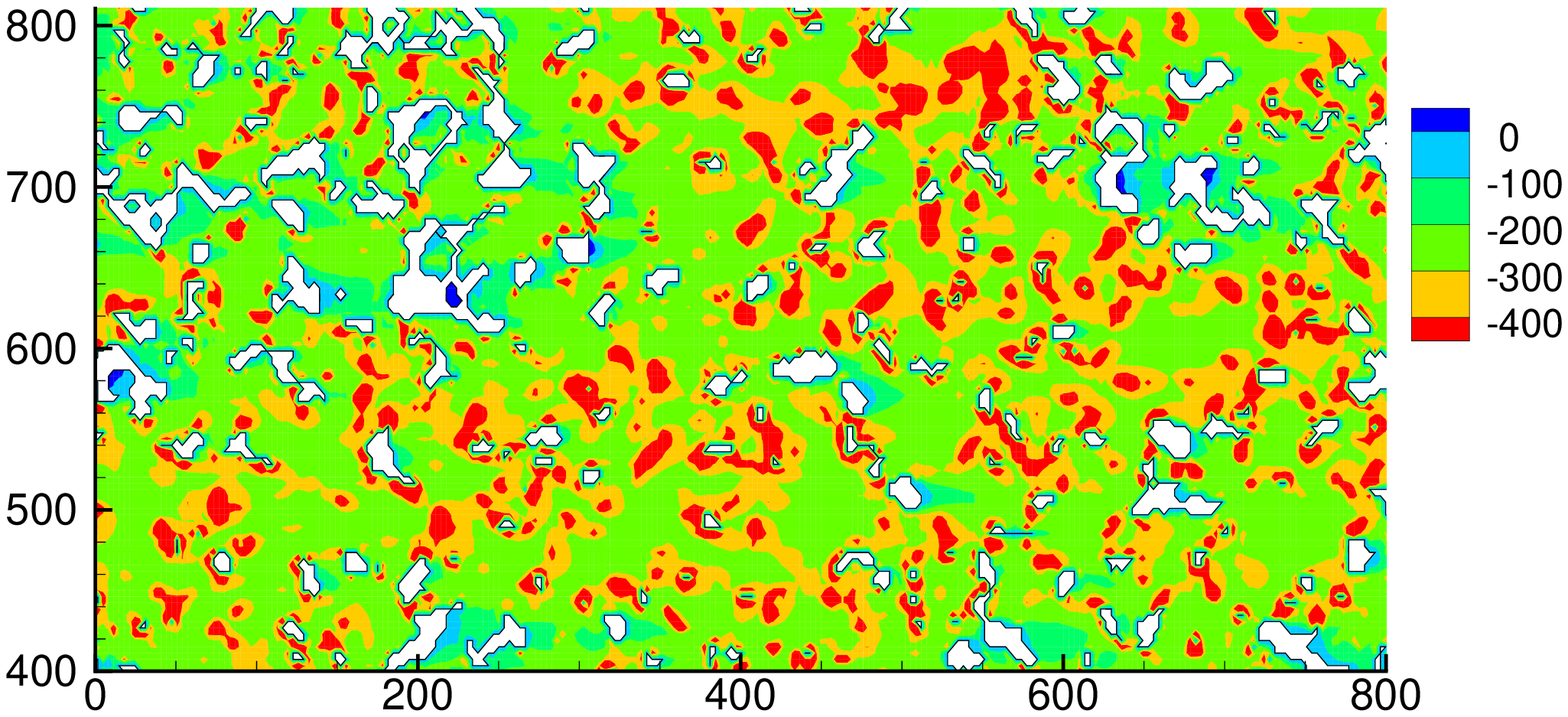}
\put(-195,47){\rotatebox{90}{$z^+$}}
\put(-100,-3){$x^+$}
\put(-20,80){$\overline{\omega_z}^+$}
 \put(-190,90){$(c)$}
 \hspace{3mm}
\includegraphics[width=70mm]{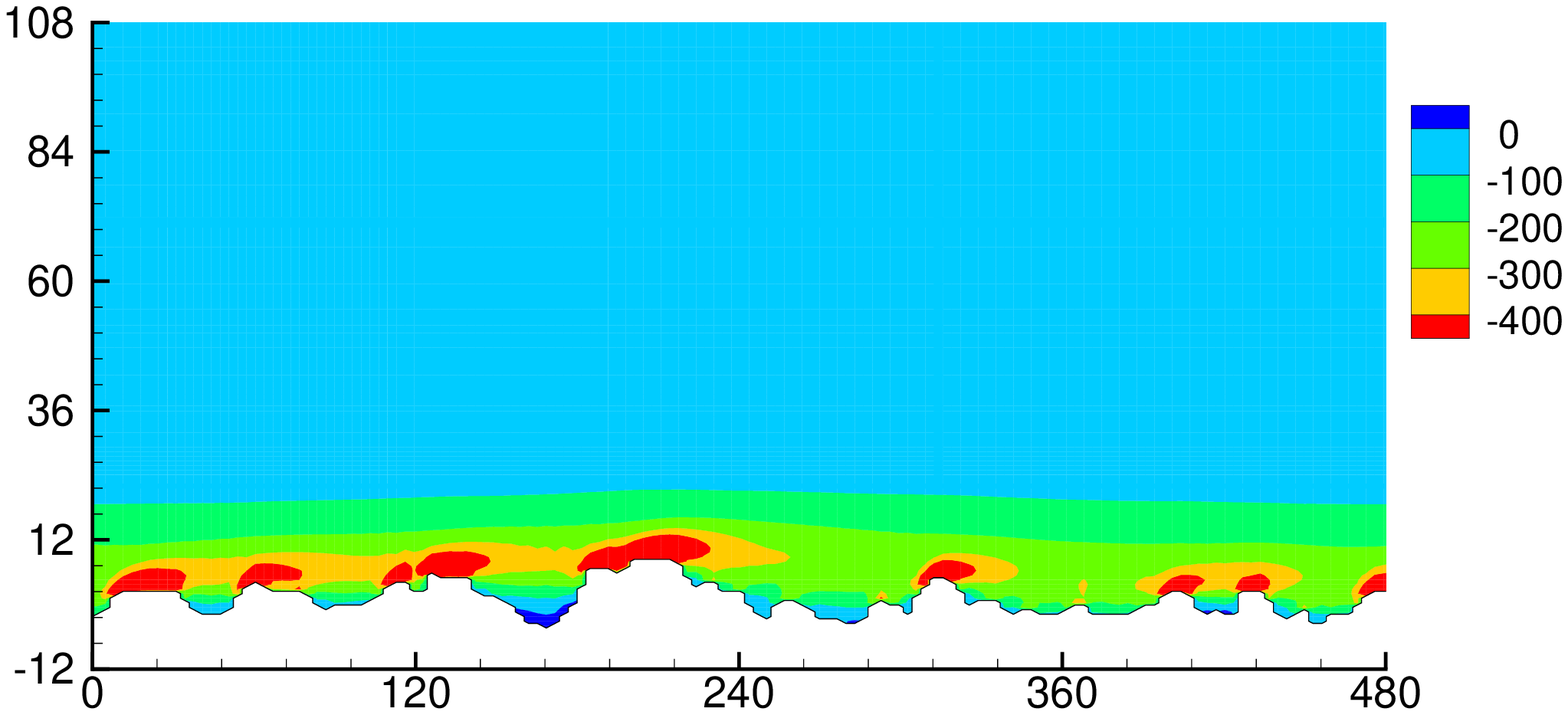}
\put(-195,47){\rotatebox{90}{$y^+$}}
\put(-100,-3){$x^+$}
\put(-20,80){$\overline{\omega_z}^+$}
 \put(-190,90){$(d)$}
\caption{Spatial variation of root-mean-square $MS_{12}$ for Case R400 at $(a)$ $y^+=3$, $(b)$ $z^+=628$. Time-averaged spanwise vorticity $\overline{\omega_z}$ normalized by $u_{\tau}/\delta$ at $(c)$ $y^+=3$, $(d)$ $z^+=628$. The dashed line in $(a)$ denotes the $x^+$--$y^+$ plane observed in $(b)$ and $(d)$, while the dashed line in $(b)$ denotes the edge of the roughness layer. } \label{fig:MS}
\end{figure}

Figures \ref{fig:MS}$(a)$ and \ref{fig:MS}$(b)$ show the local variation of rms $MS_{12}$ in the near-wall region for Case R400. The intense high-magnitude spots (red regions) are produced in front of the roughness protrusions and on the crests. The time-averaged spanwise vorticity in figures \ref{fig:MS}$(c)$ and \ref{fig:MS}$(d)$ shows the local mean shear layers that form at the same locations. These shear layers combine to yield the mean velocity gradient at a wall-normal location.
The higher levels of $\partial {v'} /\partial x$ within the roughness layer are due to the upward velocity in front of the protrusions and the “roll-up”
motions on the crests observed in  \S \ref{subsec:dispersive flux}.

The high-magnitude regions of rms $MS_{12}$ and $\overline{\omega_z}^+$ are consistent with the regions of high pressure fluctuations in figure \ref{fig:pp}. The more intense pressure fluctuations in front of and above the roughness elements can be attributed to the attached shear layers formed upstream of the obstacles. The low level of rms $MS_{12}$ in the wake regions is because the flow is more quiescent behind the roughness elements. 

\begin{figure}
 \includegraphics[width=70mm]{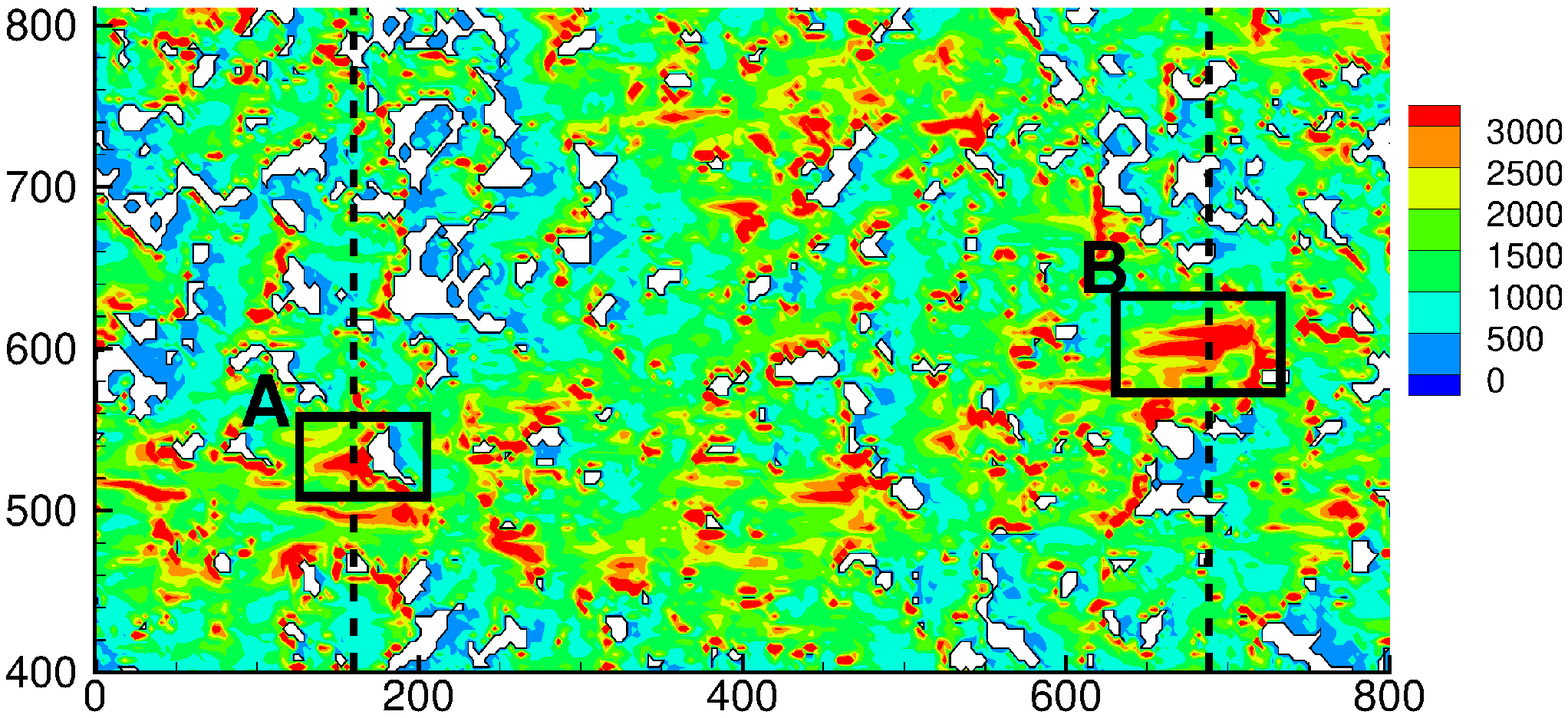}
\put(-195,47){\rotatebox{90}{$z^+$}}
\put(-100,-3){$x^+$}
\put(-25,80){\scriptsize{$TT_{23,rms}^+$}}
 \put(-190,90){$(a)$}
 \hspace{3mm}
\includegraphics[width=70mm]{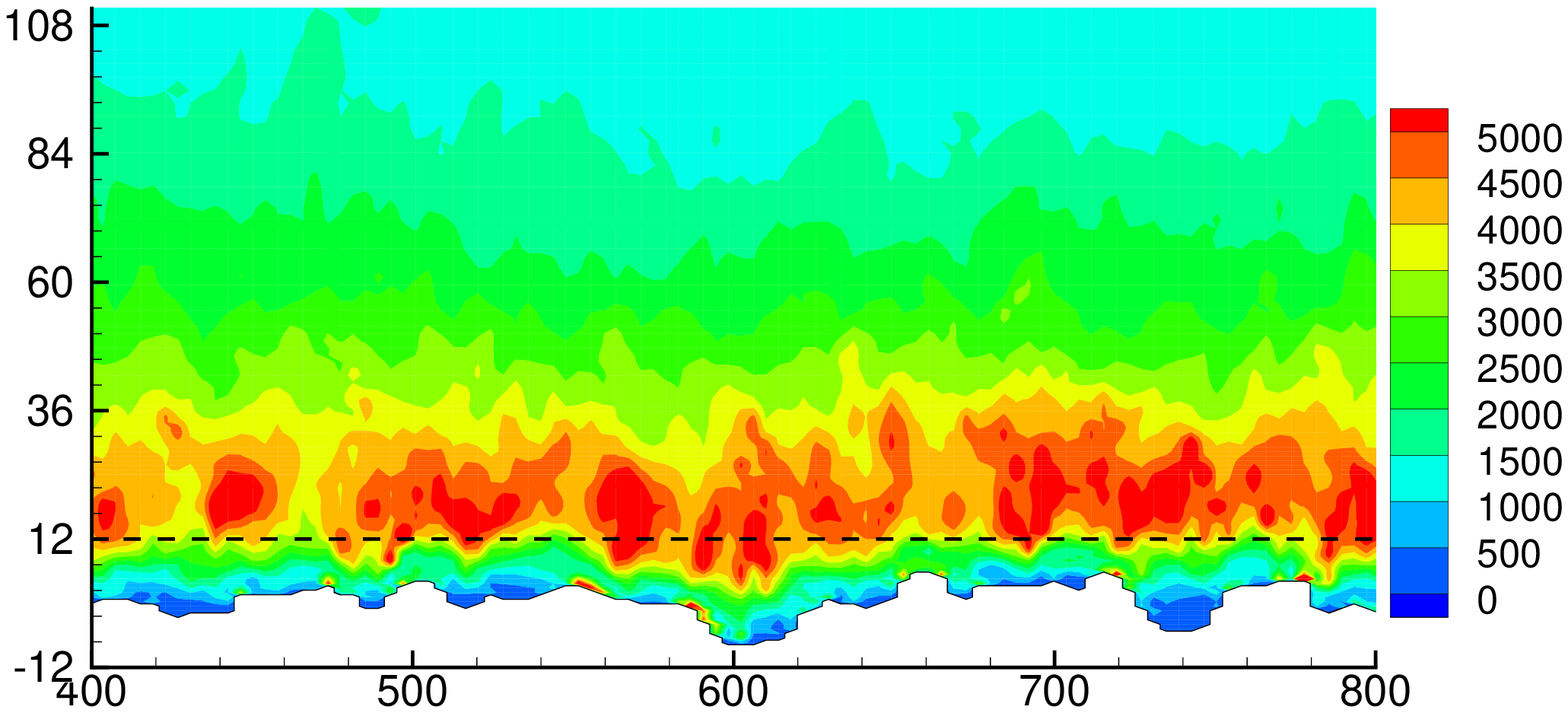}
\put(-195,47){\rotatebox{90}{$y^+$}}
\put(-100,-3){$z^+$}
\put(-25,80){\scriptsize{$TT_{23,rms}^+$}}
 \put(-190,90){$(b)$}
 \hspace{5mm}
  \includegraphics[width=70mm]{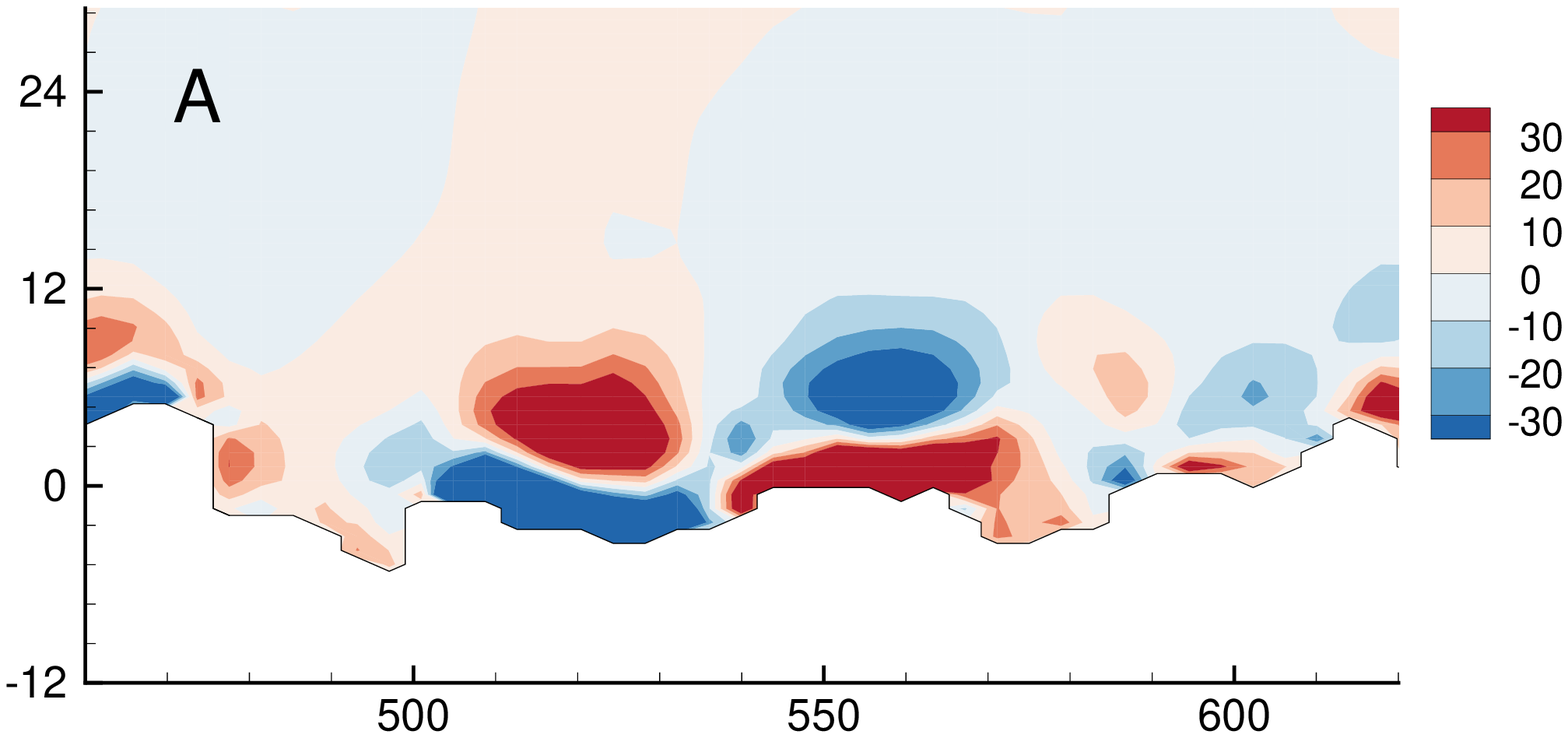}
\put(-195,47){\rotatebox{90}{$y^+$}}
\put(-100,-3){$z^+$}
\put(-20,80){$\overline{\omega_x}^+$}
 \put(-190,90){$(c)$}
 \hspace{2mm}
\includegraphics[width=70mm]{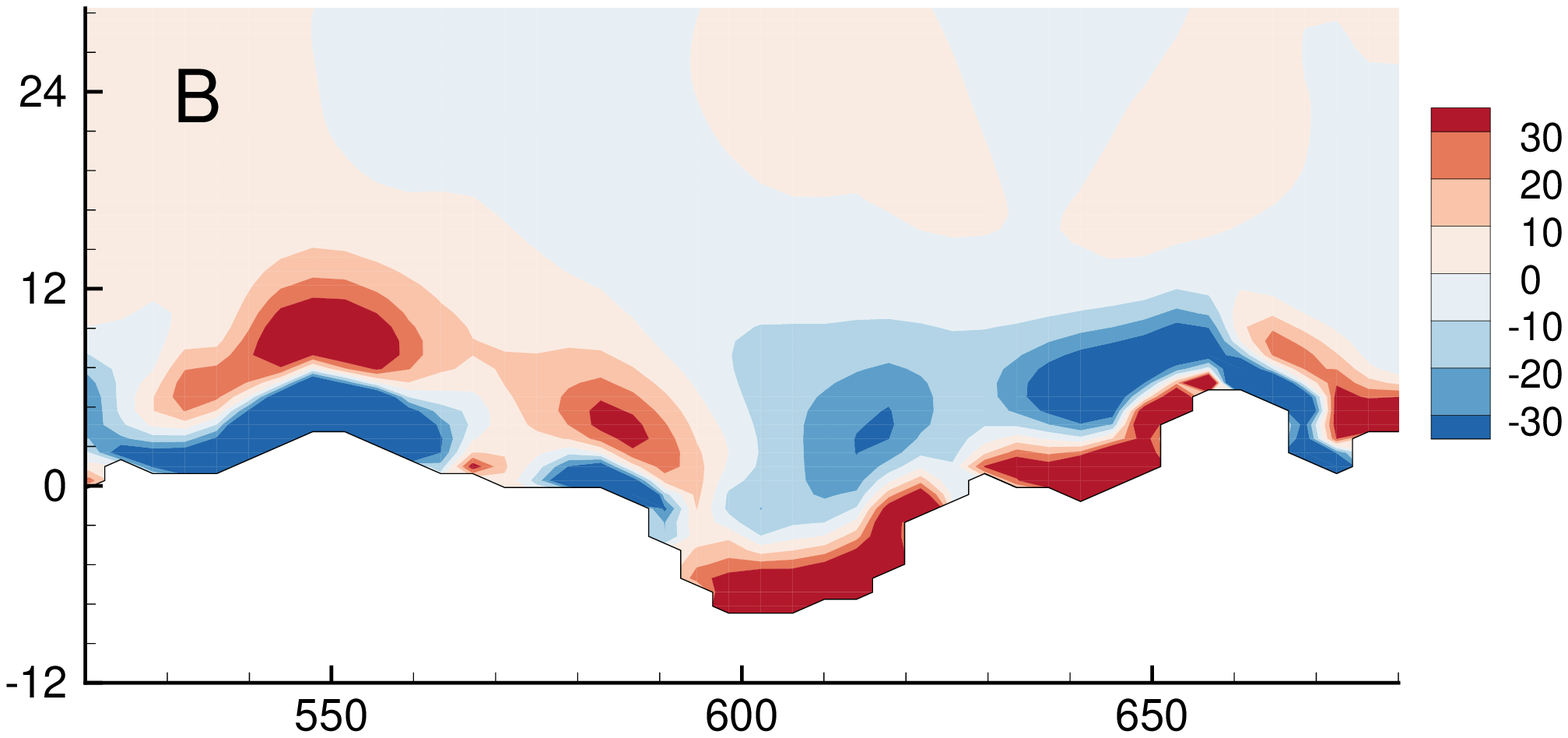}
\put(-195,47){\rotatebox{90}{$y^+$}}
\put(-100,-3){$z^+$}
\put(-20,80){$\overline{\omega_x}^+$}
 \put(-190,90){$(d)$}
\caption{Spatial variation of root-mean-square $TT_{23}$ for Case R400 at $(a)$ $y^+=3$, $(b)$ $x^+=xu_{\tau}/\nu=688$. Time-averaged streamwise vorticity $\overline{\omega_x}$ normalized by $u_{\tau}/\delta$ corresponding to $(c)$ the square region A at $x^+=160$, $(d)$ the square region B at $x^+=688$. The dashed lines in $(a)$ denote the $z^+$--$y^+$ planes observed in $(b)$-$(d)$, while the dashed line in $(b)$ denotes the edge of the roughness layer. Note for the $z^+$--$y^+$ plane, the mean flow direction points out of the page.} \label{fig:TT}
\end{figure}

The rms of $TT_{23}$ is examined in figure \ref{fig:source term}$(d)$. Note that for both cases R400 and R600, the level of $TT_{23}$ increases in the roughness layer, the peak location remains at $y^+=18$, and the profile above the roughness layer  collapses with that of the smooth wall. The spatial distribution of rms $TT_{23}$  at $y^+=3$ and $x^+=688$ for Case R400, is shown in figures \ref{fig:TT}$(a)$ and \ref{fig:TT}$(b)$, respectively. The high-magnitude regions (red) are observed as the streaky structures in the troughs as well as a large band at $y^+=18$. To specifically investigate how the rms of $TT_{23}$ is increased in the roughness layer, two typical high-magnitude streaks are identified in figure \ref{fig:TT}$(a)$ and the time-averaged streamwise vorticity in the corresponding $z$-$y$ planes are examined in figures \ref{fig:TT}$(c)$ and \ref{fig:TT}$(d)$. Two main features are identified. For location A, the level of $TT_{23}$ is increased in the immediate vicinity upstream of the roughness protrusion. Meanwhile, a pair of streamwise vortices is observed at the same location which is induced by the upward velocity in front of the protrusions. A pair of secondary vortices is observed below the induced streamwise vortices, closely attaching to the bottom wall. For location B, the high-magnitude streak is  located in the valley in the same region where the streamwise vortices occur. These observations show that $TT_{23}$ is correlated with the quasi-streamwise vortices. Since quasi-streamwise vortices are present in the troughs \citep{aghaei2019turbulence} and induced in front of the protrusions, the rms of $TT_{23}$ is accordingly increased. 

 \begin{figure}
 \centering{
	 \includegraphics[height=60mm]{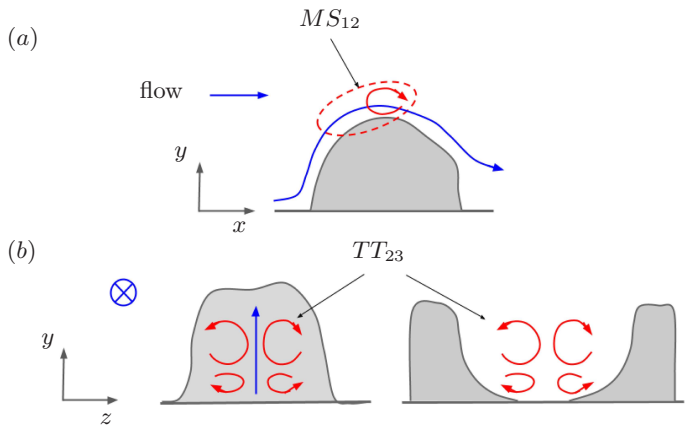}}
\put(-227,108){$y$}
\put(-205,80){$x$}
\put(-277,38){$y$}
\put(-255,8){$z$}
\put(-240,130){flow}
\put(-180,157){$MS_{12}$}
\put(-160,70){$TT_{23}$}
\put(-290,150){$(a)$}
\put(-290,70){$(b)$}
\caption{Illustration of the typical flow features contributing to  $(a)$ $MS_{12}$ and $(b)$ $TT_{23}$. The dashed ellipse in $(a)$ denotes the shear layer formed on the protuberance. The red arrows in $(b)$ represent a pair of streamwise vortices and  secondary vortices beneath them.} 
\label{fig:sketch_prms}
\end{figure}

The flow characteristics related to the contributions of $MS_{12}$ and $TT_{23}$ to the pressure fluctuations are summarized in figure \ref{fig:sketch_prms}. The shear layer along with the "roll-up" motion indicated by the dashed ellipse is generated at the upstream stagnation point of the protrusion in figure \ref{fig:sketch_prms}$(a)$, and is a primary source of the larger pressure fluctuations. Figure \ref{fig:sketch_prms}$(b)$ illustrates the streamwise vortices that occur in front of the protrusion and also in the troughs, generating secondary vortical structures below the main vortices.  \\

\subsection{Wall-shear stress fluctuations}\label{subsec:wall-shear stress}
The effects of roughness  on  wall-shear stress fluctuations are examined in this section. The wall-shear stress components $\tau_{yx}$ and $\tau_{yz}$ are calculated at the rough surface.
Data is collected for a time period $T=32\delta/u_{\tau}$ with  sampling time interval $t=2.5\times 10^{-3}\delta/u_{\tau}$. The wall-shear stress fluctuations are denoted by $\tau_{yx}'$ and $\tau_{yz}'$, the time-averaged values are expressed as $\overline{\tau_{yx}}$ and $\overline{\tau_{yz}}$, and the rms values are $\tau_{yx,rms}$ and $\tau_{yz,rms}$.


\subsubsection{Local variation of wall-shear stress}\label{sec:localshearvar}

\begin{figure}
\includegraphics[height=60mm]{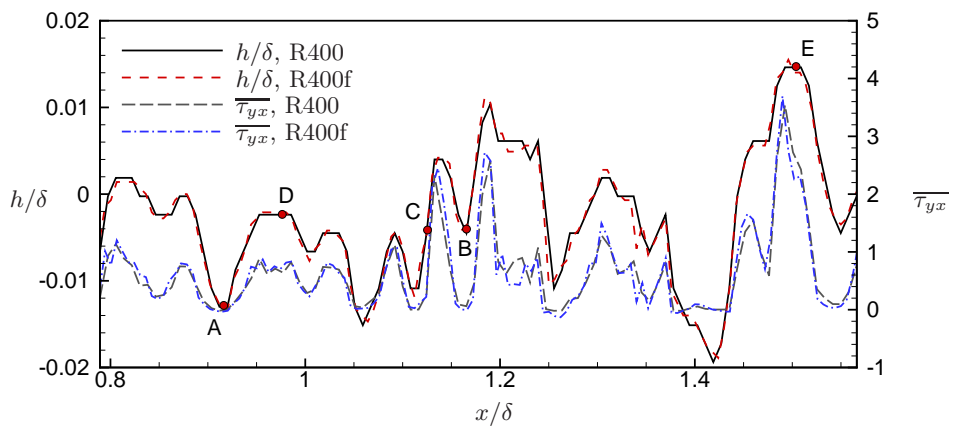}
\put(-190,0){$x/\delta$}
\put(-365,80){$h/\delta$}
\put(-25,80){$\overline{\tau_{yx}}$}
\put(-280,135){$h/\delta$, R400}
\put(-280,125){$h/\delta$, R400f}
\put(-280,115){$\overline{\tau_{yx}}$, R400}
\put(-280,105){$\overline{\tau_{yx}}$, R400f}
\caption{Roughness height and  time-averaged streamwise wall-shear stress for a portion of the rough surface. The red circles denote  locations A-E where wall-shear stress is probed.} 
\label{fig:tauyx_grid_ind}
\end{figure}

 $\overline{\tau_{yx}}$ and the roughness height for a portion of the surface are shown in figure \ref{fig:tauyx_grid_ind}. The results of Case R400f are  compared to Case R400 and found to be in acceptable agreement. 
 Figure \ref{fig:tauyx_grid_ind} shows that the variation of $\overline{\tau_{yx}}$ in the rough case is highly correlated with the roughness height. $\overline{\tau_{yx}}$ has large positive values at the roughness protrusions, and is nearly zero or has small negative values in the valleys.

\begin{table}
 \begin{center}
 \def~{\hphantom{0}}
\begin{tabular}{lccccc}
Point & Location  &  $y^+$  & $\overline{\tau_{yx}}$ & $\tau_{yx,rms}$ \\
\hline
A    &   trough  &  -5.2   &  -0.023   &  0.051   \\
B    &   trough  &  -1.8   &   0.078   &  0.048   \\
C    &   ascent  &  -1.8   &   0.349   &  0.136  \\
D    &   crest   &  -1.0   &   0.953   &  0.475   \\
E    &   crest   &   5.8   &   3.849   &  1.673   \\
\hline
\end{tabular}
 \caption{Locations, time-averaged and rms values of $\tau_{yx}$ for the probed points in figure \ref{fig:tauyx_grid_ind}. }
\label{tab:tauyx_point}
\end{center}
\end{table}

Five different locations (see table \ref{tab:tauyx_point}) on the roughness profile in figure \ref{fig:tauyx_grid_ind} are probed to show the dependence of wall shear on the surface topography. 
Point $A$ is located at a trough, where $\overline{\tau_{yx}}$ is negative, due to the reverse flow that occurs in the trough. Points $B$, $C$, and $D$ have a similar $y^+$. However,  $B$ is at a trough where $\overline{\tau_{yx}}$ is close to zero. Point $C$ is located at the ascent of the protrusion, which shows a higher $\overline{\tau_{yx}}$ and $\tau_{yx,rms}$. In contrast to point $B$ and $C$, point $D$ is located at a crest,  leading to even larger mean and rms of wall-shear stress. The wall-shear stress at the crest for an even higher $y^+$ is examined at point $E$. The mean and rms values are much higher than other probed locations. 


\subsubsection{p.d.f of wall-shear stress}


\begin{table}
 \begin{center}
 \def~{\hphantom{0}}
\begin{tabular}{lcccccccccccc}
Case & region & $Re_{\tau}$ & $\mu(\tau_{yx})$ & $\sigma(\tau_{yx})$ & $Sk(\tau_{yx}^{'})$ & $Ku(\tau_{yx}^{'})$ & $\mu(\tau_{yz})$ & $\sigma(\tau_{yz})$ & $Sk(\tau_{yz}^{'})$ & $Ku(\tau_{yz}^{'})$ & $\sigma(\phi_{\tau})$ \\
\hline
SW    &  -   & 400 & 1.09 & 0.40 & 0.97 & 4.50 & 1.45e-2 & 0.25 & -0.18 & 8.79 & 14.54 \\
R400 & peak   & 400  & 1.34 & 1.25 & 2.21 & 11.00 & 3.95e-3 & 0.54 & -1.28e-2 & 10.09 & 32.52\\
R400 & valley  & 400  & 0.21 & 0.37 & 2.22 & 13.09 & -7.11e-3 & 0.29 & -0.17 & 14.72 & 80.65  \\
R600 & peak   & 600  & 1.42 & 1.43 & 2.08 & 9.85 & -7.23e-4 & 0.71 & -1.82e-2 & 11.48 & 41.94\\
R600 & valley  & 600  & 0.20 & 0.50 & 2.49 & 16.66 & -1.42e-2 & 0.40 & -0.40 & 19.47 & 93.78  \\
\hline
\end{tabular}
 \caption{Statistics of the wall-shear stress components $\tau_{yx}$, $\tau_{yz}$ and yaw angle $\phi_{\tau}$: mean $\mu(\cdot)$, standard deviation $\sigma(\cdot)$, skewness $Sk(\cdot)$, kurtosis $Ku(\cdot)$.}
\label{tab:stats_pdf}
\end{center}
\end{table}

\begin{figure}
 \includegraphics[height=40mm]{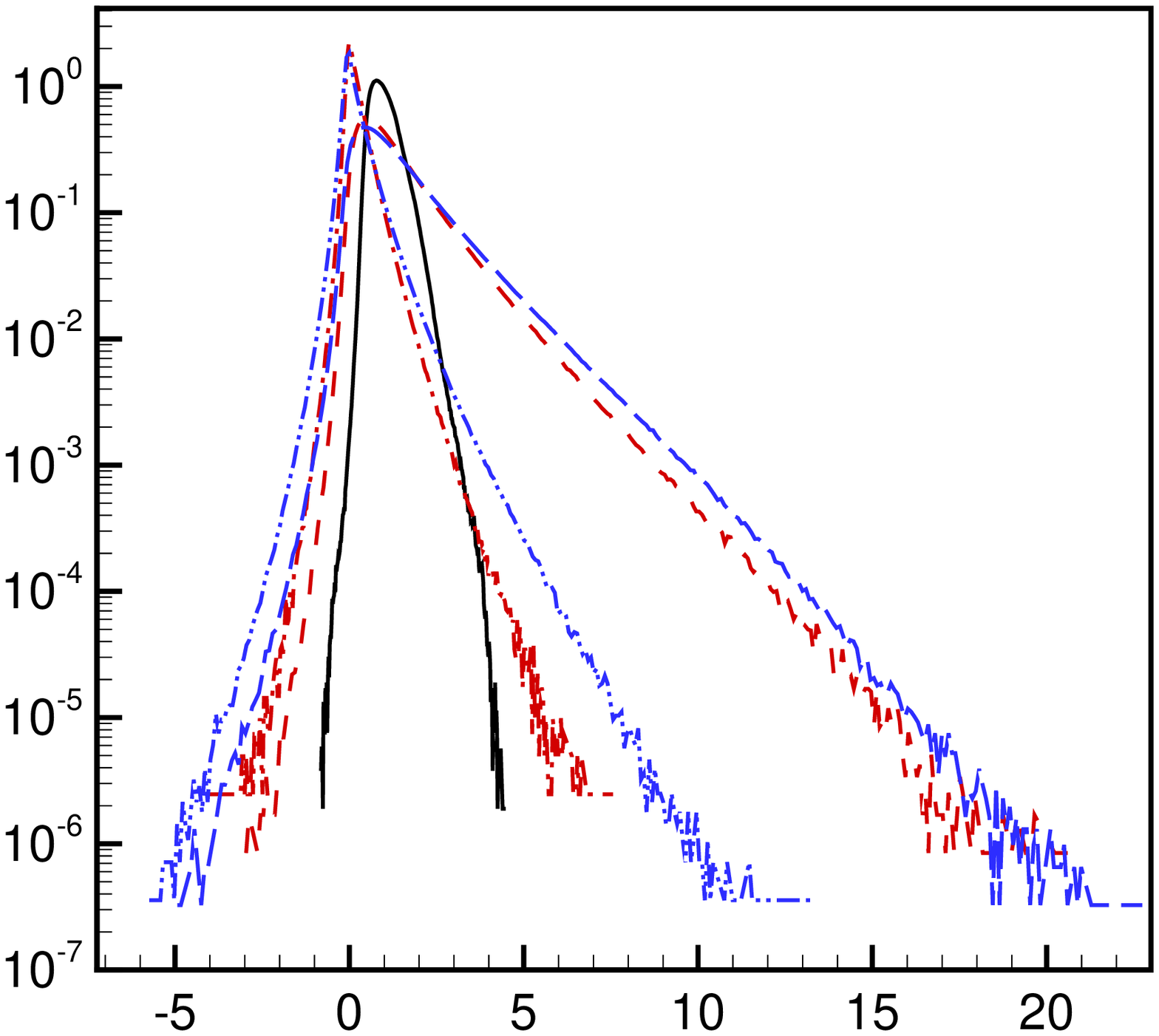}
 \put(-72,-3){$\tau_{yx}$}
 \put(-132,43){\rotatebox{90}{$P(\tau_{yx})$}}
 \put(-70,110){$(a)$}
 \includegraphics[height=40mm]{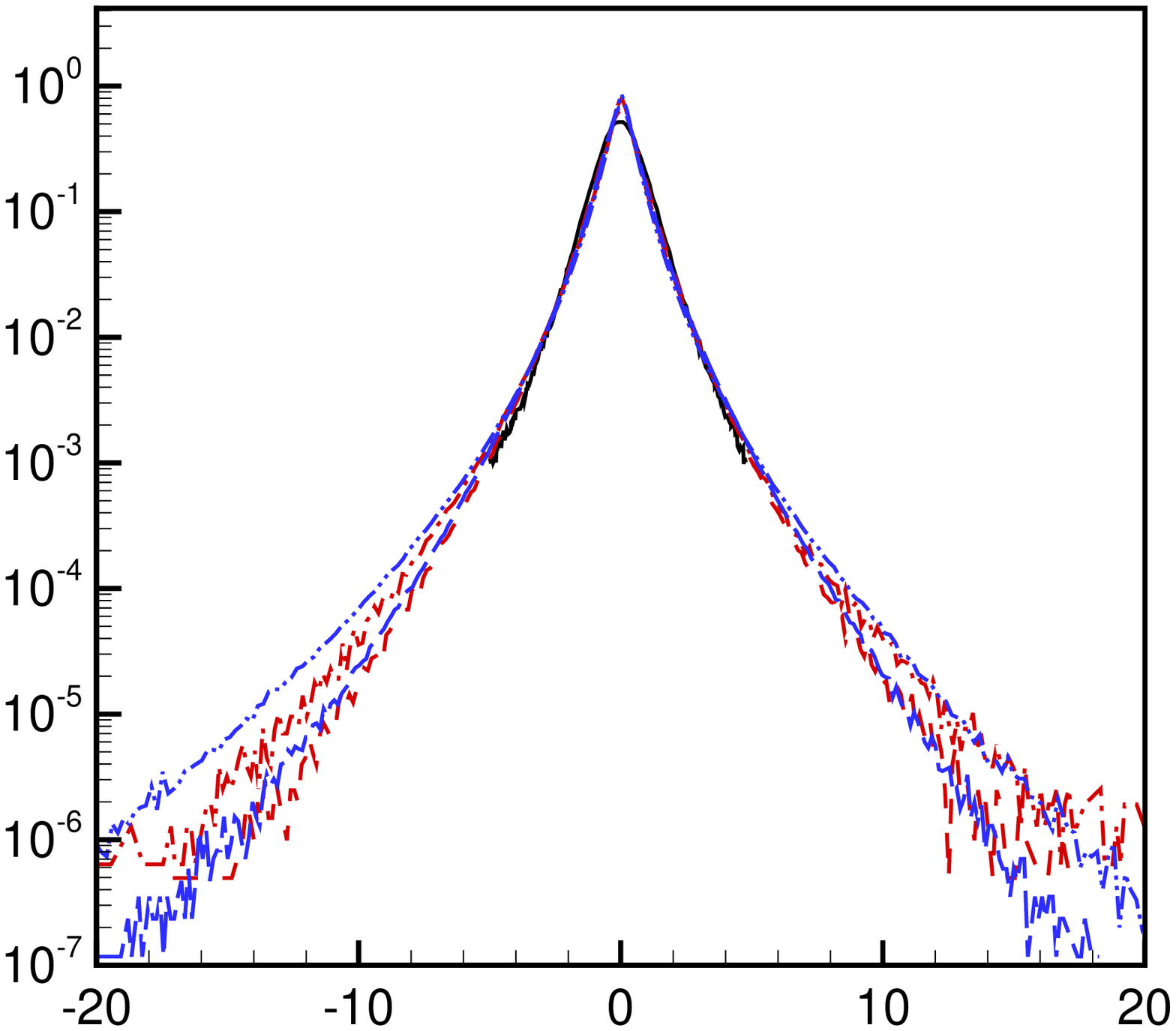}
 \put(-80,-3){$\tau_{yz}'/\tau_{yz,rms}$}
 \put(-132,33){\rotatebox{90}{$P(\tau_{yz}'/\tau_{yz,rms})$}}
 \put(-70,110){$(b)$}
\includegraphics[height=40mm]{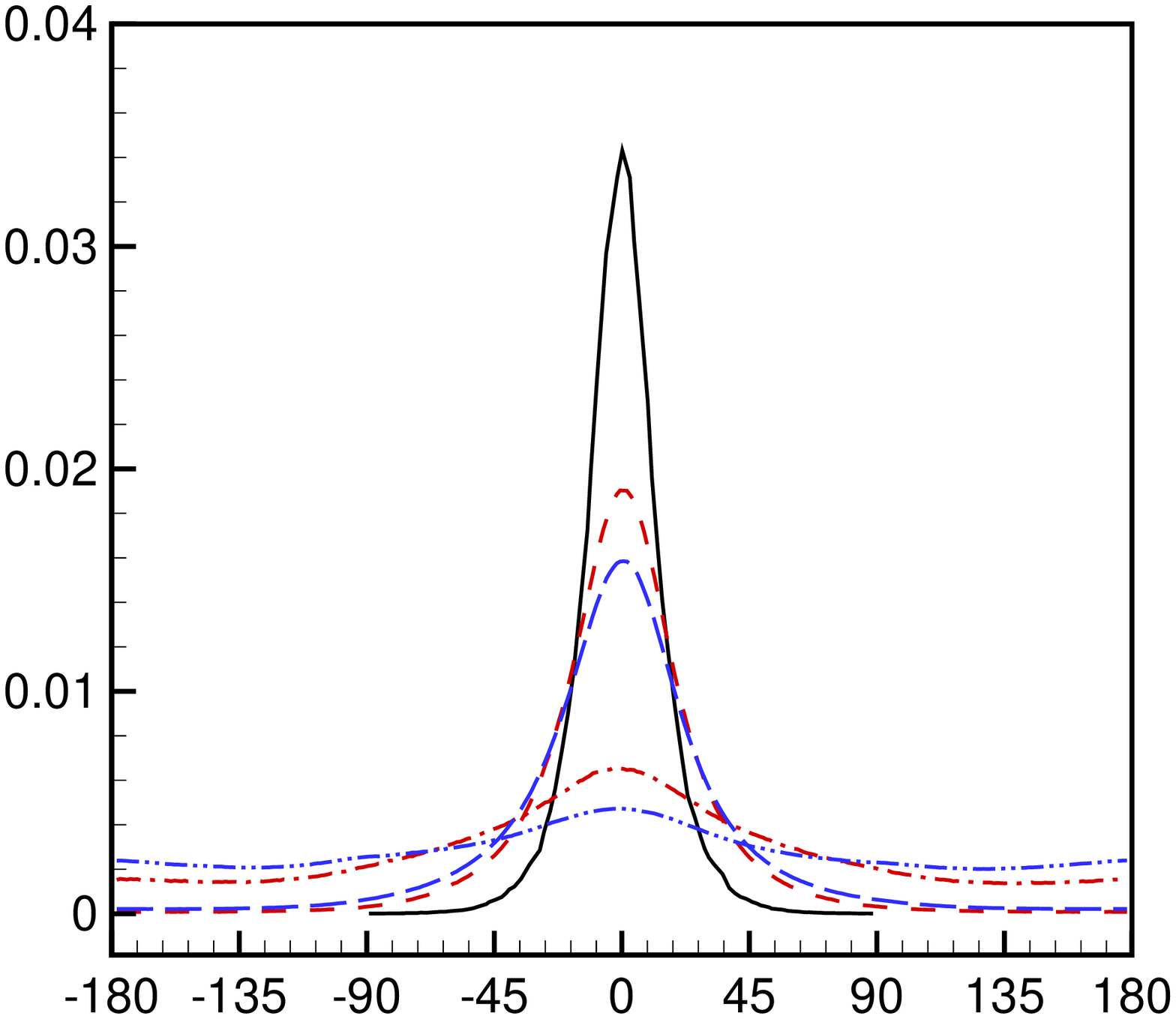}
\put(-68,-3){$\phi_{\tau}$}
\put(-132,43){\rotatebox{90}{$P(\phi_{\tau})$}}
\put(-70,110){$(c)$}
\caption{p.d.f of $(a)$ streamwise,  $(b)$  spanwise wall-shear stress  and $(c)$  angle between  shear-stress vector and  streamwise direction. Case SW (solid black), R400 peak (dashed red), R400 valley (dash dot red), R600 peak (long dash blue) and R600 valley (dash dot dot blue). }
\label{fig:pdf_tauyx}
\end{figure}

\begin{figure}
\begin{center}
\includegraphics[height=25mm,trim={0cm 0cm 0cm 0cm},clip]{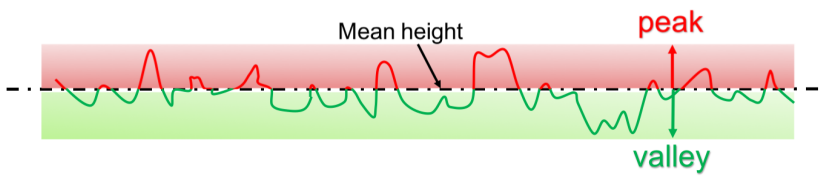}
\hspace{10mm}
\end{center}
\caption{
 Illustration of how wall shear stress is divided into peak regions (above the mean height location) and valley regions (below the mean height location).}
\label{fig:sketch}
\end{figure}

The effect of roughness on the p.d.f of wall shear stress is examined. The p.d.f of the streamwise shear stress component $\tau_{yx}$, spanwise shear stress component $\tau_{yz}$, and the shear-stress yaw angle $\phi_{\tau} = \tan^{-1}(\tau_{yz}(t)/\tau_{yx}(t))$ is shown in figure \ref{fig:pdf_tauyx}, and tabulated in table \ref{tab:stats_pdf}. 
Case SW is the baseline, whose statistics of wall shear-stress fluctuations show good agreement with the results for a smooth-wall turbulent boundary layer \citep{diaz2017wall}. Motivated by the influence of topography in \S \ref{sec:localshearvar}, the shear stress is probed separately in two regions - the `peak region' (above the mean height location), and the `valley region' (below the mean height location). 
A schematic of the decomposition is shown in figure \ref{fig:sketch}.  

\begin{figure}
\includegraphics[height=50mm,trim={1cm 0cm 0cm 0cm},clip]{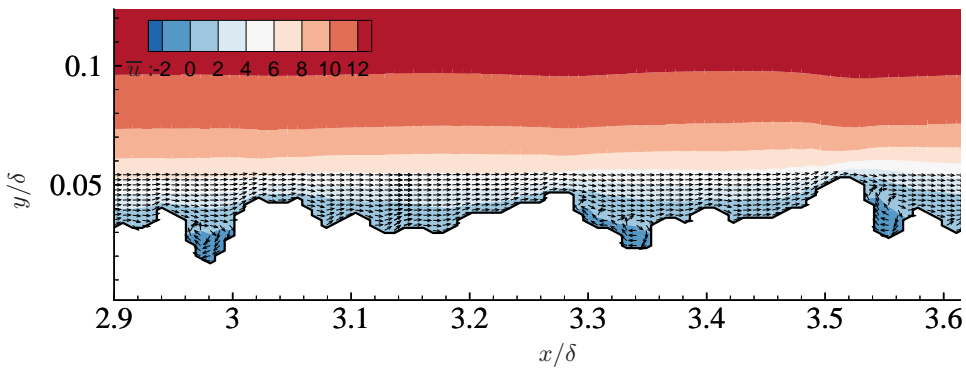}
\put(-410,50){\rotatebox{90}{$y/\delta$}}
\put(-210,-7){$x/\delta$}
\put(-365,100){$\overline{u}:$}
\caption{Contours of time-averaged streamwise velocity and vectors of ($\overline{u}$,$\overline{v}$) in the vicinity of the rough surface for Case R400.}
\label{fig:reversal}
\end{figure}

The rms of the streamwise shear stress for Case SW (table \ref{tab:stats_pdf}) is in good agreement with the correlation $\tau_{yx,rms}^{+}=\tau_{yx,rms}/\tau_{w}=0.298+0.018\ln Re_{\tau}$, proposed by \cite{orlu2011fluctuating}. 
Comparison of the rough walls to Case SW at the same $Re_{\tau}$ reveals the following features. A significant increase in the probability of events with positive $\tau_{yx}$ is observed for the peak region  in figure \ref{fig:pdf_tauyx}$(a)$   since $\tau_{yx}$ is associated with the streamwise velocity which generally increases as $y$ increases. The probability of events with negative $\tau_{yx}$ is higher in the valleys, corresponding to  figure \ref{fig:reversal} that shows how reverse flows occur mainly in the valleys. The p.d.f for rough surfaces has longer tails (larger kurtosis), since intermittent events in the flow can interact randomly with peaks or valleys in the rough surface, resulting in more extreme events. 

The p.d.f of spanwise wall-shear stress fluctuations $\tau_{yz}$, shown in figure \ref{fig:pdf_tauyx}$(b)$, has zero mean and zero skewness for both smooth and rough cases. The normalization by the rms values collapses the peak of the p.d.f, however, the tails are longer for the rough walls. This behavior is more pronounced at the higher $Re_\tau$.

The probability distribution of $\phi_{\tau}$ in figure \ref{fig:pdf_tauyx}$(c)$ examines the correlation between $\tau_{yx}$ and $\tau_{yz}$. \citet{jeon1999space} found that the probability for events with $|\phi_{\tau}|>45\degree$ is very small in smooth channel flows; i.e. the flow is dominated by events where  large positive $\tau_{yx}$ is associated with small $\tau_{yz}$. The standard deviation of $\phi_{\tau}$ is much higher for the rough wall, compared to Case SW. The probability of events with $|\phi_{\tau}| >45\degree$ is enhanced by the roughness, suggesting that the events with comparable magnitudes of $\tau_{yx}$ and $\tau_{yz}$ occur more frequently. This behavior can be explained by the three-dimensionalization of the vortical motions in the roughness layer. The roughness elements break up the directional bias of the near-wall streaks and generate more isotropic vorticity fields \citep{mehdi2010mean, aghaei2019turbulence}. In the valley region, events with $90\degree < |\phi_{\tau}| <180\degree$ (corresponding to negative $\tau_{yx}$) are more probable, consistent with reverse flow in the valleys. The p.d.f of $|\phi_{\tau}|$ tends to be more evenly distributed for Case R600, implying that the reverse flow is  enhanced as $Re_{\tau}$ increases. \\




    
 
\section{Summary}\label{sec:conclusions}
DNS of turbulent channel flow with an irregular rough wall is performed at $Re_{\tau}=400$ and $600$. The rough wall is generated from  experimental line scans \citep{flack2019skin}. Statistics, spectra and p.d.f of the computational surface are ensured to be in good agreement with the experiment. No-slip Dirichlet boundary conditions are imposed on the rough surface in the simulations, and grid convergence studies are conducted to establish  that the flow is adequately resolved. The roughness height has Gaussian p.d.f, and $k_s^+$ is equal to 6.4 and 9.6 for $Re_\tau=400$ and $600$ respectively. The roughness height is smaller than most past rough-wall studies and the flow is in the transitionally rough regime.

The computed skin friction coefficients and roughness functions agree with those measured by \cite{flack2019skin}.
When scaled with the global friction velocity, the streamwise fluctuations decrease from their smooth wall values, while wall--normal and spanwise fluctuations, and the Reynolds shear stress increase in magnitude at the rough wall. Scaling using the local friction velocity reveals that the peak value of streamwise velocity fluctuations  decreases while the spanwise velocity fluctuations increase. The wall--normal velocity fluctuations and Reynolds shear stress show relatively weaker variation. The near-wall motions and momentum transfer due to the roughness are studied using double-averaged statistics.  "Ejection" and "roll-up" motions on the roughness crests are found to alter the wall-normal momentum transfer. Form drag is produced mainly by the roughness protrusions, and the negative and positive $\Tilde p$ contributing to the form drag are correlated with negative $\Tilde u$ and positive $\Tilde v$ respectively. The mean momentum is transferred downward by the viscous force and turbulence inertia, to balance the total drag induced by the roughness. 

Roughness effects on pressure fluctuations are investigated. Pressure fluctuation amplitudes  increase in the roughness layer, and collapse with the smooth wall levels away from the wall. Their spatial variation suggests that the higher pressure fluctuations are due to the unsteady stagnation and wake regions induced by the roughness elements. As $Re_{\tau}$ increases, roughness effects on pressure fluctuations are enhanced. The dominant source terms in the pressure Poisson equation, $MS_{12}$ and $TT_{23}$, are examined. Roughness causes the mean shear to peak within the roughness layer, while also increasing $\partial v^\prime/\partial x$, both of which combine to increase $MS_{12}$ within the roughness layer. The local variation of $MS_{12}$ suggests  the mean shear layers on the roughness protrusions as the primary source of increased pressure fluctuations. $TT_{23}$ is enhanced in the troughs and upstream of the protrusions, and is associated with the streamwise vortices at these locations. 

The effects of roughness on wall-shear stress are examined. The time-average and rms of $\tau_{yx}$ is positively correlated with the roughness height, and highly dependent on the local roughness topography. The p.d.fs of wall-shear stress fluctuations are examined for the influence of topography.  The probability of events with positive $\tau_{yx}$ is found to be higher in the peak regions since $\tau_{yx}$ is associated with the streamwise velocity which generally increases as $y$ increases. The probability of events with negative $\tau_{yx}$ is higher in the valley regions, since reverse flow occurs mainly in the valleys. Surface roughness increases the probability of extreme events since intermittent events in the flow can interact randomly with peaks or valleys in the rough surface.  Events with comparable magnitudes of $\tau_{yx}$ and $\tau_{yz}$ occur more frequently since the roughness elements break up the directional bias of near-wall streaks. These effects are more pronounced at the higher $Re_\tau$.



\section*{Acknowledgements}
This work was supported by the United States Office of Naval Research (ONR) Grant N00014-17-1-2308 managed by Drs. J. Gorski, T. Fu and P. Chang. Computing resources were provided by the Minnesota Supercomputing Institute (MSI). We are grateful to Professor K. Flack at the United States Naval Academy for providing us with the scanned surface data used in the present work. The authors would like to thank Mr. S. Anantharamu, Dr. P. Kumar and Dr. Y. Li for their helpful discussions and suggestions.  

\section*{Declaration of Interests}
The authors report no conflict of interest.

\bibliographystyle{jfm}
\bibliography{jfm-instructions}

\end{document}